\documentclass[twocolumn]{aastex63}
\usepackage{verbatim}
\usepackage{CJKutf8}

\shorttitle{V892 Tau}
\shortauthors{Long et al.}

\graphicspath{{./}{figures/}}
\begin{document}
\begin{CJK*}{UTF8}{gbsn}

\title{The Architecture of the V892 Tau System: the Binary and its Circumbinary Disk}

\correspondingauthor{Feng Long}
\email{feng.long@cfa.harvard.edu}

\author[0000-0002-7607-719X]{Feng Long(龙凤)}
\affiliation{Center for Astrophysics \textbar\, Harvard \& Smithsonian, 60 Garden St., Cambridge, MA 02138, USA}

\author[0000-0003-2253-2270]{Sean M. Andrews}
\affiliation{Center for Astrophysics \textbar\, Harvard \& Smithsonian, 60 Garden St., Cambridge, MA 02138, USA}

\author[0000-0003-1481-8076]{Justin Vega}
\affiliation{Northeastern University, 360 Huntington Ave, Boston, MA 02115, USA}

\author[0000-0003-1526-7587]{David J. Wilner}
\affiliation{Center for Astrophysics \textbar\, Harvard \& Smithsonian, 60 Garden St., Cambridge, MA 02138, USA}

\author{Claire J. Chandler}
\affiliation{National Radio Astronomy Observatory, P.O. Box O, Socorro, NM 87801, USA}

\author[0000-0001-5378-7749]{Enrico Ragusa}
\affiliation{School of Physics and Astronomy, University of Leicester, Leicester LE1 7RH, UK}

\author[0000-0003-1534-5186]{Richard Teague}
\affiliation{Center for Astrophysics \textbar\, Harvard \& Smithsonian, 60 Garden St., Cambridge, MA 02138, USA}

\author[0000-0002-1199-9564]{Laura M. P\'erez} 
\affiliation{Departamento de Astronom\'ia, Universidad de Chile, Camino El Observatorio 1515, Las Condes, Santiago, Chile}

\author[0000-0002-3950-5386]{Nuria Calvet}
\affiliation{Department of Astronomy, University of Michigan, 314 West Hall, 1085 S. University, Ann Arbor, MI 48109, USA}

\author[0000-0003-2251-0602]{John M. Carpenter}
\affiliation{Joint ALMA Observatory, Avenida Alonso de C\'ordova 3107, Vitacura, Santiago, Chile}

\author{Thomas Henning}
\affiliation{Max-Planck-Institut f\"{u}r Astronomie, K\"{o}nigstuhl 17, 69117, Heidelberg, Germany}

\author[0000-0003-4022-4132]{Woojin Kwon}
\affiliation{Department of Earth Science Education, Seoul National University, 1 Gwanak-ro, Gwanak-gu, Seoul 08826, Republic of Korea}
\affiliation{SNU Astronomy Research Center, Seoul National University, 1 Gwanak-ro, Gwanak-gu, Seoul 08826, Republic of Korea}

\author[0000-0002-8115-8437]{Hendrik Linz}
\affiliation{Max-Planck-Institut f\"{u}r Astronomie, K\"{o}nigstuhl 17, 69117, Heidelberg, Germany}

\author{Luca Ricci}
\affiliation{Department of Physics and Astronomy, California State University Northridge, 18111 Nordhoff Street, Northridge, CA 91330, USA}

\begin{abstract}
%The evolution of circumstellar disks in a multiple stellar system can be significantly affected by the binary and disk interactions, which have direct consequences on the formation and survival of the associated planetary system. 
We present high resolution millimeter continuum and CO line observations for the circumbinary disk around V892 Tau to constrain the stellar and disk properties.  The total mass of the two near-equal-mass A stars is estimated to be $6.0\pm0.2\,M_{\odot}$ based on our models of the Keplerian-dominated gas disk rotation. The detection of strong ionized gas emission associated with the two stars at 8\,mm, when combined with previous astrometric measurements in the near-infrared, provides an updated view of the binary orbit with $a=7.1\pm0.1$\,au, $e=0.27\pm0.1$, and $P=7.7\pm0.2$\,yr, which is about half of a previously reported orbital period.
The binary orbital plane is proposed to be near coplanar to the circumbinary disk plane (with a mutual inclination of only $\Delta=8\pm4.2\degr$; another solution with $\Delta=113\degr$ is less likely given the short re-alignment timescale). An asymmetric dust disk ring peaking at a radius of 0$\farcs$2 is detected at 1.3\,mm and its fainter counterparts are also detected at the longer 8 and 9.8\,mm. The CO gas disk, though dominated by Keplerian rotation, presents a mild inner and outer disk misalignment. %, such that the inner disk to the SW and outer disk to the NE appear brighter than their counterparts at the opposite disk sides. 
The radial extension of the disk, its asymmetric dust ring, and the presence of a disk warp could all be explained by the interaction between the eccentric binary and the circumbinary disk, which we assume were formed with non-zero mutual inclination. 
Some tentatively detected gas spirals in the outer disk are likely produced by interactions with the low mass tertiary component located 4\arcsec\ to the northeast.
Our analyses demonstrate the promising usage of V892 Tau as an excellent benchmark system to study the details of binary--disk interactions.

\end{abstract}


\section{Introduction} \label{sec:intro}
%stellar multiplicity -- planet formation in multiple stellar systems 
Multiple stellar systems are a common outcome of the star formation process. More than half of Sun-like stars in the field are members of binary or higher-order multiple systems \citep{Duquennoy1991, Raghavan2010}. This multiplicity fraction is likely even higher at earlier evolutionary stages and around more massive stars \citep{Duchene2013, Tobin2016Natur}.  The dynamical interplay between stars and their circumstellar disks can dramatically influence the distribution and evolution of disk materials, and consequently the assembly of planets. Given the prevalence of stellar multiplicity, a comprehensive understanding of the star and planet formation process therefore largely depends on these dynamical interactions, for which observational constraints are still limited.

Depending on the stellar configuration, multiple disks could be present in a binary system, including a circumbinary disk orbiting around the binary and disks surrounding individual stellar components. Dynamical interactions between stars and disks tend to truncate the disks (e.g., creating an inner cavity in a circumbinary disk), reduce their masses, and shorten their lifetimes \citep[e.g.,][]{Artymowicz1994, Breslau2014, Hirsh2020}. For example, disks in close binaries (separation $<30-40$\,au) have fainter millimeter emission \citep{Jensen1996, Harris2012, Akeson2019}, with a lower frequency presenting accretion signatures and/or near-infrared excess emission \citep{Bouwman2006, Cieza2009, Kraus2012},  compared to disks around single stars. This indicates a rapid disk dispersal 
%both locally (in the inner disk) and globally 
in multiple systems. For systems with even shorter separations ($<10$\,au), formation of such binaries is inhibited through direct fragmentation by the opacity limit and requires subsequent dynamical evolution involving orbital decay and/or exchange interactions \citep{Artymowicz1983, Bate2002, Tokovinin2020}. All these processes can significantly affect the presence and evolution of circumstellar disks, thus providing challenging conditions for planets to form and survive \citep{Rafikov2015, Moe2019}.

Indeed, only a small fraction of currently known exoplanets resides in multiple systems \citep{Schwarz2016,Martin2018haex.bookE.156M}\footnote{The catalogue of exoplanets in binary star systems can be found in \url{https://www.univie.ac.at/adg/schwarz/multiple.html}}, although this rate may be associated with detection and survey design biases. Both ``S-type" (orbiting around a single star) and ``P-type" (orbiting around the binary) planets have been discovered. There is a severe paucity of ``S-type" planets in close binaries ($<50$\,au) compared to single stars \citep{Wang2014,Kraus2016}, consistent with the view that less disk material is available in such systems. Such effects on wider pairs are minimal. For ``P-type" planets in the \textit{Kepler} sample of eclipsing binaries, the giant planet occurrence rate is comparable to that of single stars \citep{Armstrong2014}. This finding is also in line with disk observations that some circumbinary disks have masses and sizes comparable to disks around single stars \citep{Harris2012, Long2019}.

The fate of individual disks and their potential in nurturing planets in a multiple stellar system depend on the binary properties, including stellar mass ratio ($q$), orbital semi-major axis ($a$), and eccentricity ($e$), as well as the mutual inclination between the binary plane and the disk plane ($\Delta$). Besides tidal truncation, disk precession, warping, or breaking could occur under certain conditions \citep[e.g.,][]{Facchini2013, Dogan2018}.
A quantitative test on the theory of star-disk interactions would therefore require the determinations of these relevant parameters. 
Most previous statistical studies have so far focused on the effects of $q$ and $a$, because of their availability in large samples (e.g., \citealt{Harris2012, Manara2019}). The measurement of $e$ requires long-term monitoring of the binary, and the constraint on $\Delta$ needs additional high resolution disk observations for a precise determination of the disk geometry. Such measurements are rare (e.g., HD 142527, \citealt{Price2018}; GW Ori, \citealt{Kraus2020}).

%source properties
The young Herbig Ae stellar system V892 Tau (also named Elias 1, \citealt{Elias1978}), serves as an interesting observational laboratory to study such dynamical interactions. 
The system, located in the L1495 dark cloud of the nearby Taurus star-forming region, consists of two nearly equal mass stars, with projected separations of 0.04--0.06$''$ \citep{Smith2005, Monnier2008}. The spectral type has some ambiguities in the literature, including classifications as B8, A0, and A6, with $T_{\rm eff}$ from 8350 to 11900\,K, associated with very high optical extinction of $A_V$=4--10\,mag \citep[e.g.,][]{Cohen1979, Hillenbrand1992, Hernandez2004, Herczeg2014}. The massive circumbinary disk surrounding the two stars has been extensively targeted in observations from the infrared to centimeter wavelengths \citep[e.g.,][]{Beckwith1991, Skinner1993, Henning1998, Andrews2005, Furlan2006, Hamidouche2010}. A near-infrared dip in the spectral energy distribution suggests an inner dust cavity, which
was first confirmed and resolved with mid-infrared imaging using Keck observations at 10.7\,$\mu$m \citep{Monnier2008} and later with high resolution ALMA observations at 1.3\,mm \citep{Pinilla2018}. A low mass stellar companion is found $\sim$4$''$ away to the northeast at PA=22$\degr$, with a recent spectral type assignment of M3 as a Taurus member (V892 Tau NE,  \citealt{Skinner1993, Leinert1997, Esplin2019}). 
Parallax measurements from the \textit{Gaia} early DR3 catalog \citep{Gaia2020} provide distances of $134.5\pm2$\,pc\footnote{This is further away than the distance measurement from \textit{Gaia} DR2 at 117$\pm$2\,pc. The new distance is also more consistent with the average value of the nearby Taurus members.} and  $131.3\pm5$\,pc for V892 Tau and V892 Tau NE, respectively, which further suggests that the low mass star V892 Tau NE is associated with the close binary V892 Tau, constituting a triple stellar system. We adopt the new distance of 134.5\,pc for this triple system throughout the paper.

In this paper, we offer an updated view of the stellar and disk properties for V892 Tau system using high resolution ALMA and VLA observations, in which the ALMA CO data and VLA data are presented here for the first time. The observational setup and data reduction details are described in Section~\ref{sec:obs}. We present the properties of the dust and gas disk, the estimate of the total stellar mass from gas disk rotation, and new constraints on the inner binary orbit in Section~\ref{sec:results}. A comprehensive picture of stellar and disk architecture is then provided and discussed under binary--disk interactions in Section~\ref{sec:diss}. Finally, we summarize our findings in Section~\ref{sec:summ}.

\section{Observations} \label{sec:obs}

\subsection{ALMA Observations at Band 6}
V892 Tau was observed on 2015 September 29 as part of the ALMA program 2013.1.00498.S (PI. L. P{\'e}rez), using the Band 6 receivers. These observations included 30 antennas with projected baseline lengths between 36 and 1860\,m (27--1400\,k$\lambda$). Two spectral windows (SPWs) were configured to record the continuum data, centered at 217 and 232.4\,GHz, each with 128 channels spanning a bandwidth of 1.875\,GHz. The $^{12}$CO $J=2-1$ line (rest frequency at 230.538\,GHz) was targeted in a separate SPW, with a channel spacing of 244\,kHz (velocity spacing of 0.32\,km\,s$^{-1}$).  Two isotopologue lines, $^{13}$CO and C$^{18}$O $J=2-1$, were covered in the remaining SPW at a channel spacing of 488\,kHz (velocity spacing of 0.66\,km\,s$^{-1}$). The quasar J0510+180 was observed at the start of the observing program, serving as both bandpass and flux calibrator. V892 Tau was then observed interleaving with the gain calibrator J0429+2724, reaching a total on-source time of $\sim$8\,min. Only the continuum data were published previously in \citet{Pinilla2018}.

The raw visibility data downloaded from the ALMA archive were calibrated following the scripts provided by ALMA staff using the required \texttt{CASA} \citep{McMullin2007} version of 4.5.0. The bandpass shape and absolute amplitude scale were determined and applied to the science target using the observations of J0510+180. The phase variations were corrected using the repeated observations of J0429+2724. Given the bright continuum emission, we then performed a series of self-calibration iterations on the continuum visibility data (including the line-free channels in the line SPWs) using \texttt{CASA v.5.6.0}. Three rounds of phase-only self-calibration (stepping down the solution interval of 90, 60, and 30\,s) provide a peak SNR increase by a factor of $\sim$3, and one iteration of amplitude self-calibration increases the peak SNR again by 30\%. The calibration tables were then applied to the line channels.

The fully calibrated visibility data are used to generate images with the \texttt{tclean} task in \texttt{CASA}. The continuum image was produced using Briggs weighting with robust=0.5 to optimize the image SNR in the outer disk, which results in a synthesized beam with an FWHM of 0$\farcs$23 $\times$ 0$\farcs$16 and an rms noise level of 76\,$\mu$Jy\,beam$^{-1}$ at a central frequency of 224\,GHz (1.3\,mm). For spectral lines, the continuum baseline was first subtracted from the line visibilities using \texttt{uvcontsub} before making the image cubes. The $^{12}$CO images were created with Briggs weighting (robust=0.5) in a velocity resolution of 0.5\,km\,s$^{-1}$. For the weaker $^{13}$CO and C$^{18}$O emission, we generated the images with a Gaussian \textit{uv} taper and robust parameter of 2 in 1\,km\,s$^{-1}$ channels, resulting in a larger beam size of 0$\farcs$40 $\times$ 0$\farcs$34. The image details are summarized in Table~\ref{tab:obs-results}.

\subsection{VLA Observations at Ka Band}
V892 Tau was observed with the Karl G.~Jansky Very Large Array (VLA) using the Ka band receivers as part of the ``Disks@EVLA'' large program (project code AC982, PI.~C.~Chandler). Two 1\,GHz-wide basebands were tuned at central frequencies of 30.5\,GHz (9.8\,mm) and 37.5\,GHz (8.0\,mm). Data were collected in the A, B, and C configurations on 2011 July 15, 2011 March 19, and 2010 November 01, 
%October 28, 
respectively. These observations cover projected $uv$-spacings from 10 to %2733
4024\,k$\lambda$. Frequent monitoring of J0429+2724 was used to track the complex gain variations. The bright quasars 3C 84 and 3C 147 were observed as bandpass and flux calibrators, respectively. These VLA data are presented here for the first time.

The raw visibilities were calibrated in CASA using the pipeline\footnote{see \url{https://science.nrao.edu/facilities/vla/data-processing/pipeline/scripted-pipeline}} created by the ``Disks@EVLA'' team. Because the observations were performed spanning half a year, data from individual configurations were shifted to a common phase center before using \texttt{concat}, to account for the source proper motion. The continuum images were then generated with \texttt{tclean} using Briggs weighting with robust=0 to better resolve the inner emission.
%and compromise the detection of the faint dust disk. 
We have also created the 30.5 and 37.5\,GHz combined continuum image using the multiterm, multifrequency synthesis algorithm (\texttt{mtmfs}) with \texttt{nterms=2}. The image properties are summarized in Table~\ref{tab:obs-results}.

\begin{deluxetable*}{lccccc}
%[!t]
\tabletypesize{\scriptsize}
\tablecaption{Observational Results\label{tab:obs-results}}
\tablewidth{0pt}
\tablehead{
\colhead{Line/Cont} & \colhead{Freq.} & \colhead{Beam (PA)} & \colhead{Disk-integrated Flux} & \colhead{image rms} & \colhead{$R_{90\%}$}  \\ 
\colhead{} & \colhead{[GHz]} & \colhead{} & \colhead{[Jy km/s] or [mJy]} & \colhead{[mJy/beam]} & \colhead{[arcsec]} \\
} 
\colnumbers
\startdata
$^{12}$CO $J=2-1$ & 230.538 & $0\farcs23\times0\farcs16(1.8\degr)$ & 14.7$\pm$0.4\tablenotemark{a} & 6.4 (0.5\,km/s channel) & 1.45$\pm$0.02 \\ 
$^{13}$CO $J=2-1$ & 220.399 & $0\farcs40\times0\farcs34(1.8\degr)$ & 3.3$\pm$0.3 & 6.3 (1\,km/s channel) & 0.97$\pm$0.03 \\ 
C$^{18}$O $J=2-1$ & 219.560 & $0\farcs40\times0\farcs34(-0.4\degr)$ & 1.1$\pm$0.1 & 5.0 (1\,km/s channel) & 0.68$\pm$0.03 \\ 
\hline
continuum-disk & 224 (1.3\,mm) & $0\farcs23\times0\farcs16(1.8\degr)$ & 290.6$\pm$0.2 & 0.076 & 0.46$\pm$0.01 \\ 
%continuum-point & 224 (1.3\,mm) & ... & 4.07$\pm$0.1 & ... & ... \\ 
continuum-disk & 37.5 (8.0\,mm) & $0\farcs08\times0\farcs05(-60.7\degr)$ & 3.1$\pm$0.4\tablenotemark{b} & 0.035 & .. \\ 
continuum-disk & 30.5 (9.8\,mm) & $0\farcs09\times0\farcs06(-70.2\degr)$ & 1.8$\pm$0.3\tablenotemark{b} & 0.025 & .. \\
\enddata
\tablenotetext{a}{Disk integrated line fluxes are measured with the Keplerian mask that encompasses all emission in a velocity range from $-$7 to 23\,km\,s$^{-1}$,  excluding the cloud contaminated velocity range from 6.5 to 8\,km\,s$^{-1}$. The reported line fluxes are therefore lower limits, due to the exclusion of these central channels. Line flux uncertainty is estimated as the rms noise level in the integrated intensity map generated without a sigma clip, multiplied by the square root of beam numbers in a mask size with an approximate radius of 2 arcsec (rms * $\sqrt{\rm Mask\,area / beam\,area}$).}
\tablenotetext{b}{The disk emission fluxes at 8 and 9.8\,mm are estimated by subtracting the two point sources as modeled in Section~\ref{sec:orbit} from the total flux (4.3 and 3.0 \,mJy, respectively) measured in the images. The uncertainty includes a $10\%$ systematic flux calibration error.}
\end{deluxetable*}

\begin{figure*}[!t]
\centering
    \includegraphics[width=0.32\textwidth]{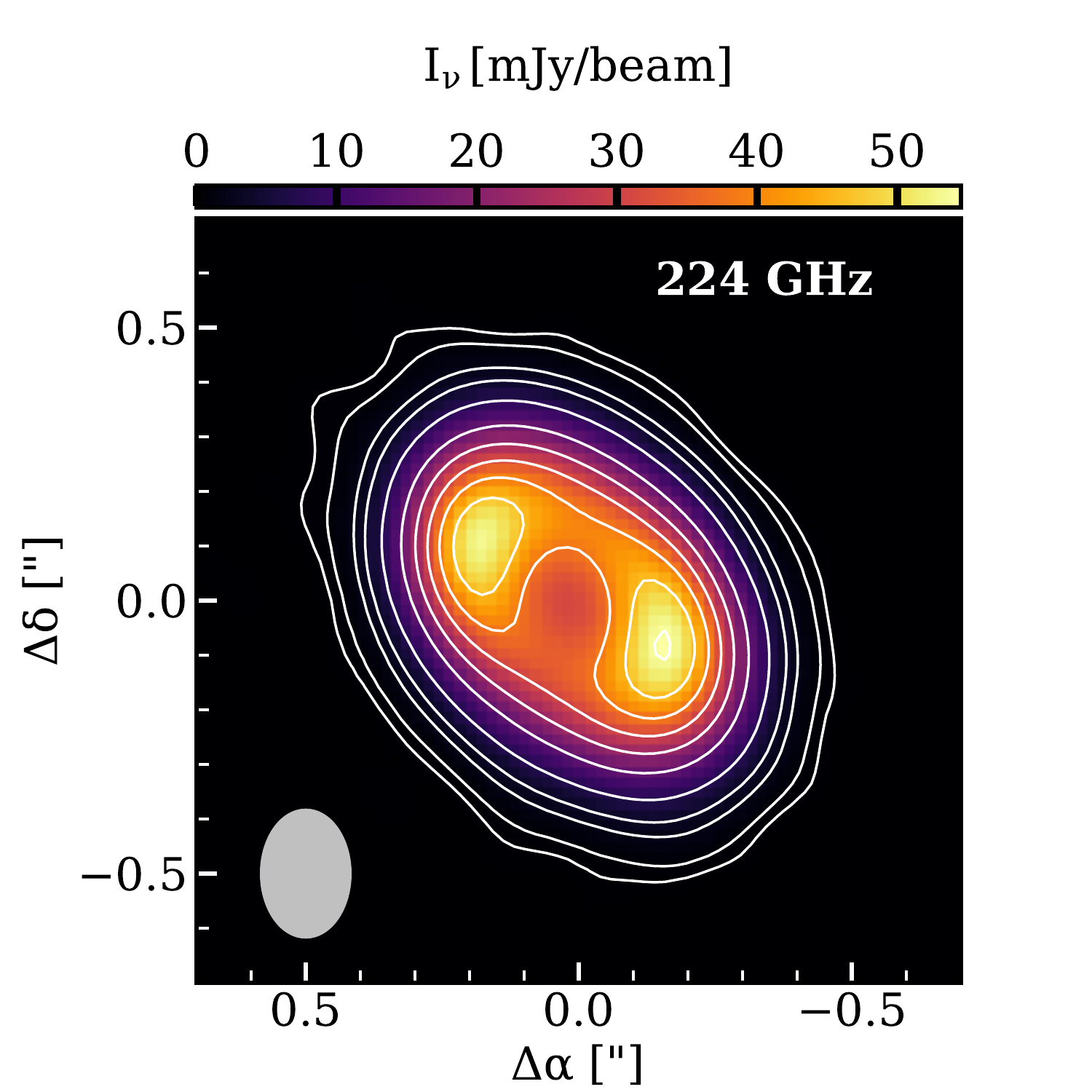} 
    \includegraphics[width=0.32\textwidth]{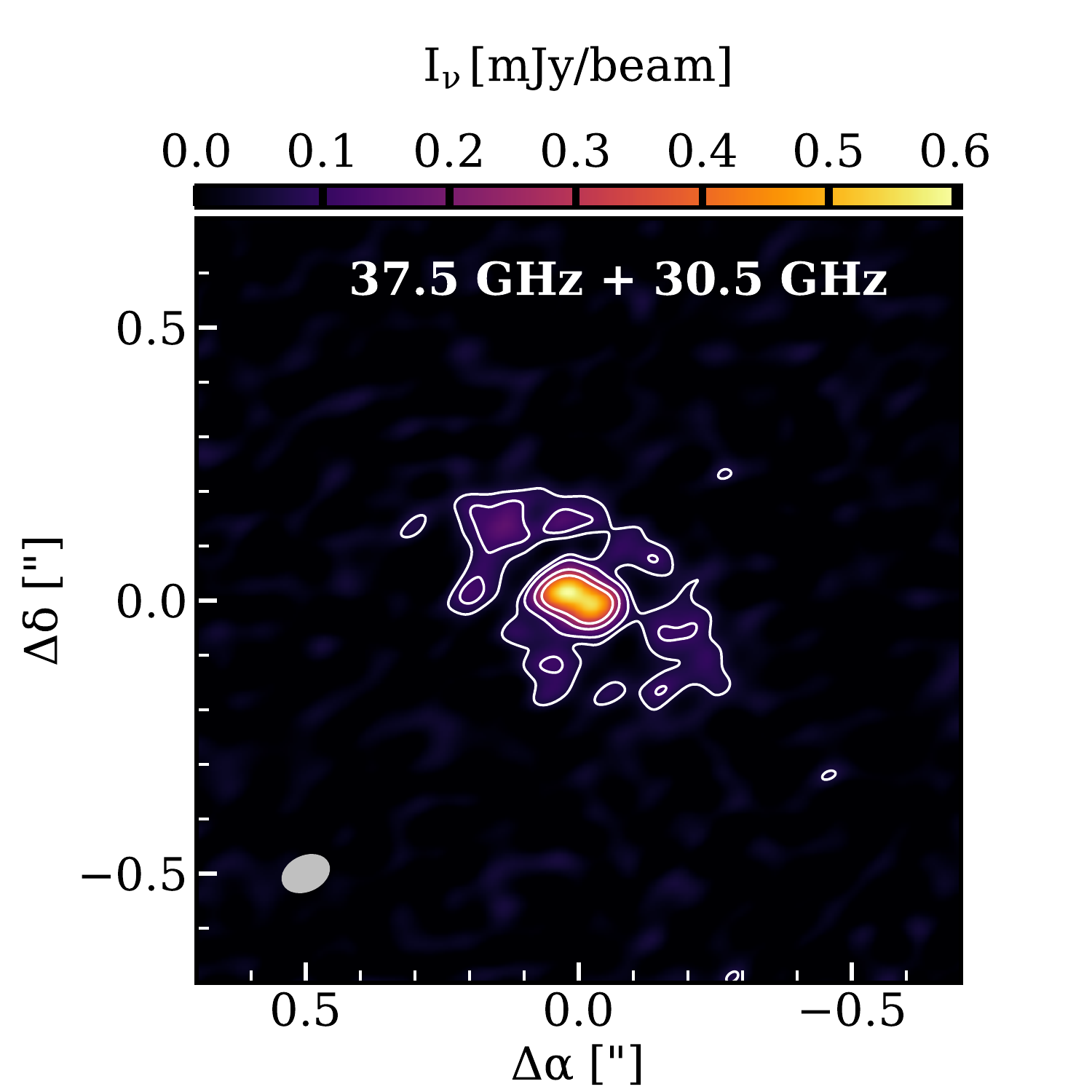}
    \includegraphics[width=0.32\textwidth]{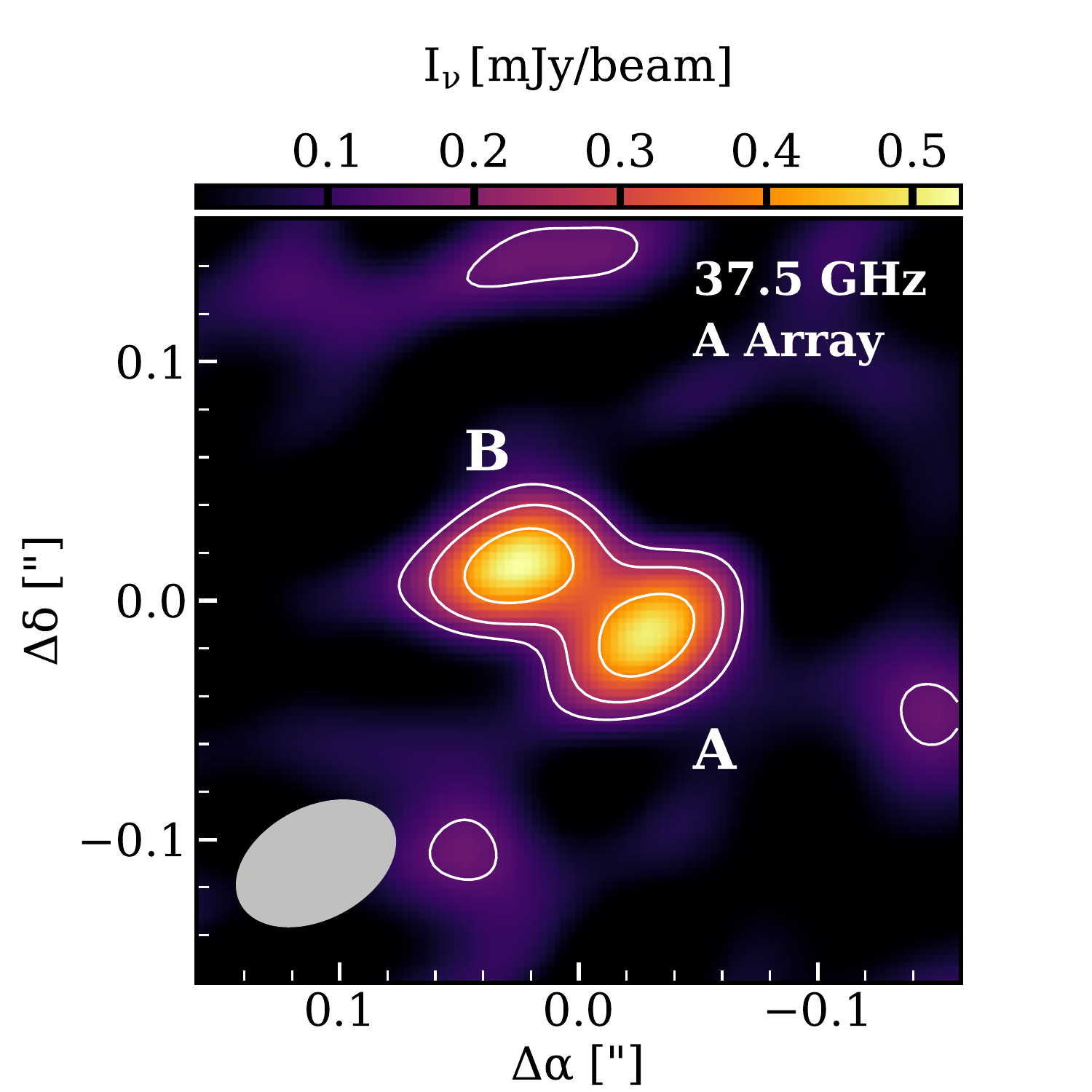} \\
    \includegraphics[width=0.32\textwidth]{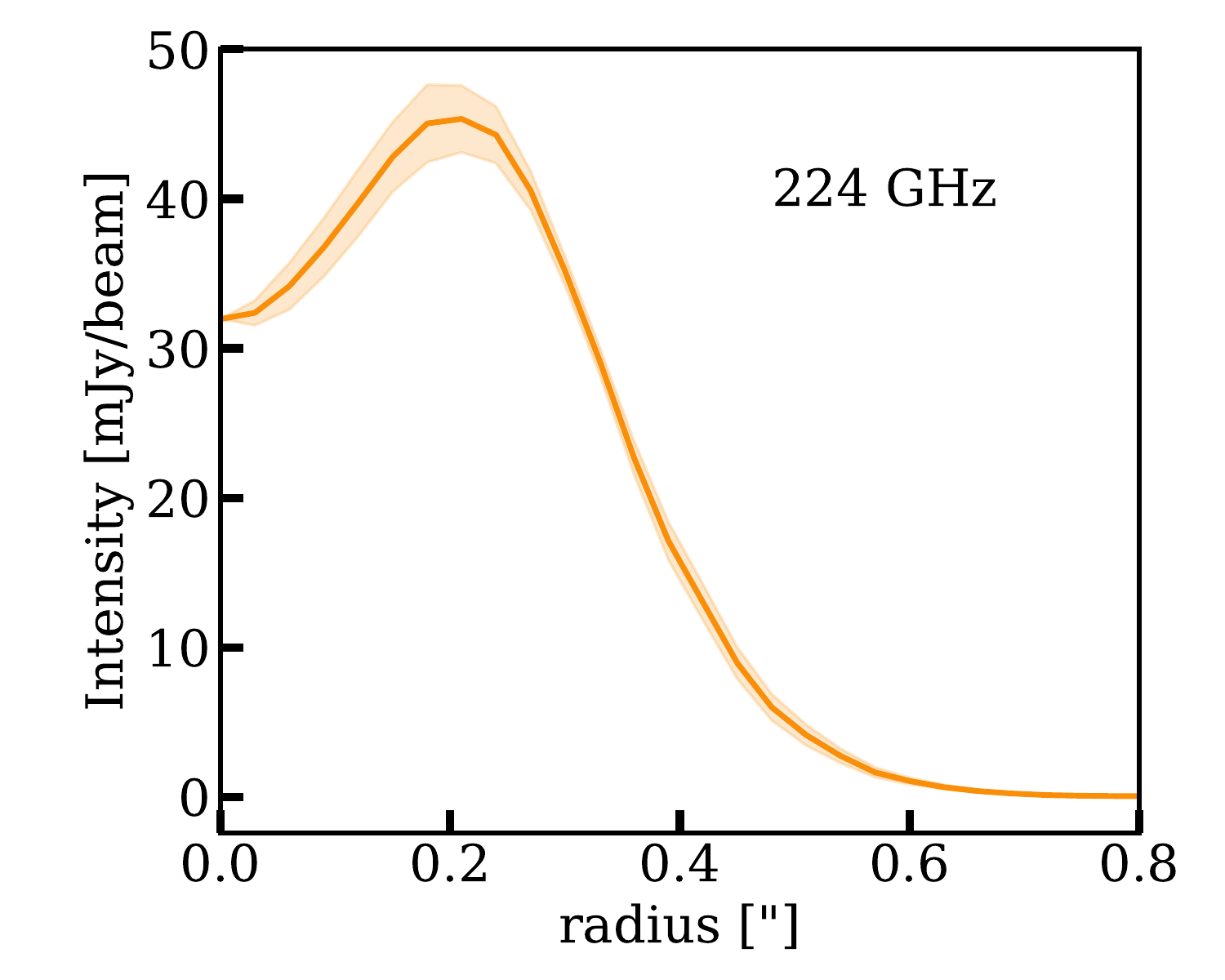} 
    \includegraphics[width=0.32\textwidth]{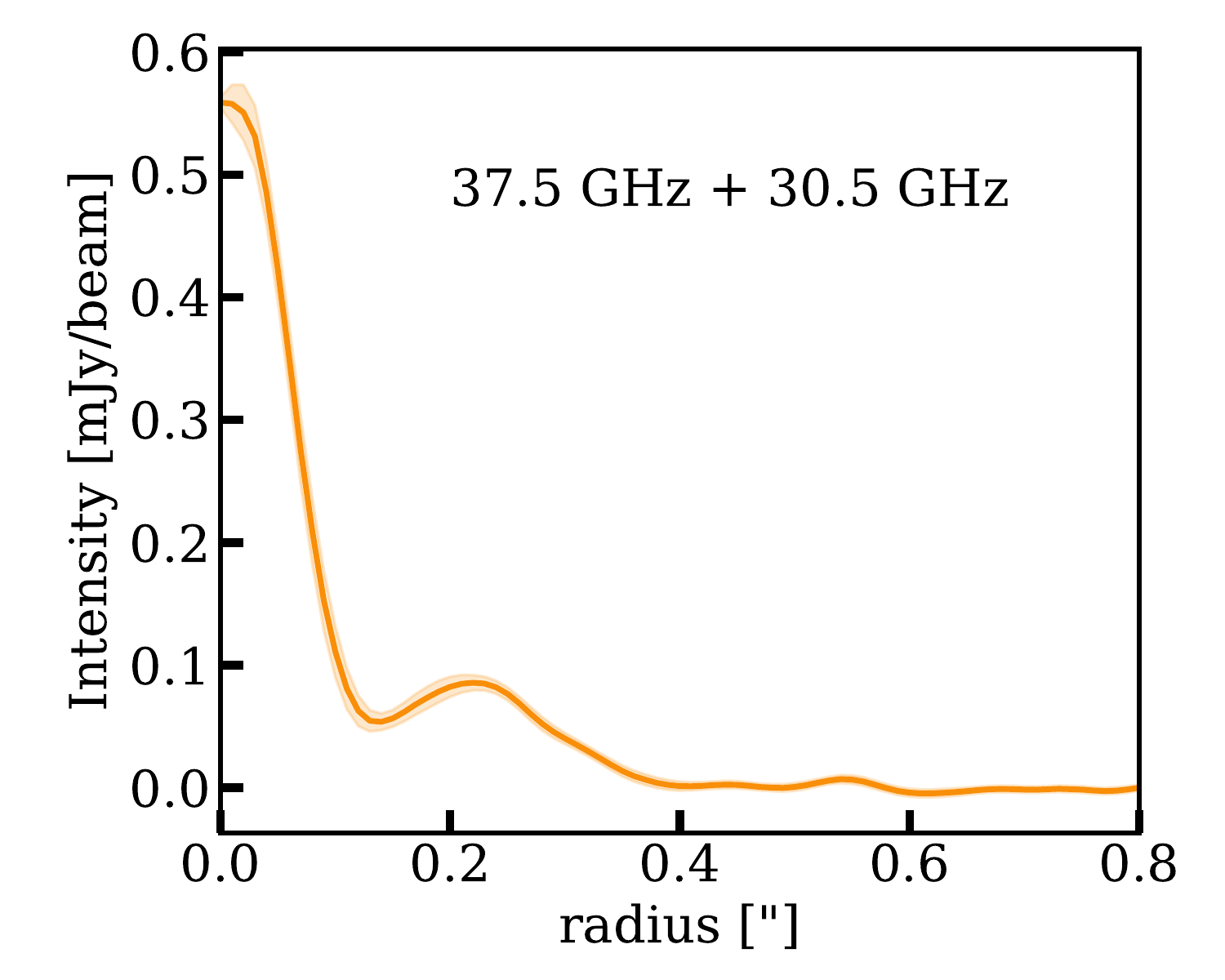}
    \includegraphics[width=0.32\textwidth]{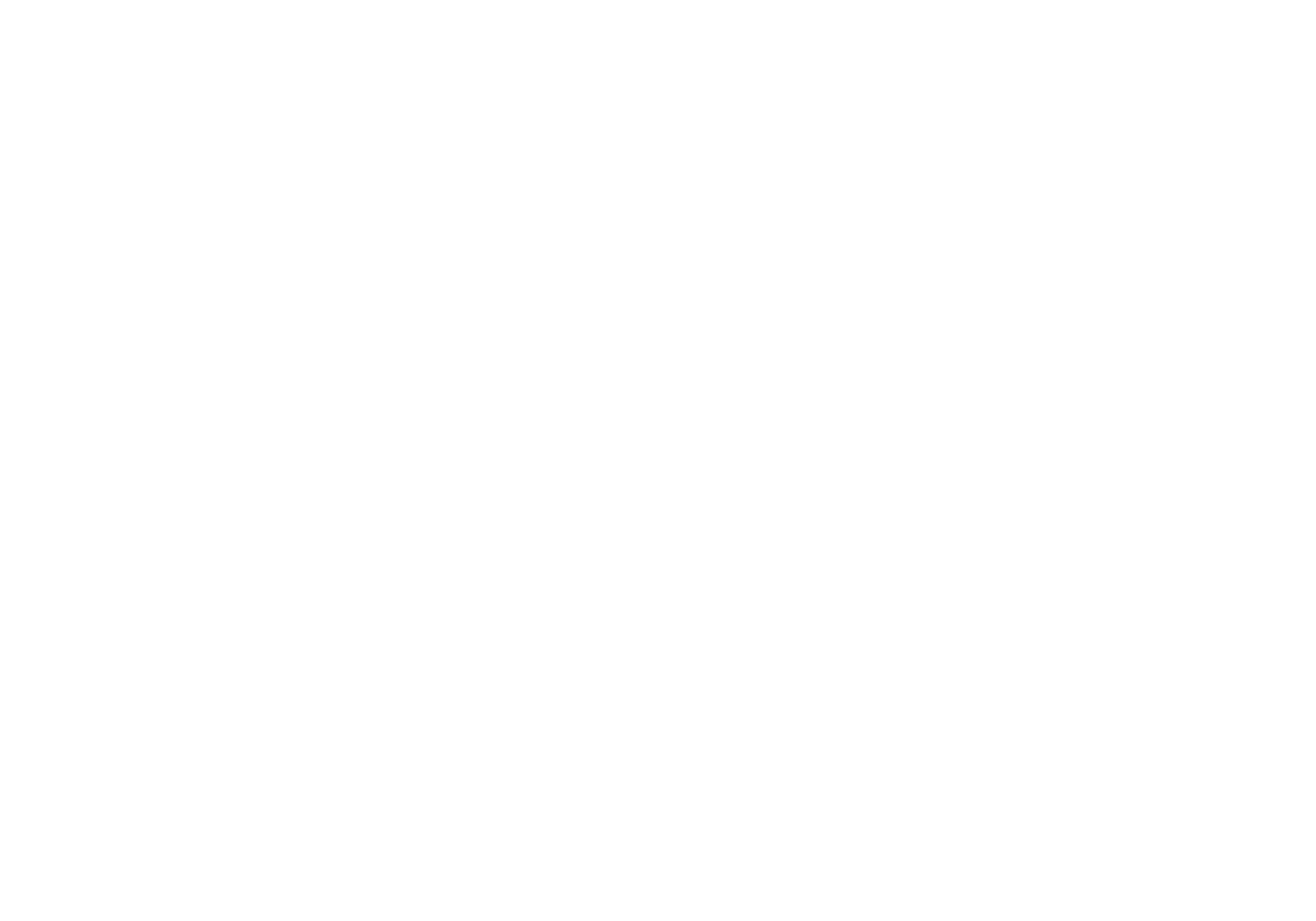}
   \\
    \caption{Continuum images and associated radial intensity profiles for V892 Tau. \textbf{From left to right in the upper panels:} ALMA 224\,GHz\,(1.3\,mm) continuum image. Contour levels are at 5, 10, 30, 50, 100, 200, 300, 400, 500, 600, 700$\sigma$; VLA 30.5\,GHz+37.5\,GHz\,(8.8\,mm) continuum image with three antenna configurations combined. Contour levels are at 3, 5, 10, 15$\sigma$; The zoom-in VLA 37.5\,GHz (8\,mm) continuum image from only A array data, highlighting the emission associated with the two stars. Contour levels are at 3, 5, 8$\sigma$. The beam sizes are displayed in the left corners. \textbf{From left to right in the lower panels:} deprojected and azimuthally averaged radial intensity profiles for emission at 1.3 and 8.8\,mm, respectively. The shaded region shows the 1$\sigma$ scatter divided by the square root of beam numbers along the annulus at each radial bin.  \label{fig:dust-images} }
\end{figure*}

\section{Results}~\label{sec:results}
% \ref{sec:results}
The V892 Tau disk is detected and spatially resolved in both mm continuum and CO line emission. In this section, we present the general emission properties, including constraints on disk geometries, radial gas temperature profiles, and disk vertical structures. The Keplerian-dominated gas rotation provides an independent estimate on the total stellar mass of the system. We then describe the remaining non-Keplerian gas emission features by comparing the Keplerian model with the data. Finally, we provide an updated binary orbit of V892 Tau using new astrometric measurements from VLA observations, in combination with previous astrometric measurements, the determined stellar mass, and the precise \textit{Gaia} distance.

% brief overview of the data result
% the two stars resolved in VLA data along with the dust ring
% ALMA observation reveals a dust ring 

\begin{figure*}[!t]
\centering
    \includegraphics[width=0.98\textwidth]{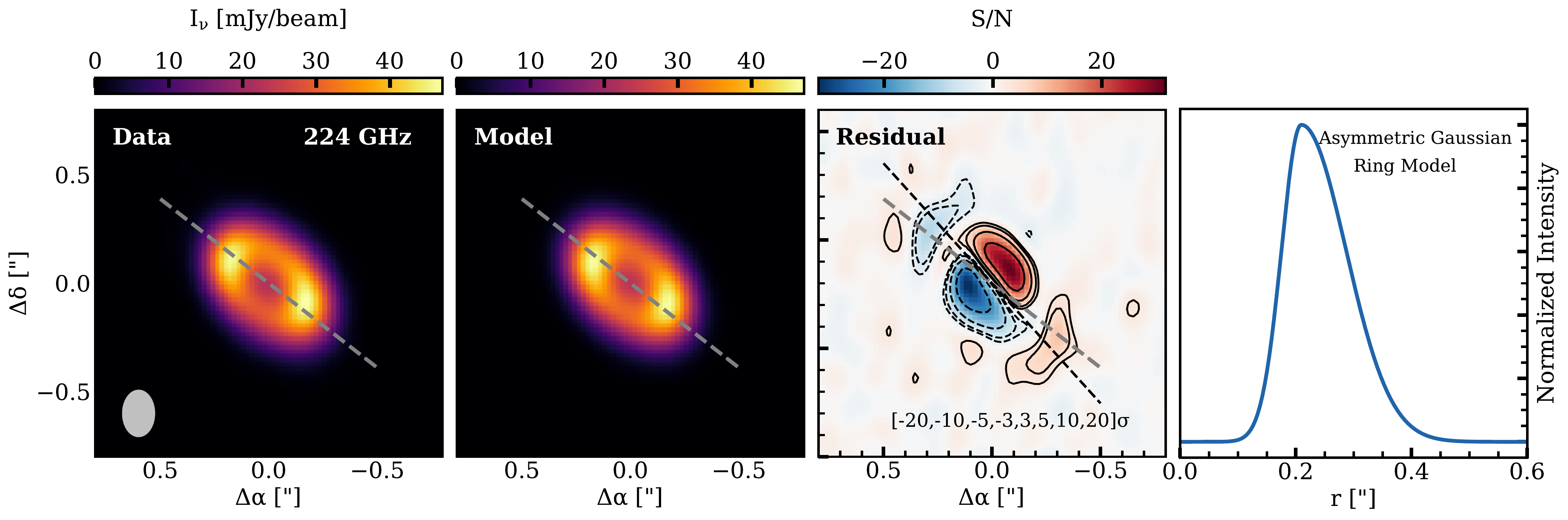} 
    \caption{\textbf{From left to right:} The data, model, and residual images at 1.3\,mm, and the adopted radial brightness profile of the asymmetric Gaussian ring model. Color scales in the data and model images are shown in emission intensity as mJy\,beam$^{-1}$, and color scale in the residual image is however shown in signal-to-noise ratio. The grey dashed line marks the disk position angle ($52.1\degr$) as derived from $uv$-plane fitting, while the black dashed line is rotated back by 10$\degr$ for a better alignment with the residual emission in the outer disk. \label{fig:model-B6} }
\end{figure*}

\subsection{Dust Disk} \label{sec:dust-model}
The ALMA 1.3\,mm and VLA (combined)\footnote{For illustrative purposes, the 30.5\,GHz (9.8\,mm) and 37.5\,GHz (8.0\,mm) combined image with better visualization of the dust ring is shown in Figure~\ref{fig:dust-images}. Our analysis mainly uses the 8\,mm image because emission in the center is better resolved. The 8.0\, and 9.8\,mm images are presented in Appendix~\ref{sec:vla}.} 8\,mm continuum images, as well as their deprojected and azimuthally averaged radial intensity profiles, are shown in Figure~\ref{fig:dust-images}. The deprojection adopts an inclination angle ($i$) of 54.5$\degr$ and a position angle (PA) of 52.1$\degr$, derived from the \textit{uv}-plane fitting as described below.  Emission at 1.3\,mm presents a prominent ring at a radius of $\sim$0$\farcs$2, with an inner disk cavity most likely carved by the binary and disk interaction. Subtle azimuthal asymmetry is seen across the dust ring, such that the northern part along the disk minor axis is about 20\% brighter. This asymmetry is reflected in the radial profile as the relatively larger uncertainties around the ring peak regions. The dust disk ring is also detected at the longer 8\,mm wavelength at similar radii, but only at 3--5$\sigma$ significance, in which any asymmetry would be difficult to quantify. The clumpy distribution of 8\,mm emission along the dust ring is likely associated with noise fluctuations 
%during the deconvolution process 
\citep[e.g.,][]{Macias2017}.
A more striking feature in the 8\,mm image is the two bright ($\sim$10$\sigma$) emission blobs inside the disk cavity. The flat spectral slopes ($\alpha\approx0$) between 8 and 9.8\,mm for these two point sources suggest that the bright emission is dominated by ionized gas emission (e.g., \citealt{Rodmann2006, Pascucci2012}). 
This emission, if present, is largely unresolved in the 1.3\,mm image. We measure a dust disk radius of $0\farcs46$ at 1.3\,mm (defined as the radius that encircles 90\%\ of the flux; e.g., \citealt{Tripathi2017}). 
The dust disk around V892 Tau NE is undetected at 1.3\,mm. Taking the $3\sigma$ upper limit of 0.2\,mJy, it is more than 10 times fainter than typical disks around M3 stars (e.g., \citealt{Andrews2013}).

To quantify the structure of the circumbinary disk of V892 Tau, we fit the 1.3\,mm dust continuum emission in the visibility domain. As the azimuthal asymmetry is only modest at the resolution of our observation, we adopt a modified Gaussian ring model, which can be expressed as: 
\begin{equation} \label{eq1}
I(r) = A \exp \left[ -\frac{ (r- R_{\rm ring} )^{2} }{2  \sigma_{\rm ring} ^{2} } \right],
\end{equation} 
where $R_{\rm ring}$ is the ring peak location and the Gaussian ring width $\sigma_{\rm ring}$ is described as
\begin{equation}
\sigma_{\rm ring} = \left\{
        \begin{array}{ll}
          \sigma_{\rm in},  & if \quad  r \leq R_{\rm ring}\\
            \sigma_{\rm out}, & if \quad r > R_{\rm ring}.
        \end{array}
    \right.
\end{equation}
This model with different inner and outer ring width is motivated by the fact that the inner ring edge would be severely truncated by the binary-disk interaction (e.g., \citealt{Cazzoletti2017}). A local pressure bump is also expected exterior to the binary orbit. In dust evolution models with a local pressure maximum, radial asymmetry in the grain distribution is expected as particles in the outer disk take longer to grow and drift to the pressure peak.  Such a modified Gaussian ring model has been successfully applied in interpreting dust emission in transition disks \citep[e.g.,][]{Pinilla2018}. Previous observations of the V892 Tau disk suggest that emission even at 2.7\,mm has a substantial non-dust contribution \citep{DiFrancesco1997}.  Such an emission component, originating close to the stars, may also be present at 1.3\,mm. As the two stars are largely unresolved in the 1.3\,mm data, we include one point source ($F_{\rm p}$) at the disk center to account for any possible non-dust emission, as well as thermal dust emission from unresolved circumstellar disks. We therefore have five parameters in describing the continuum emission distribution: $F_{\rm p}, F_{\rm disk}, R_{\rm ring}, \sigma_{\rm in}, \sigma_{\rm out}$, in which $F_{\rm disk}$ is the integral of Eq~\ref{eq1}. Four additional disk geometry parameters ($i$, PA, phase center offsets in RA and Dec as $\delta_{\rm \alpha}$ and $\delta_{\rm \beta}$) are also included in the fit.

The Fourier transform of this azimuthally symmetric brightness distribution (i.e., the model visibilities) is then calculated using the Hankel transformation \citep{Pearson1999} and sampled at the same observed spatial frequencies. The point source contribution is directly added as a constant flux in the phase-shifted model visibilities.  The comparison of the model and data visibilities adopts a Gaussian likelihood $\mathcal{L}\propto \exp(-\chi^2/2)$, where $\chi^{2}= \sum \left| V_{\mathrm{obs}}(u_{k}, v_{k})-V_{\mathrm{mod}}(u_{k}, v_{k}) \right|^{2} w_{k}$, with $w_{k}$ as the observed visibility weights. %\sum_{k=1}^{M}
Uniform priors are assumed for the fitted parameters, with bounds set based on image inspections (see Table~\ref{tab:dust-disk-model}). We then explore the parameter space with a Markov Chain Monte Carlo (MCMC) method (\texttt{emcee}, \citealt{ForemanMackey2013}), sampled with 45 walkers and 5000 steps per walker. The autocorrelation length is on the order of 100 steps. The adopted parameters and the associated uncertainties are summarized in Table~\ref{tab:dust-disk-model} (Model 1), which are estimated as the peaks and 68\% confidence intervals of the posterior distributions, constructed with the last 1000 steps of the chains.

\begin{deluxetable}{cccc}
%[!t]
\tabletypesize{\scriptsize}
\tablecaption{Dust Disk Model Results\label{tab:dust-disk-model}}
\tablewidth{0pt}
\tablehead{
\colhead{Parameter} & \colhead{Prior} &\colhead{Model 1} & \colhead{Model 2}  } 
\colnumbers
\startdata
F$_{\rm p}$ & [0,100] mJy &  $4.07\pm0.10$ & $6.65\pm0.08$  \\ 
$\delta_{\rm \alpha, Point}$ & [-0.2,0.2] arcsec &  .. & $-0.171\pm0.001$  \\ 
$\delta_{\rm \delta, Point}$ & [-0.2,0.2] arcsec &  .. & $-0.197\pm0.001$  \\
\hline
F$_{\rm disk}$ & [0,1000] mJy & $290.59\pm0.18$ & $288.15\pm0.16$ \\ 
$\sigma_{\rm in}$ &  [0,0.2] arcsec & $0.034\pm0.001$ & $0.027\pm0.001$   \\ 
$\sigma_{\rm out}$ &  [0,1.0] arcsec & $0.077\pm0.001$ & $0.081\pm0.001$   \\ 
$R_{\rm ring}$ &  [0.1,0.5] arcsec & $0.209\pm0.001$ & $0.203\pm0.001$   \\ 
$i$ &  [20,90] degree & $54.51\pm0.03$ & $54.76\pm0.03$      \\ 
PA &  [10,70] degree & $52.10\pm0.04$ & $52.16\pm0.04$   \\ 
$\delta_{\rm \alpha}$ & [-0.2,0.2] arcsec & $-0.128\pm0.001$   & $-0.129\pm0.001$    \\ 
$\delta_{\rm \delta}$ & [-0.2,0.2] arcsec & $-0.146\pm0.001$  & $-0.146\pm0.001$   \\
\enddata
\tablecomments{In Model 1, we assume the point source emission for any non-dust contribution and circumstellar dust emission at 1.3\,mm is located at the disk center, while Model 2 includes two additional parameters to account for phase center offsets of the point source.}
%\tablerefs{}
\end{deluxetable}

The right panel of Figure~\ref{fig:model-B6} shows the adopted model radial brightness profile, with a truncated inner disk described with $\sigma_{\rm out}/\sigma_{\rm in}=2.26$. The analysis by \citet{Pinilla2018} using this same ALMA dataset only found a subtle asymmetry, with $\sigma_{\rm out}/\sigma_{\rm in}=1.06$. This difference could be explained by the inclusion of the interior point source model in our fitting, which, if not properly accounted, could lead to an overestimation of the ring width. As shown in Figure~\ref{fig:model-B6}, our model describes the main dust ring reasonably well.
%(compared to the peak SNR$>700$, ). 
The residual image, created using the residual visibilities ($V_{\mathrm{obs}}-V_{\mathrm{mod}}$) with the same \texttt{tclean} parameters as the data and model images, highlights the remaining features that are not reproduced by the simple model. Firstly, the azimuthal asymmetry noted above could not be accounted with an axisymmetric model; this is seen as positive and negative residuals across the disk. Secondly, nearly symmetric residual emission is observed in the outer disk, which is offset by $\sim10\degr$ from the disk major axis as indicated by the black and grey dashed lines in Figure~\ref{fig:model-B6}. This residual pattern suggests that the outer disk may have a different geometry with respect to the main dust ring; such structure might be expected from the binary-disk dynamical interactions. However, the resolution and sensitivity of the current data preclude us from revealing the precise disk morphology variation of the dust content.

\begin{figure}[!t]
\centering
    \includegraphics[width=0.45\textwidth]{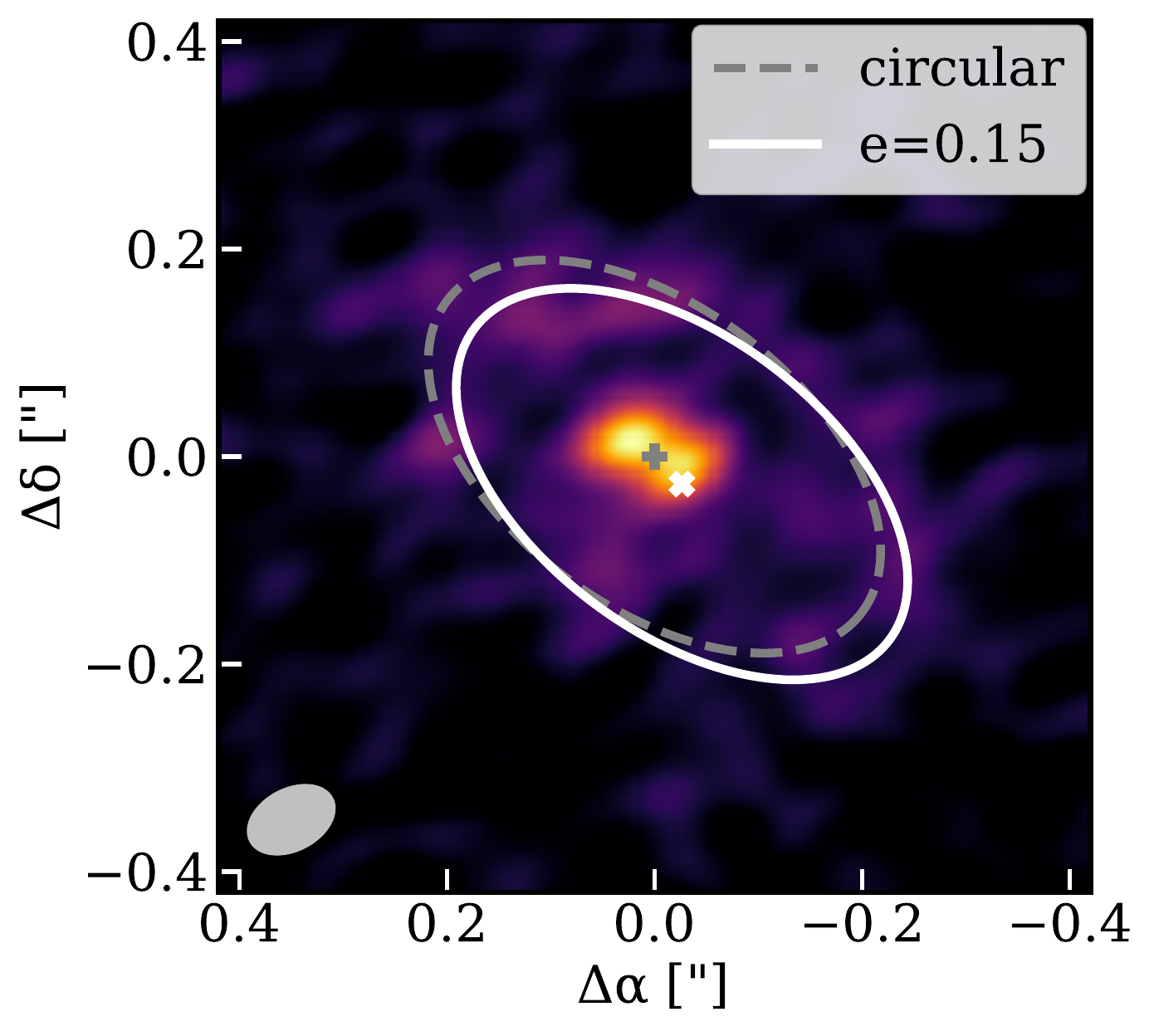} 
    \caption{The continuum image at 8\,mm, with a circular ring centered at the center of stellar mass (grey dashed curve centered at the grey plus), and a ring with $e=0.15$ (white solid curve centered at the white cross, assuming the binary center as one focus of the ellipse). Both rings have semi-major axis of 0$\farcs$25 and are projected with $i=54.5\degr$ and PA=$52.1\degr$. The image is created with natural weighting to highlight the faint dust ring. \label{fig:disk-eccentricity} }
\end{figure}

In the above modeling, we assumed that the point source contribution is located at the disk center, with the same phase center offsets as the dust ring. This simple assumption is reasonable given our large beam size. The point source model is introduced to describe emission coming from close to the two stars. The emission center therefore depends on the nature of the radiation (relative brightness of the emission from each component),  and could be offset from the disk center. We performed another fitting with two additional parameters included to allow the location of the point source to vary within the dust cavity. Following the same modeling procedure as outlined above, the results are derived and summarized in Table~\ref{tab:dust-disk-model} (Model 2). This new model provides similar dust ring properties to Model 1, but prefers a brighter point source separated from the disk center by $0\farcs06$\footnote{Based on the orbital solution obtained in Section~\ref{sec:orbit}, this offset is about twice the binary separation in 2015 September, the ALMA observation date.}.
This offset could be explained if the dust ring is eccentric, a feature naturally produced through binary-disk interaction (e.g., GW Ori, \citealt{Bi2020, Kraus2020}).
In the 8\,mm image, emission from (or close to) the two stars is marginally resolved and well separated from the dust ring. This allows us to investigate the possibility of an eccentric ring, though a direct modeling in the $uv$-plane is hindered by the low SNR. Figure~\ref{fig:disk-eccentricity} shows our by-eye estimate in the image. Assuming the mass center of the binary is one focus of the ellipse, we find that our data are not able to distinguish a dust ring with low eccentricity ($e<0.15$) from a circular ring. Given the brightness (4--6\,mJy) of the point source emission at 1.3\,mm, future ALMA observations with resolution better than the binary separation (30--60\,mas) could easily constrain the disk eccentricity.

If we assume the disk mm emission is optically thin, the available dust mass can be simply estimated. The observed fluxes in the dust ring translate into dust masses of 97--160\,$M_{\oplus}$ and 207--315\,$M_{\oplus}$ for 1.3 and 8\,mm measurements, respectively. The upper-end mass corresponds to $T_{\rm dust}=20$\,K, typical of Class II disks, and the lower value assumes a higher dust temperature of $T_{\rm dust}=30$\,K, which should be more representative for the warm disk condition of V892 Tau (see Section~\ref{sec:gas-structure}). In this calculation, we adopt the dust opacity $\kappa_{\rm 1.3mm}=2.3\,\rm cm^{2}\,g^{-1}$ (e.g., \citealt{Andrews2013}) and an opacity power-law index $\beta=1$, which extrapolates to $\kappa_{\rm 8mm}=0.37\,\rm cm^{2}\,g^{-1}$.
We note that a lower $\beta$, often observed in dust rings (e.g., \citealt{Long2020}), will lead to an even higher dust mass at the longer wavelength. The dust mass differences between the two wavelengths should therefore be largely attributed to the high optical depth of the 1.3\,mm emission.

%asymmetry should be more pronounced if with lower optical depth?

\begin{figure}[!t]
\centering
    \includegraphics[width=0.4\textwidth]{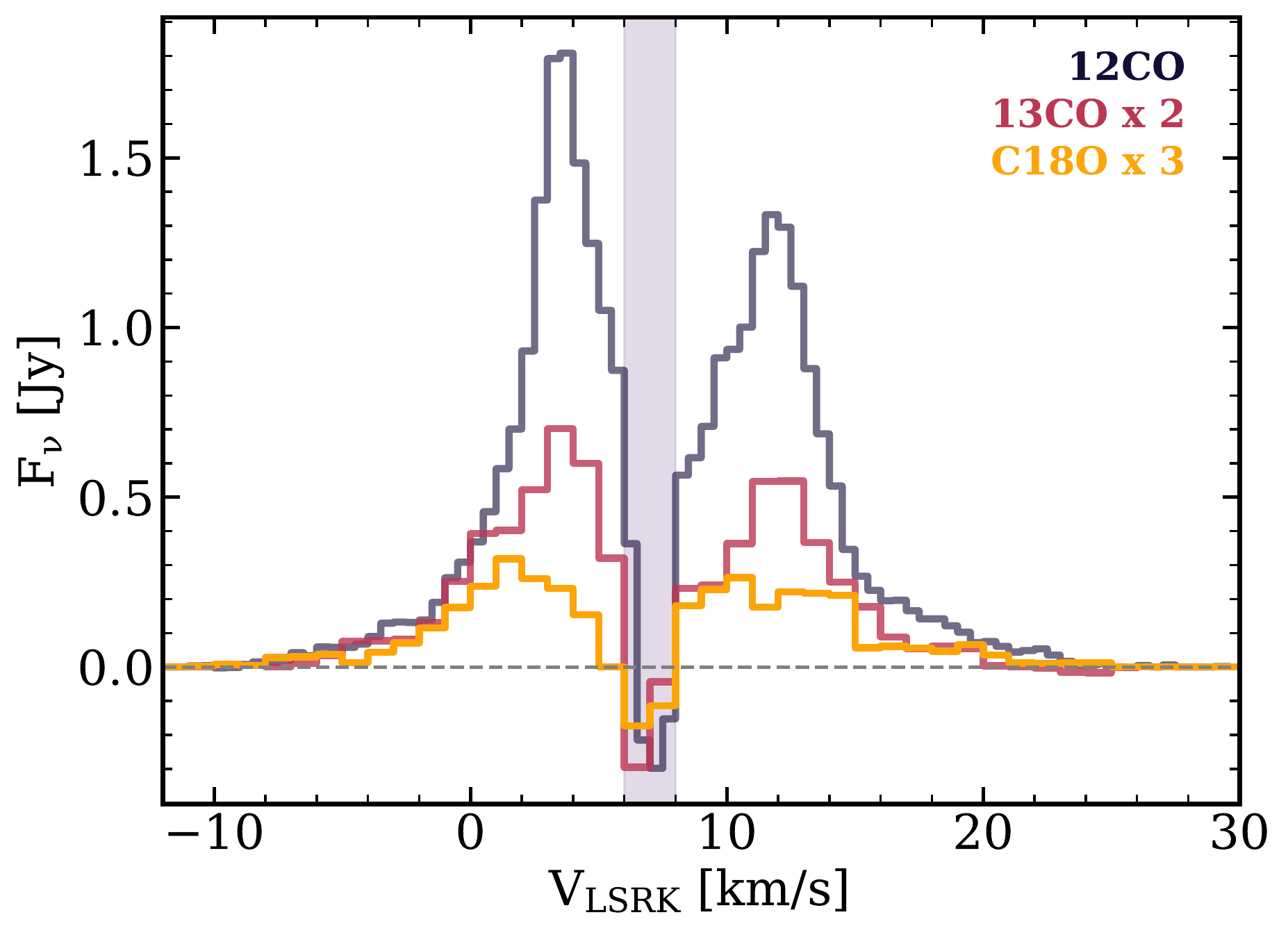} \\ 
    \includegraphics[width=0.4\textwidth]{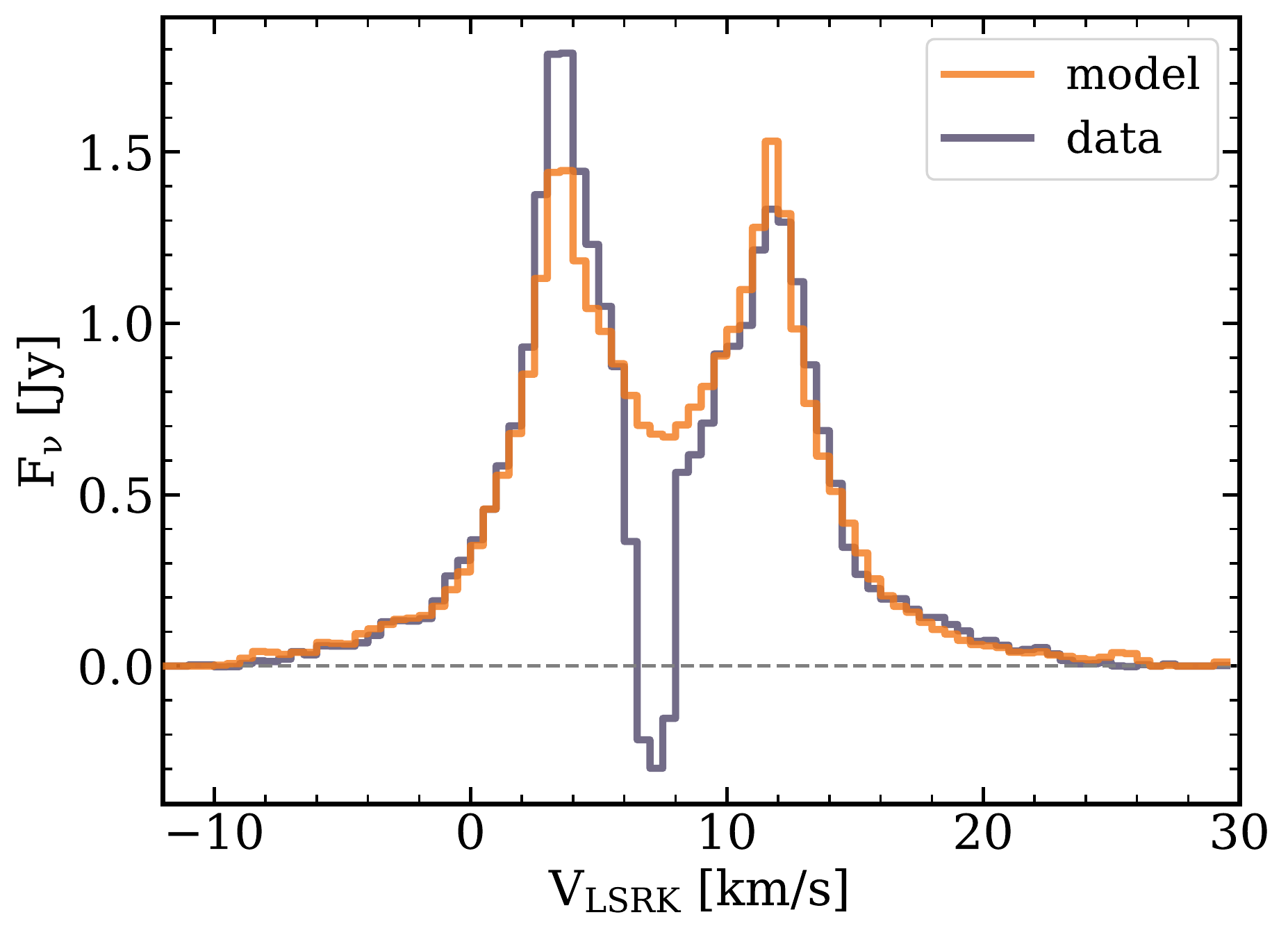} 
    \caption{\textbf{Upper panel:} Spatially integrated line spectra for the CO isotopologue $J=2-1$ emission, extracted with a customized Keplerian mask that covers the full emission regions of $^{12}$CO. The same mask is applied to extract $^{13}$CO and C$^{18}$O line spectra. The velocity range with negative emission is marked with the shaded region.  \textbf{Lower panel:} Comparison of $^{12}$CO line spectra from the data and the adopted Keplerian disk model as described in Section~\ref{sec:dyn-mass}. \label{fig:linespec} }
\end{figure}

\begin{figure*}[!t]
\centering
    \includegraphics[width=0.3\textwidth, trim=0 0 50 0, clip]{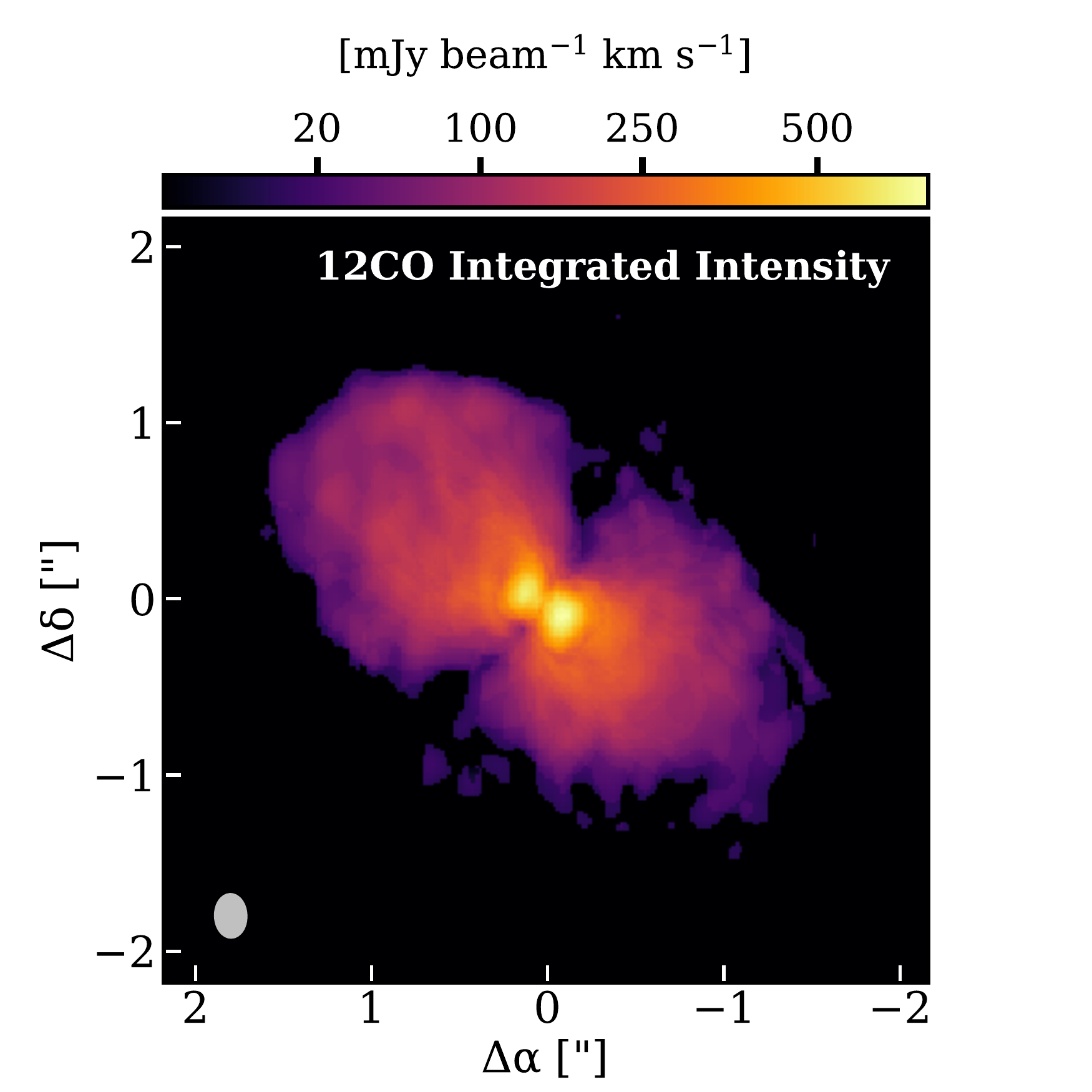} 
    \includegraphics[width=0.3\textwidth, trim=0 0 50 0, clip]{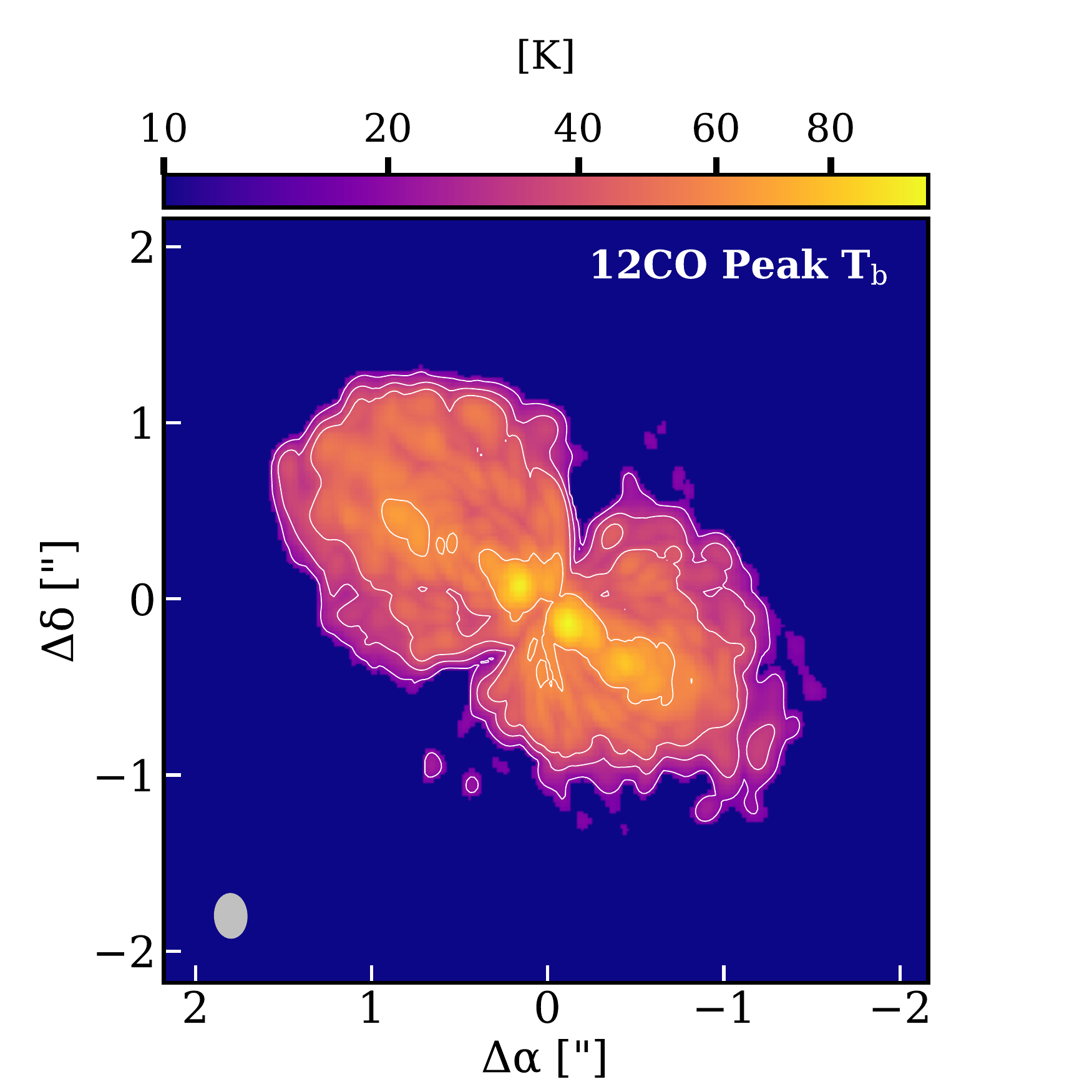} 
    \includegraphics[width=0.3\textwidth, trim=0 0 50 0, clip]{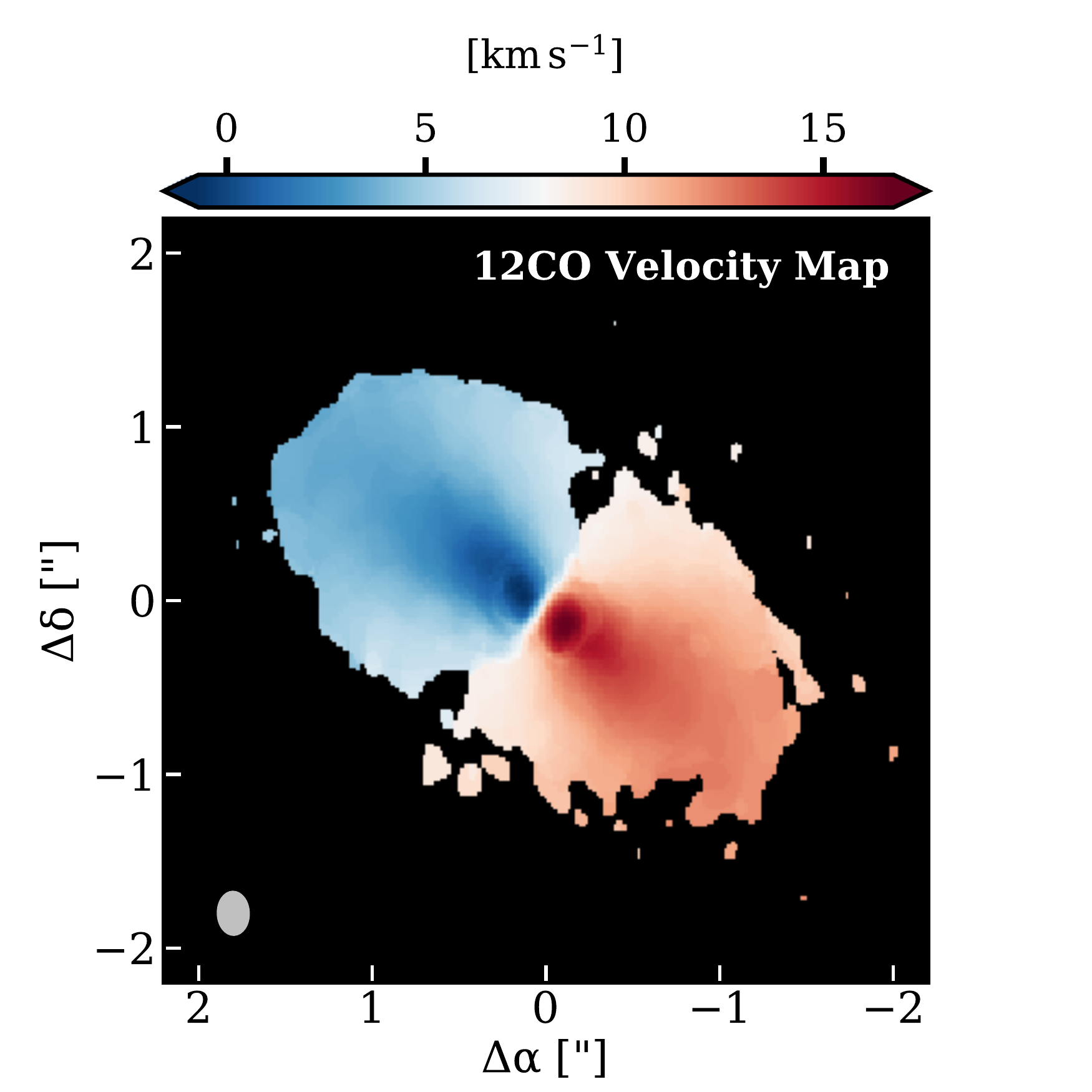}  \\
    \includegraphics[width=0.3\textwidth, trim=0 0 50 50, clip]{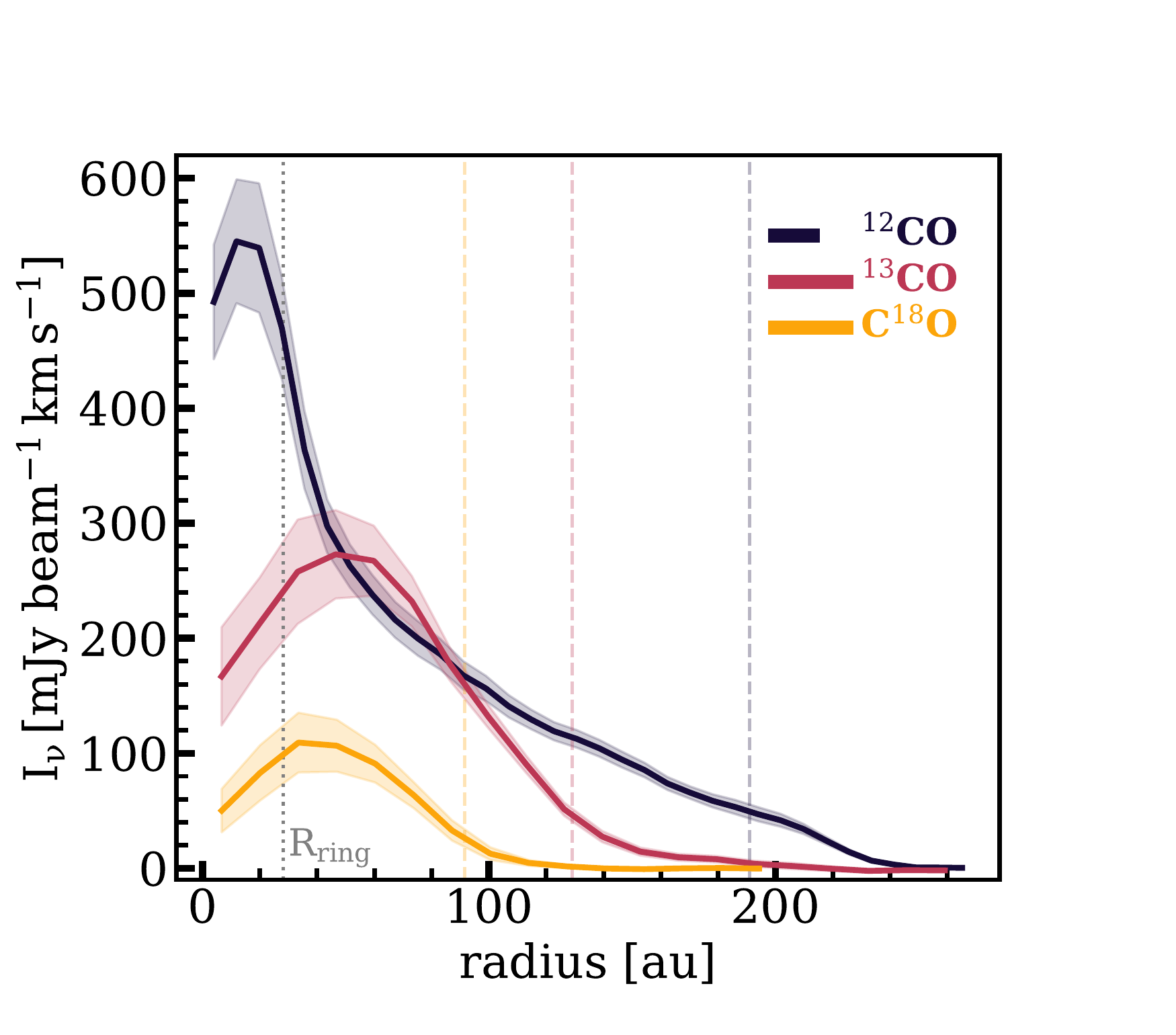} 
    \includegraphics[width=0.3\textwidth, trim=0 0 50 50, clip]{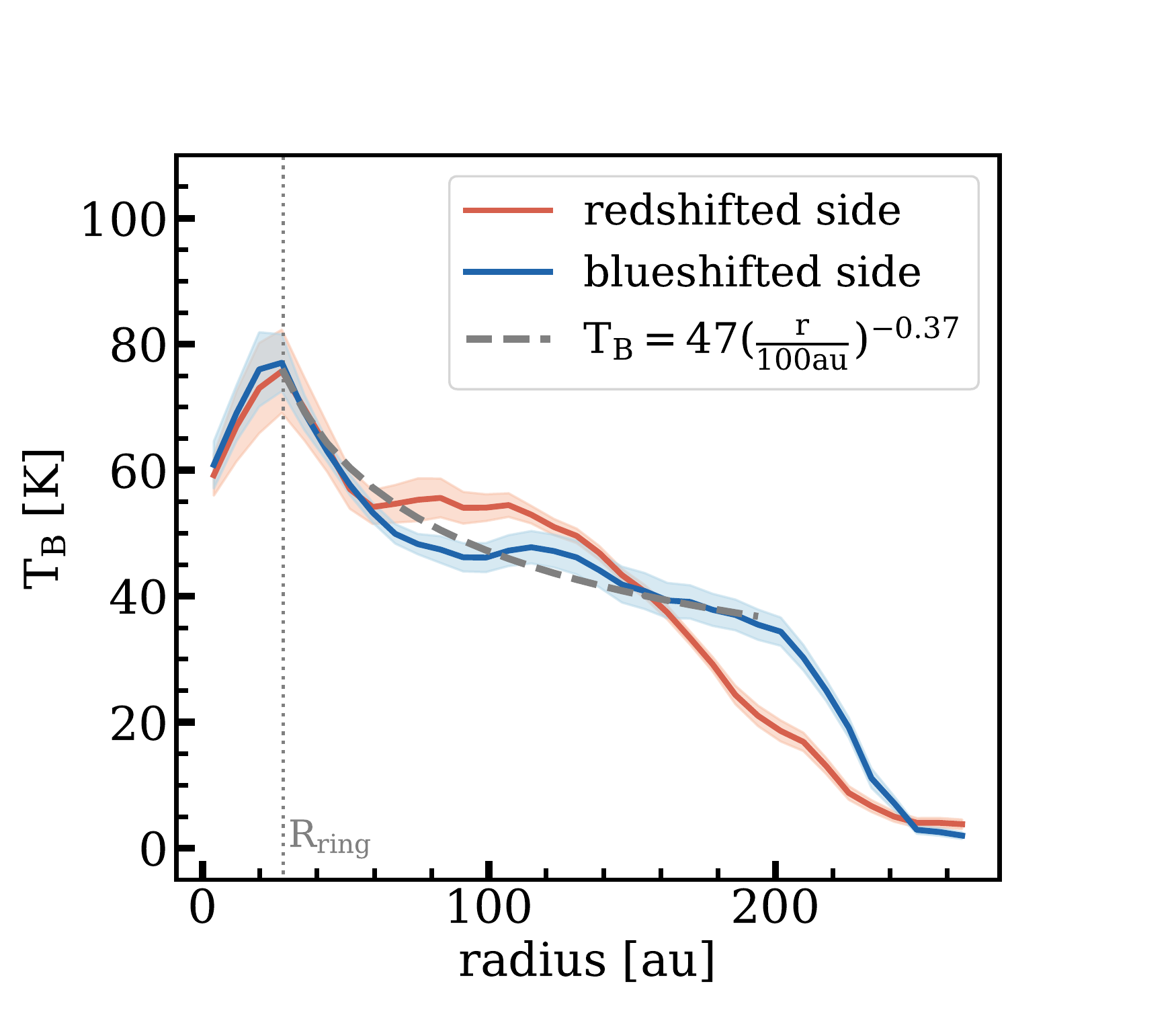} 
    \includegraphics[width=0.3\textwidth]{place_holder.pdf} 
    \caption{\textbf{Upper panels:} Integrated intensity map, peak brightness temperature map, and velocity map of $^{12}$CO $J=2-1$ emission in V892 Tau. Contours for $T_{b}$ are at [20, 30, 40, 60]\,K. All images are made with a 3$\sigma$ clip. Synthesized beams are shown in the lower left corner of each panel. \textbf{Lower left panel:} The azimuthally averaged radial intensity profiles for the CO lines. The grey dotted line indicates the location of the ring peak at 1.3\,mm. The dashed vertical lines mark the $R_{90\%}$ for the corresponding CO lines. The thick bars in the upper right corner represent the average beam radius for each line. \textbf{Lower right panel:} Radial profiles of brightness temperature in the blueshifted and redshifted sides of the disk. The grey dashed curve represents the fitted power-law function of the averaged brightness temperature profile between 28 (ring peak location) and 200\,au.  \label{fig:COmaps} }
\end{figure*}

\subsection{CO Gas Disk}
The CO emission around V892 Tau is detected and spatially resolved with high significance in all three isotopologues, spanning a very broad LSRK velocity range from $-$7 to 23\,km\,s$^{-1}$. The channel maps for the three lines with this full velocity range are shown in the Appendix. Strong cloud absorption is seen near the systemic velocity around 8\,km\,s$^{-1}$, even in C$^{18}$O. All emission is observed within a radius of 2$\farcs$0, and the C$^{18}$O emission only emerges from within 1$\farcs$0 in radius. 
The maximum recoverable scale (MRS)\footnote{\url{https://almascience.nrao.edu/documents-and-tools/cycle7/alma-technical-handbook/view}} for this observation is only $1\farcs7$ if defined based on the 5th percentile of the baseline lengths. 
%The maximum recoverable scale (MRS)\footnote{\url{https://almascience.nrao.edu/documents-and-tools/cycle7/alma-technical-handbook/view}} for this observation defined by the shortest baseline ($0.6\lambda_{\rm obs}/L_{\rm min}$) is $\sim4''$, or $1\farcs7$ if defined based on the 5th percentile of the baseline lengths as a more realistic estimate. 
Interferometric observations tend to filter out emission with scales larger than the MRS, therefore these data are likely insufficient to reveal the full gas distribution in the outer disk. Figure~\ref{fig:linespec} shows the spatially integrated line spectra, extracted using a Keplerian mask\footnote{\url{https://github.com/richteague/keplerian_mask}} designed to encompass all emission in individual channels. Though the CO emission pattern is dominated by the disk rotation, the brighter blue-shifted emission suggests asymmetric features along the disk major axis. We estimate the line fluxes by integrating over the velocity range from $-$7 to 23\,km\,s$^{-1}$, excluding the range from 6.5 to 8\,km\,s$^{-1}$ where the line emission is severely contaminated by cloud absorption. These lower limits to the line fluxes are summarized in Table~\ref{tab:obs-results}.

% describe the mom0 map first and the radial profile 
% also the Rgas. 
The $^{12}$CO integrated intensity (moment 0) map is shown in  Figure~\ref{fig:COmaps}, including only pixels in individual channels above the 3$\sigma$ level. The deprojected and azimuthally averaged radial profile extracted from the moment 0 map exhibits a central dip and reaches its maximum brightness interior to the dust ring peak.  Radial profiles for the $^{13}$CO and C$^{18}$O lines show similar morphologies, with a deficit of line emission inside the dust disk cavity but with peak locations slightly shifted outward (Figure~\ref{fig:COmaps}). These signs of a gas density drop in the inner dust disk cavity have been observed in many transition disks (e.g., \citealt{vanderMarel2021}), usually with a smaller cavity radius and lower degree of depletion when compared to the dust component. In the V892 Tau disk, gas depletion is expected from the binary-disk interaction.  However, quantifying the gas depletion level and the gas cavity radius are challenged by the cloud absorption. Other molecules with lower abundances, whose lines are less susceptible to cloud contamination, should be employed to trace the gas structure inside the dust cavity.

Based on the derived radial intensity profile, we measure the CO gas disk radius using a similar method as for the dust disk size (the radius where 90\% of the integrated line flux is encircled). This estimation provides a good approximation of the gas disk extensions and can be directly compared with literature measurements (e.g., \citealt{Ansdell2018}). The $R_{\rm 90\%, gas}$ values are summarized in Table~\ref{tab:obs-results}.

The intensity-weighted velocity (moment 1) map is also shown in Figure~\ref{fig:COmaps}. The rotation pattern suggests that the northeast side of the disk is blueshifted with respect to the systemic velocity of 8\,km\,s$^{-1}$. Irregular emission features are also visible, mostly from the redshifted side of the disk, emerging from the disk edge. In the following subsections, we will first use the main feature of Keplerian rotation to constrain the total mass of the central stars, then explore the thermal and vertical structures of the disk, and finally describe the non-Keplerian features associated with the system by comparing the data images with our established Keplerian model.

\subsubsection{Stellar Mass from Gas Disk Rotation} \label{sec:dyn-mass}
To determine the total stellar mass of the V892 Tau system, we model the spatially and spectrally resolved Keplerian rotation of the circumbinary gas disk. As seen from the line spectrum and velocity map (Figure~\ref{fig:linespec} and \ref{fig:COmaps}), our assumption that the motion of the gas disk is primarily governed by the gravitational force of the central stars should be valid. This disk-based stellar mass measurement is independent of the observed stellar properties, could be used as a benchmark to calibrate pre-main sequence stellar evolution models, and has been validated and employed in previous works \citep[e.g.,][]{Simon2000,Rosenfeld2012,Guilloteau2014,Czekala2015}. Those studies often adopt prescriptions of disk physical structures, and perform radiative transfer calculations to simulate the line emission. It is therefore computationally expensive to obtain a robust parameter inference.

We instead employ an empirical approach to model the emission line surface brightness directly (Andrews et al.~in preparation). Here, a brief overview is provided. Firstly, we assume an azimuthally symmetric disk, in which the height of the line emission surface, the line brightness temperature, and the (pseudo-)optical depth, are all taken to follow a truncated power-law distribution in the radial direction. The gas velocity field is set as Keplerian rotation determined by the central stellar mass. A 3D emission cube could then be produced by projecting these emission prescriptions onto the sky frame, taking into account the disk inclination and position angle, as well as image center offsets. This model image cube is later Fourier transformed to generate the model visibilities using the python code \texttt{vis\_sample}\footnote{\url{https://github.com/AstroChem/vis_sample}}, sampled at the observed \textit{uv}-spacings, and then compared with the data visibilities using a standard Gaussian likelihood function. We assumed uniform priors for all parameters, and sampled the posterior distributions with \texttt{emcee} \citep{ForemanMackey2013} using 65 walkers for 10,000 iterations (discarding the initial 2000 as burn-in, based on the saturation of the integrated autocorrelation time).

The inferred total stellar mass is $M_{\rm tot}=6.0\pm0.2\,M_{\odot}$. In the fitting, we have fixed the source distance to 134.5\,pc. A closer distance of 130\,pc, more consistent with the \textit{Gaia} EDR3 measurement for V892 Tau NE, would lead to a lower $M_{\rm tot}$ of 5.8\,$M_{\odot}$. Another significant source of uncertainty in the inference of stellar mass comes from the disk inclination, as the observed line-of-sight velocity, $v_{\rm obs} \propto \sqrt{M_{\rm tot}}$sin$(i_{\rm disk})$. 
Our fitting to the $^{12}$CO emission provides a best-fit inclination of 55.1$\pm$1.3$\degr$ and position angle of 53.9$\pm$0.7$\degr$, which are in excellent agreement with the dust disk modeling results from Section~\ref{sec:dust-model}. An uncertainty of 1$\degr$ in $i_{\rm disk}$ could lead to a stellar mass uncertainty of $\sim0.1\,M_{\odot}$. By convention, we define the disk inclination angle varying between 0 and 90$\degr$ as a measure of the deviation of the disk plane from the sky plane. However, there is an ambiguity of the sign of the inclination\footnote{The inclination angle and position angle together define the true disk plane, so this could also be taken as an ambiguity of 180$\degr$ on position angle, if we fix the disk inclination.}, depending on which side of the disk is closer to the observer. For a disk with an intermediate inclination angle, gas observations with high spatial resolution could in principle separate the brighter front disk from the dimmer back side (e.g., \citealt{Huang2020_GMAur}), and robustly set the disk rotation axis. Such features should be more easily identified at velocity channels close to the systemic velocity. 
Unfortunately, our data are not sufficient to break the degeneracy. Future high resolution gas observations, or near-infrared scattered light images \citep[e.g.,][]{Garufi2020}, will be necessary to determine the near and far side of the disk, for a final determination of the disk geometry.

We show in the Appendix (Figure~\ref{fig:12co-model-channel}) the model and residual channel maps for the $^{12}$CO emission, which are made with the same \texttt{tclean} parameters as the data cube. The adopted model describes the main Keplerian rotation pattern well, although considerable residuals are also seen. We will investigate these non-Keplerian features in detail use the residual emission in Section~\ref{sec:non-Keplerian}.

\subsubsection{Disk Thermal and Vertical Structure} \label{sec:gas-structure}
\begin{figure}[!t]
\centering
    \includegraphics[width=0.45\textwidth]{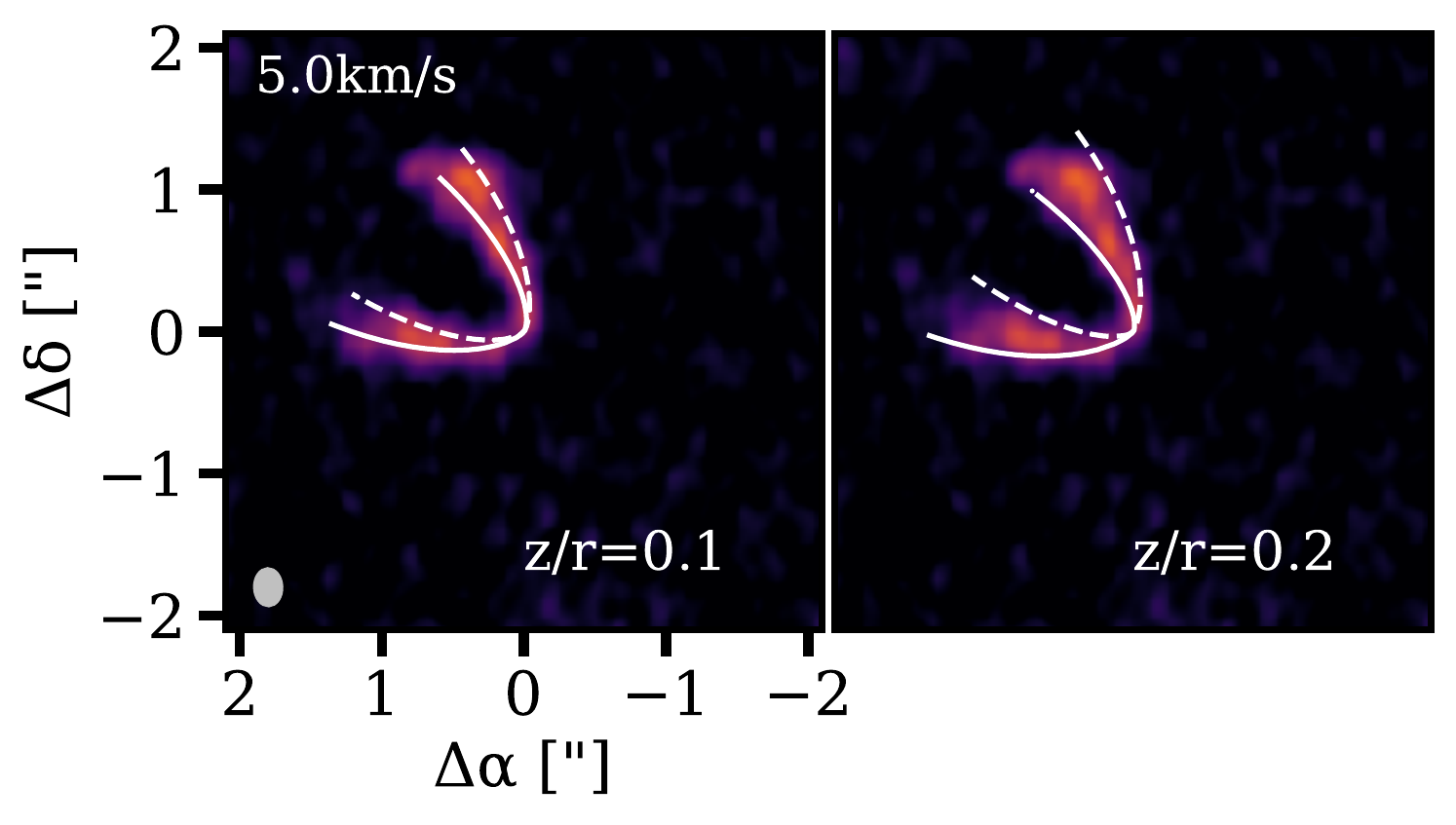} 
    \caption{The channel map of $^{12}$CO emission at an LSRK velocity of 5\,km\,s$^{-1}$, overlaid with isovelocity contours assuming different emitting heights ($z/r$), as indicated in each panel. Solid and dashed curves denote the front and back side of the disk, respectively. \label{fig:zr} }
\end{figure}

Optically thick lines emit from elevated layers above the disk mid-plane, and therefore can be employed to explore the disk vertical structure (e.g., \citealt{deGregorio-Monsalvo2013, Rosenfeld2013}). The comparison of predicted isovelocity contours to line emission in Figure~\ref{fig:zr} suggests that $^{12}$CO $J=2-1$ emission in the V892 Tau disk emerges from a surface with an aspect ratio ($z/r$) of $\sim0.1$. This is consistent with the fact that emission at $\sim1$\arcsec\ is only marginally spatially resolved as seen from the channel maps, given the beam size of $0\farcs2$. The CO emitting surfaces from previous determinations for both T Tauri and Herbig Ae disks are often found at $z/r>0.3$ (e.g., \citealt{Pinte2018, law20a, Paneque-Carreno2021arXiv}). The emission surfaces for small dust grains probed by near-infrared scattered light images are slightly lower than the CO surfaces, with $z/r \approx 0.15-0.3$ \citep[e.g.,][]{Ginski2016,Avenhaus2018}. Thus the disk around V892 Tau is an especially ``flat" example. The gas disk vertical extension depends on the balance between disk thermal pressure and stellar gravity, in which the pressure scale height $h \propto c_{s} / \Omega \propto \sqrt{T_{\rm gas} / M_\ast}$, where the surface gas temperature could be assumed to scale with $L_\ast^{1/4}$. The disk scale height would be 30\% lower for a system with two equal-mass, coeval stars compared to a single star. Assuming the CO emitting height scales with the disk pressure height, the flat disk geometry in V892 Tau therefore seems consistent with its nature as an equal-mass binary.

The brightness temperatures for spatially resolved, optically thick lines in local thermodynamic equilibrium are roughly equivalent to the gas temperature of the emitting disk layer (e.g., \citealt{Pinte2018}). Using $^{12}$CO, we are therefore able to probe the disk surface temperature at $z/r \approx 0.1$. 
As continuum subtraction in optically thick line regions would underestimate the true line fluxes (e.g., \citealt{Weaver2018}), the peak brightness temperature map (Figure~\ref{fig:COmaps}) is produced using the $^{12}$CO image cubes generated from the line+continuum visibilities. Our conversion from peak intensity into brightness temperature adopts the full Planck function. As seen in Figure~\ref{fig:COmaps}, the CO emitting gas has temperatures spanning a range of 20--80\,K, which reaches its peak value at the dust ring location. The drop of gas temperature inside the dust ring is likely due to the combined effects of gas depletion and beam dilution. The gas disk around V892 Tau is warmer than disks around T Tauri stars (e.g., IM Lup, \citealt{Pinte2018}; GM Aur, \citealt{Huang2020_GMAur}), but has comparable gas temperatures to the disks around Herbig Ae stars in single stellar systems (e.g., HD 163296, MWC 480, \citealt{law20a}). Note that the cloud absorption could lead to an underestimate of the gas temperature.

The radial distribution of the surface temperature is usually approximated by a power law. Our fitting of the azimuthally averaged temperature in the radius range of 28--200\,au results in a radial profile of $T(r)=47(r/\rm 100au)^{-0.37}$\,K. As shown in Figure~\ref{fig:COmaps}, there are significant brightness temperature variations in the two sides of the disk. The emission profile from the blueshifted side of the disk resembles the fitted averaged profile in a power-law manner within the fitting radial range, which however exhibits a steep drop outside 200\,au.  
The line emission from the redshifted side of the disk is $\sim$10\,K brighter than the blueshifted side at around 100\,au, and then decreases more sharply when moving outwards to produce up to a 20\,K difference in the outer disk. The comparison between these two sides of the disk therefore demonstrates strong azimuthal asymmetries in the V892 Tau gas disk (see also Elias 2-27 in \citealt{Paneque-Carreno2021arXiv}). Such asymmetry is also seen in the mid-infrared image at 10.7$\mu$m, in which the inner disk toward the southwest (redshifted) side is brighter \citep{Monnier2008}.

% 12CO uncertainty from vertical structure 
% test with also 13CO and C18O, though with lower SNR

\begin{figure*}[!t]
\centering
    \includegraphics[width=0.9\textwidth]{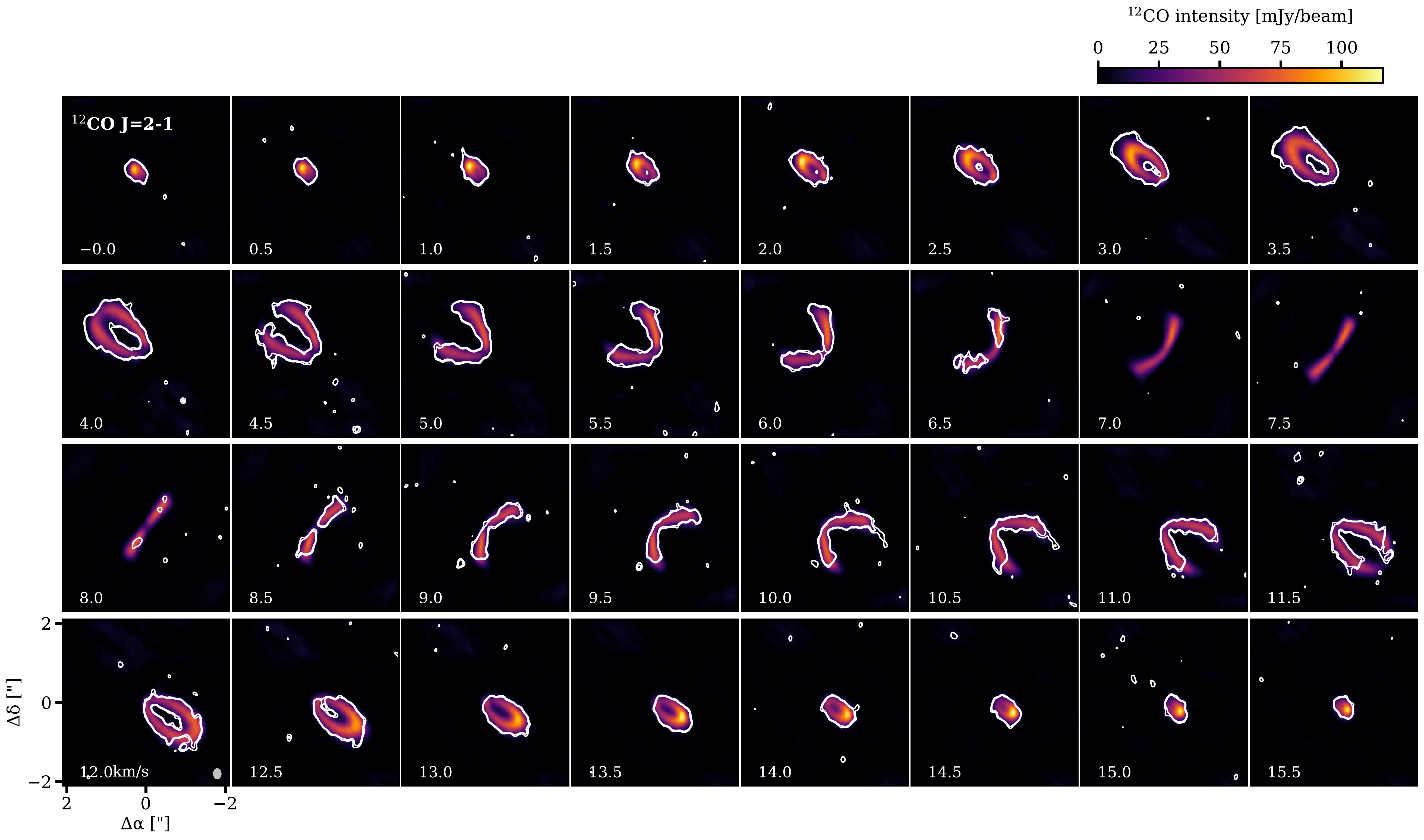} \\
    \includegraphics[width=0.45\textwidth]{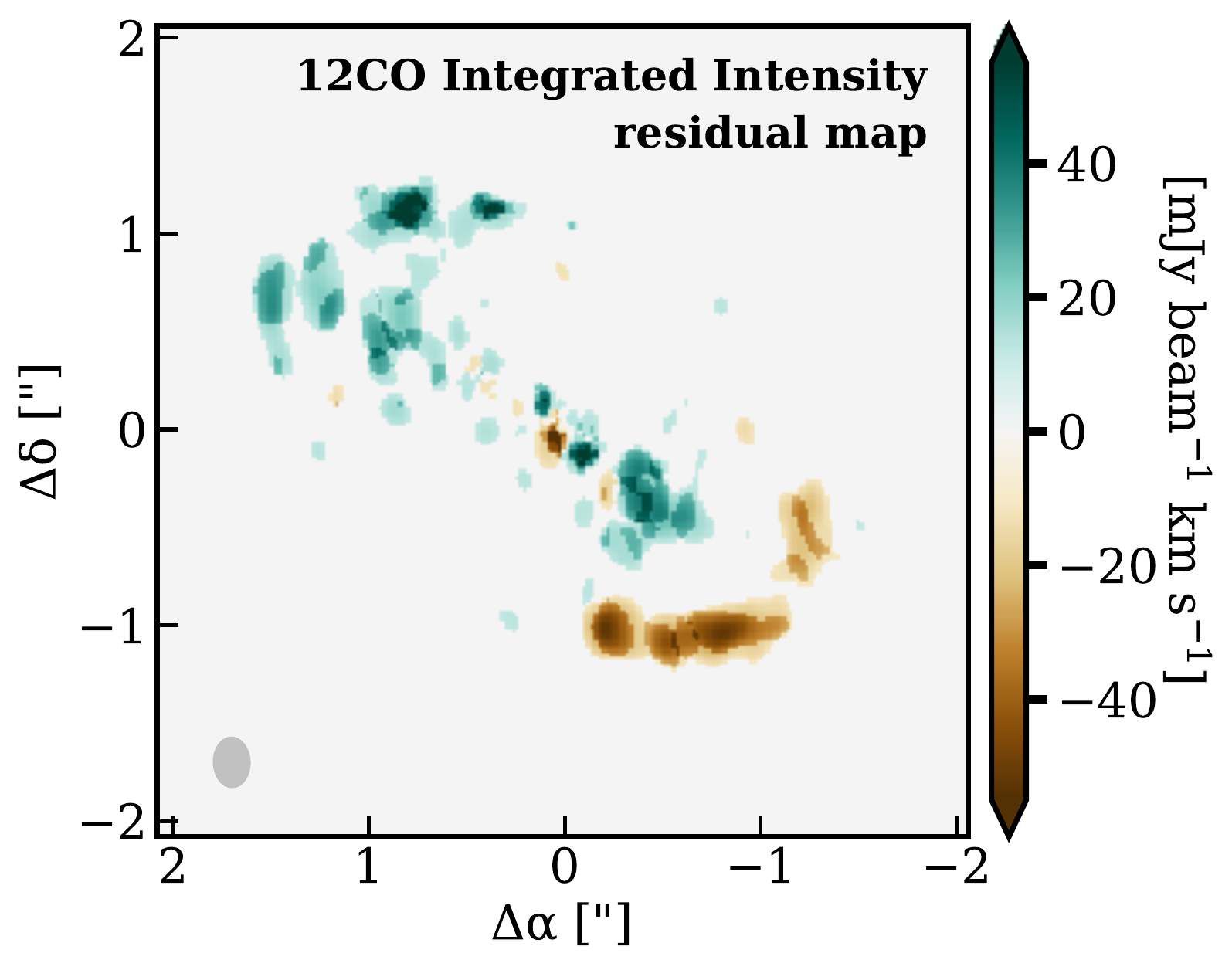}
    \includegraphics[width=0.45\textwidth]{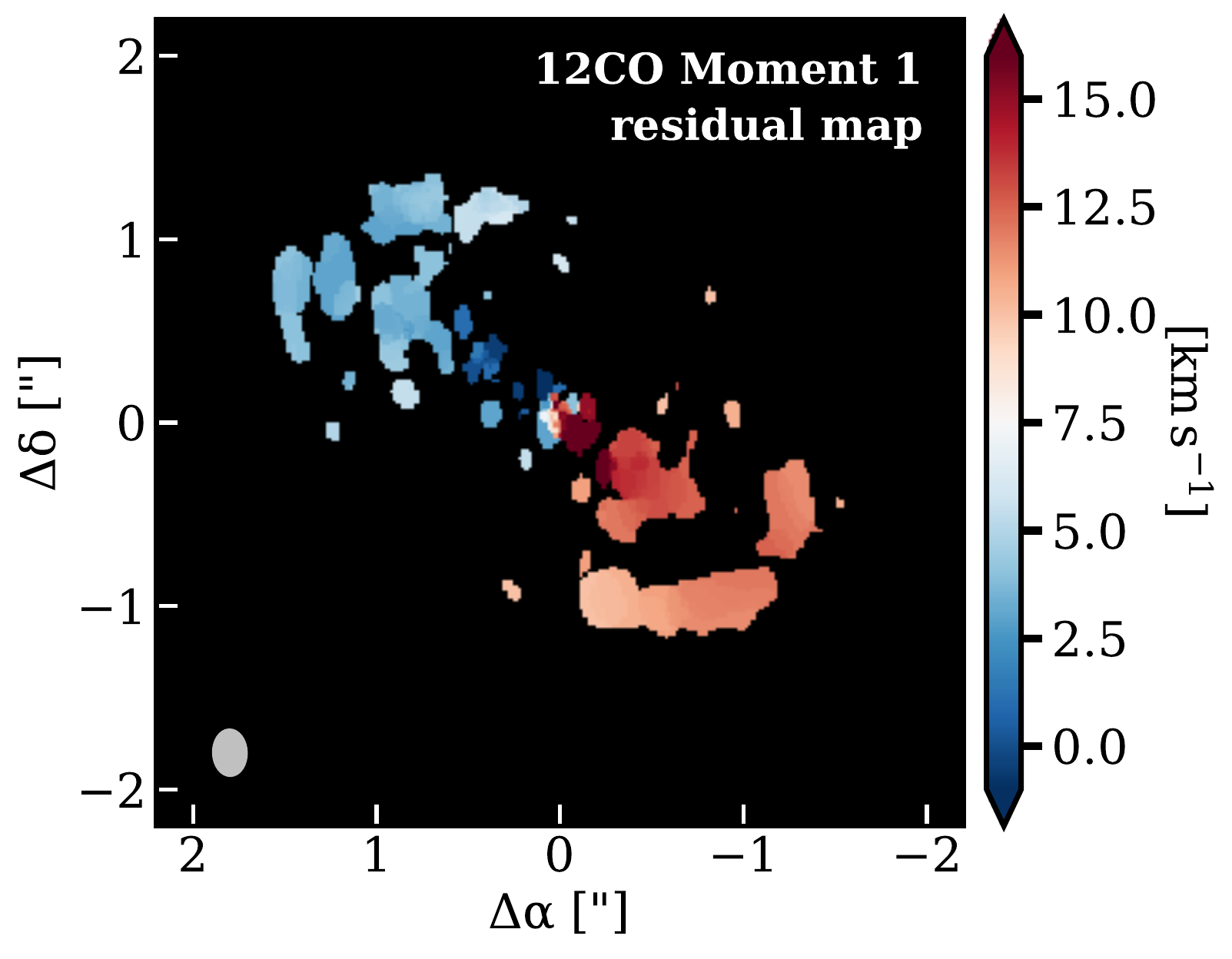} \\
    \caption{\textbf{Upper panels:} The channel maps of $^{12}$CO emission for a limited velocity range of 0--16\,km\,s$^{-1}$, with emission from the adopted model shown in color scale and data emission in [3,4]$\sigma$ contours. \textbf{Lower panels:} The intensity integrated map and velocity map for the residual $^{12}$CO emission, excluding the contaminated velocity range of 6--10\,km\,s$^{-1}$ (symmetric around the systemic velocity), with a  3.5$\sigma$ clip applied. The model and residual image cubes are made with the same \texttt{tclean} parameter as the data, using the model and residual visibilities ($V_{\rm obs}-V_{\rm mod}$), respectively. 
    \label{fig:CO_model_res} }
\end{figure*}

\subsubsection{Non-Keplerian Gas Emission Features} \label{sec:non-Keplerian}
To explore the asymmetric nature of the CO emission, we compare in Figure~\ref{fig:CO_model_res} the observed CO images with the model images (as obtained in Section~\ref{sec:dyn-mass}) channel by channel. This detailed comparison helps us to examine how CO line emission in the V892 Tau system deviates from a symmetric disk model. Excess line emission outside of the Keplerian disk edge is observed between 3--6\,km\,s$^{-1}$ in the blueshifted disk side. While in the opposite side of the disk with redshifted velocities from 9--12\,km\,s$^{-1}$, the Keplerian disk model produces more emission than what data suggest. 
These reveal two main arcs (at 8--10$\sigma$ significance) along the disk outskirts in the collapsed residual maps (lower panels in Figure~\ref{fig:CO_model_res}): one residual arc with positive emission in the northeast (blueshifted) side of the disk,  and another arc with negative emission in the southwest (redshifted) side. The northeast (blueshifted) side therefore appears more extended. The two arcs are close to rotational symmetry by an angle of $\sim$180$\degr$. A gas disk with varying geometry where the inner and outer disk have different inclination and position angles, could be one possible explanation for such residual patterns.

\begin{figure}[!t]
\centering
    \includegraphics[width=0.45\textwidth]{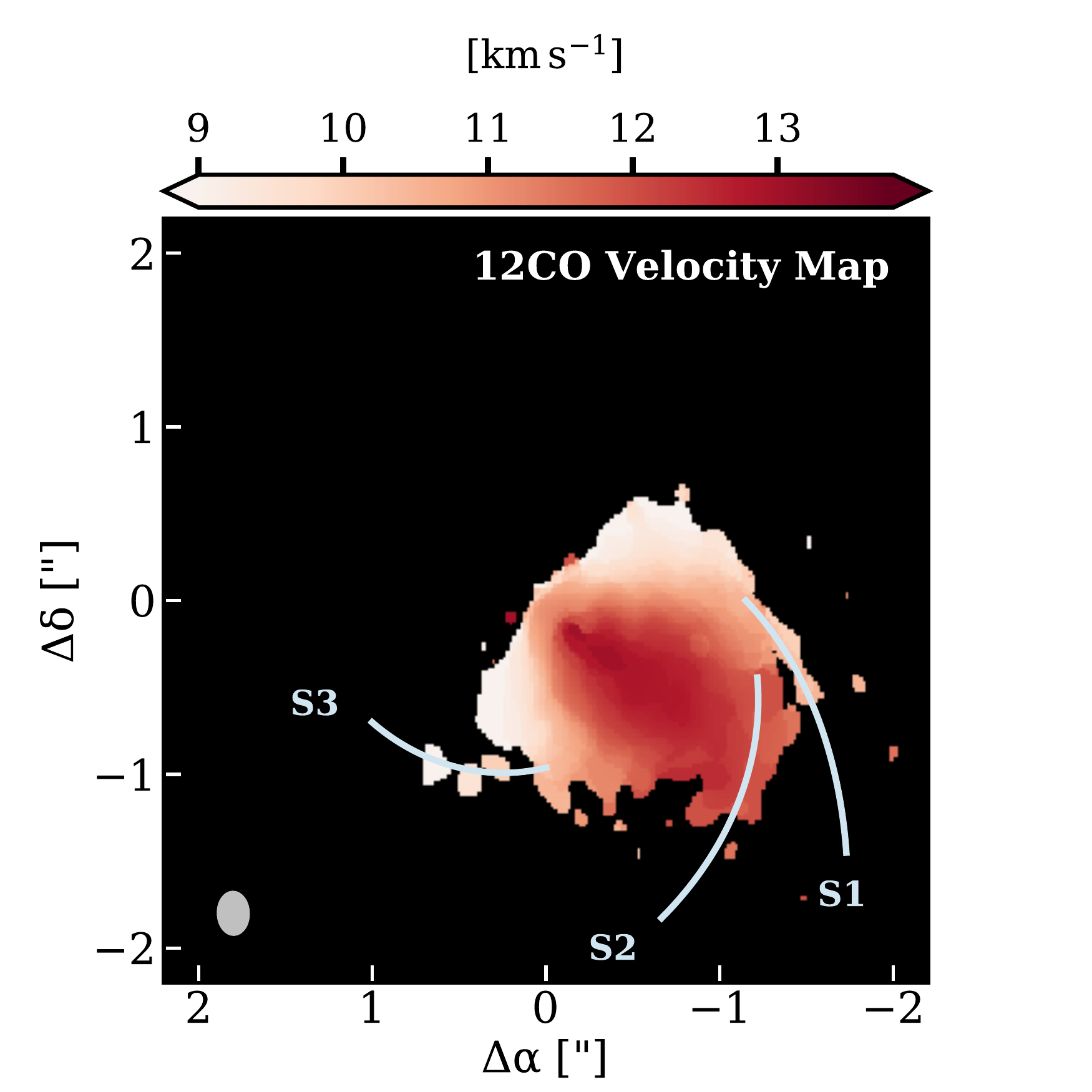} 
    \caption{The velocity map of $^{12}$CO emission in a limited velocity range of 9--13\,km\,s$^{-1}$ to highlight the possible spiral structures associated with the disk. Three proposed spiral curves are overlaid. Deeper CO observations that are sensitive to the large scale gas distribution are required to confirm or rule out these tentative detections of spiral arms.
    \label{fig:spiral} }
\end{figure}

Besides the asymmetry along the disk major axis (between the blue- and red-shifted disk sides), we also observe emission asymmetries along the disk minor axis, especially from the redshifted disk side. From the velocity range of 9--11\,km\,s$^{-1}$, emission from the northern side is more extended compared to its southern counterpart, with faint narrow arcs (3--4$\sigma$ in brightness, spanning $\sim0\farcs5$ in length) sticking out from the Keplerian disk edge. In velocity channels covering 11.5--12.5\,km\,s$^{-1}$, CO emission in the outer disk shows discontinuities and short arcs extending outside the main rotating disk. We also see emission clumps to the southeast from 9--10\,km\,s$^{-1}$, with higher velocity when moving closer to the disk. These features are only detected at relatively low significance, and might be parts of some larger scale spiral structures in the outer disk that our current data are not able to probe (e.g., \citealt{Huang2020_RUlup}). Based on visual inspection, we propose three spiral arms that roughly match the emission features described above, as shown in Figure~\ref{fig:spiral}. Our spiral identifications here are highly inconclusive given the lack of short spacing data.

Other non-Keplerian features are also present, including excess line emission around 100\,au in the 13--14\,km\,s$^{-1}$ velocity range, twisted emission in the outer disk edge at $10.5\,\rm km\,s^{-1}$, and a possible emission bridge connecting the Keplerian wings across the minor axis at 4.5\,km\,s$^{-1}$. All of the emission features discussed above could be better identified and characterized by future,  deeper observations sensitive to a broader range of spatial scales.

% inner gas flow? 

\begin{deluxetable*}{lcccc}
%[!t]
\tabletypesize{\scriptsize}
\tablecaption{V892 Tau Astrometric Catalog\label{tab:astrometric}}
\tablewidth{0pt}
\tablehead{
\colhead{Data} & \colhead{Facility/Wavelength} & \colhead{Separation(mas)} & \colhead{PA($\degr$)} & \colhead{Reference}  \\ 
} 
\colnumbers
\startdata
1996-10-01\tablenotemark{*} & SAO 6m / 2.2$\mu$m & $50.5\pm4$ & 54$\pm$3 & \citet{Smith2005} \\ 
2003-10-05 & SAO 6m / 2.2$\mu$m & $60.4\pm1$ & 59$\pm$1 & \citet{Smith2005} \\ 
2004-09-04 & Keck I / 2.2$\mu$m & $44.2\pm1$ & 79.9$\pm$1 & \citet{Monnier2008}  \\ 
2011-07-15 & VLA / 8mm & $61.3\pm3$ & 61$\pm$3 & this work\\
\hline
\enddata
%\tablecomments{}
\tablenotetext{*}{This early measurement differs significantly from other results, and is not used in the orbital determination in this work.}
%\tablerefs{}
\end{deluxetable*}

\begin{figure*}[!t]
\centering
    \includegraphics[width=0.9\textwidth]{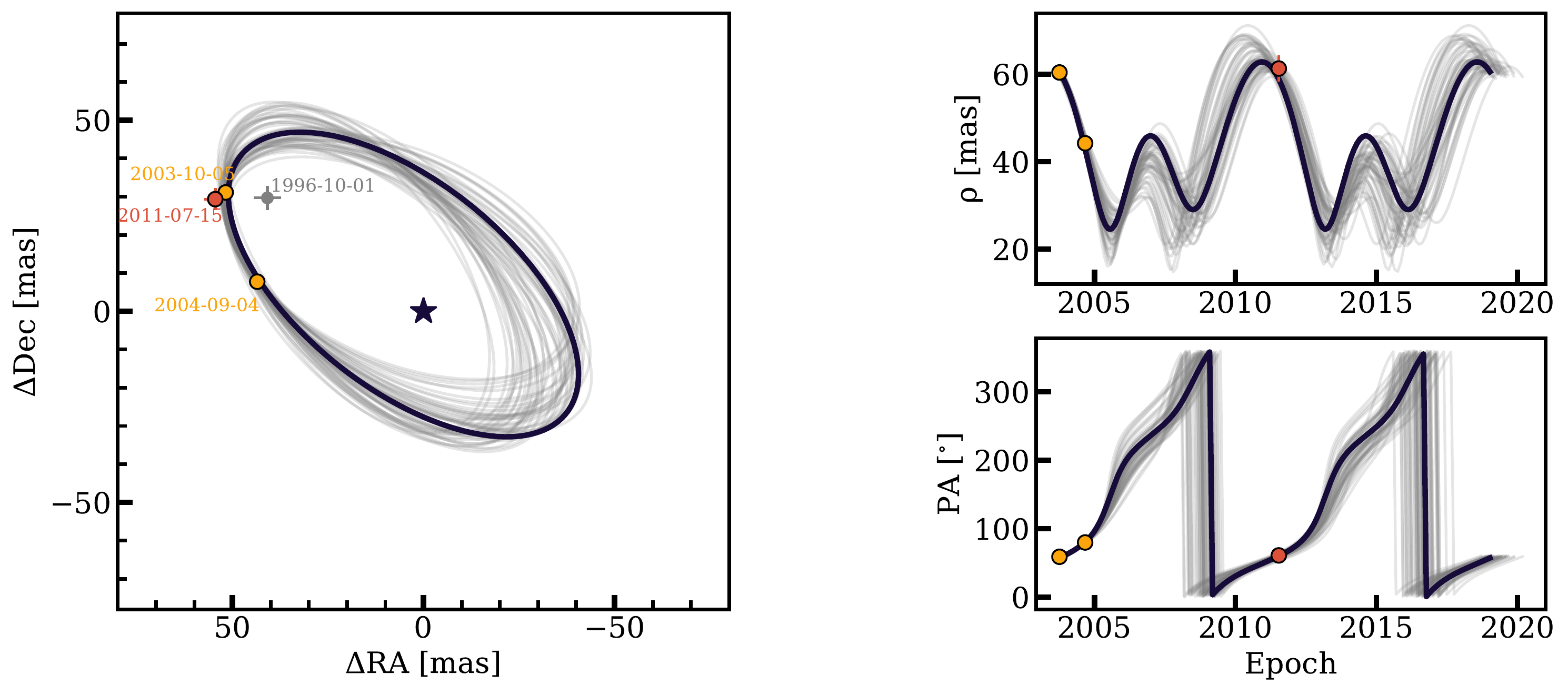} 
    \caption{The binary orbit of the V892 Tau system. The orbital solution with the adopted parameters is shown as a thick curve, with 50 randomly selected  orbits overlaid in light grey to represent the uncertainties. The primary star is placed at (0,0) and the measured positions of the secondary are plotted in colored dots. The right panels show the corresponding binary separations and position angles as a function of time. \label{fig:orbit} }
\end{figure*}

\subsection{Updated Binary Orbit using VLA Astrometry} \label{sec:orbit}
The two bright emission blobs inside the disk cavity at 8\,mm are roughly separated by one beam ($\sim0\farcs05$, Figure~\ref{fig:dust-images}), which is comparable to previous astrometric measurements of the inner binary \citep{Smith2005, Monnier2008}. It is therefore reasonable to assume that they trace the locations of the two stars. In this section, we will take this unique opportunity to measure the binary separation and position angle at the epoch of the VLA observations. With the well-established total stellar mass from models of the gas disk rotation and the source distance from \textit{Gaia}, we then provide an updated binary orbital solution by combining our new astrometric determinations with previous measurements.

The stellar locations are determined by fitting to the VLA visibilities. Given the unresolved nature of the two emission clumps, we employ two point source models, each with three free parameters (flux, centroid position offsets relative to the observed phase center in RA and DEC). Model visibilities are directly written in the form of a constant flux at the observed spatial frequencies with phase center offsets accounted. 
Following the fitting procedure\footnote{The visibility weights ($w_{k}$) in the likelihood calculation are scaled by a factor of 37 to represent the standard deviation of the data visibilities, a known relative scaling issue in the early VLA data calibration. With this correction, the uncertainties of phase center offsets are reduced by a factor of  6--8.} outlined in Section~\ref{sec:dust-model}, the parameter space is sampled with 30 walkers and 10,000 steps per walker with \texttt{emcee} \citep{ForemanMackey2013}. 
The autocorrelation length is on the order of 100 steps. We then use the chains of the last 5,000 steps to derive the best-fit parameters. Our fitting only uses the A array data, as this is the only configuration that separates the two peaks. Meanwhile, since the full VLA dataset, combining A, B, and C array configurations, were taken spanning about half a year, this choice largely reduces the uncertainties of stellar location determinations and the subsequent orbit fitting. 
The measured phase center offsets of the two peaks are $\delta_{\alpha, A}=-0\farcs031$, $\delta_{\delta, A}=-0\farcs024$ and $\delta_{\alpha, B}=0\farcs023$, $\delta_{\delta, B}=0\farcs005$, where star A and B are to the west and east, respectively (see the relative location in Figure~\ref{fig:dust-images} and discussion below). 
Uncertainties in each offset direction are $\sim$0$\farcs$002 (The corner plot of the fitting parameters is shown in the Appendix Figure~\ref{fig:vla_corner}). The derived binary separation is 61$\pm$3\,mas, with a position angle of 61$\pm$3$\degr$, measured from north to the east.  The 8\,mm peak fluxes are 0.566 and 0.606\,mJy for star A and B, respectively, with uncertainties of 0.037\,mJy. The 9.8\,mm data have slightly worse resolution and the two emission clumps are less well-resolved (see Appendix~\ref{sec:vla}). To evaluate the systematic errors on the astrometry, we have fitted the 9.8\,mm A array data in the same way and obtained $\delta_{\alpha, A}=-0\farcs035$, $\delta_{\delta, A}=-0\farcs026$ and $\delta_{\alpha, B}=0\farcs023$, $\delta_{\delta, B}=0\farcs006$  with the same uncertainties of 0$\farcs$002, and fluxes of 0.569 and 0.695\,mJy for star A and B, respectively. The consistency from both wavelengths indicates that the astrometric measurements are robust.

The binarity of V892 Tau was initially proposed by \citet{Smith2005} based on high-resolution speckle imaging at $K$-band ($\lambda=2.2\mu$m), with observations conducted in 1997 and 2003. The two stars were then resolved by Keck observations using near-infrared aperture masking in 2004 \citep{Monnier2008}. These three epochs of observations consistently found slightly brighter emission (by $<20\%$) at 2.2$\mu$m from the southwest component, which was then assigned as the primary star (star A). The binary separations and position angles from these literature measurements are summarized in Table~\ref{tab:astrometric}. The consistent astrometric measurements in 2003 and 2011 (the latter from the VLA observations) suggest an orbital period of $\sim$7--8 years. This estimate assumes that the primary star is located to the southwest at the time when the VLA data were taken. 
However, given the similar stellar properties of the two stars, there could be a 180$\degr$ ambiguity of the position angle for individual measurements. \citet{Monnier2008} preferred a flipped geometry of the earlier 1996 measurement, which resulted in an orbital period of 13.8$\pm$1.5 yr and semi-major axis of 72.4$\pm$6.3 mas (9.7\,au). If we follow this orbital solution, our 2011 measurement should also have a flipped position angle of 61+180$\degr$ so that the primary star is located in the northeast.
In either case, orbital fitting with all four measurements would be problematic. As shown in Figure~\ref{fig:orbit}, the similar projected distances along the declination axis in 1996 and 2011 are not compatible with a stationary orbital solution. We decided to exclude the earlier 1996 measurements, in which emission from the two stars was not fully resolved \citep{Smith2005}. Our final orbital fitting therefore uses the latest three measurements, and assumes the primary star is always located to the southeast. This decision is supported by long-term monitoring of the V892 Tau system that samples more than one complete orbit (A.~Kraus and A.~Rizzuto, private communication).

We derived stellar orbits using the Orbits for the Impatient (OFTI) Bayesian rejection sampling algorithm encoded in the \texttt{orbitize!} software package \citep{Blunt2017, Blunt2020}, which is well suited to constrain orbital elements for systems with data only covering a small fraction of the orbital period. Uniform priors were adopted for the following orbital elements: semi-major axis ($a$) from 0.001 to 10$^7$\,au (in log space), eccentricity ($e$) from 0 to 0.5, inclination ($i_{\rm b}$) from 0 to $\pi$, argument of periastron ($\omega_{\rm b}$) from 0 to 2$\pi$, position angle of the ascending node ($\Omega_{\rm b}$) from 0 to 2$\pi$ (counted from N to E), and time of periastron passage ($\tau$) from 0 to 1. The upper boundary of the prior on $e$ is again suggested by the long-term monitoring of V892 Tau (A.~Kraus and A.~Rizzuto, private communication). 
Experimental fitting with $e\in{[0,1]}$ returns a local peak of low eccentricity, and some possible highly eccentric orbits (see Appendix Figure~\ref{fig:orbit_corner}). 
The source distance and total stellar mass are allowed to vary but are set as Gaussian priors with parallax measurements taken from \textit{Gaia} and the dynamical mass posterior from the modeling of the gas disk rotation in Section \ref{sec:dyn-mass}.

Using the three astrometric measurements, we run a sample of 10$^5$ accepted orbits, which constrain the orbital elements as $a=7.1\pm0.1$\,au, $e=0.27\pm0.1$, $i_{\rm b}=59.3\pm2.7\degr$, and $P=7.7\pm0.2$\,yr as computed from $a$ and $M_{\rm tot}$ following Kepler's third law, which is about half of the orbital period reported in \citet{Monnier2008}.
The lack of radial velocity information introduces an ambiguity of 180$\degr$ in $\Omega_{\rm b}$ and a coupled degeneracy between $\omega_{\rm b}$ and $\Omega_{\rm b}$. Therefore we obtain two families of solutions for $\omega_{\rm b}$ and $\Omega_{\rm b}$ (in unit of degree), as 1) $\Omega_{\rm b}=50.5^{+9.6}_{-8.8}$, $\omega_{\rm b}=179.9^{+44.4}_{-30.3}$, and 2) $\Omega_{\rm b}=230.4^{+9.7}_{-8.9}$, $\omega_{\rm b}=0.2^{+40.8}_{-33.4}$ where the negative $\omega_{\rm b}$ values are taken from $\omega_{\rm b}-180\degr$. Random draws of orbital solutions are plotted in Figure~\ref{fig:orbit}.

\section{Discussion} \label{sec:diss}

\subsection{The Stellar-Disk Architecture of V892 Tau}
Our observations in Section~\ref{sec:results} show that the circumbinary disk around V892 Tau exhibits an inner dust cavity with mm emission peaking at a radius of $\sim0\farcs2$.  This ring of emission is also detected at the longer wavelengths of 8 and 9.8\,mm. The CO gas observations suggest that the disk motion is dominated by Keplerian rotation, in which emission from the northeast side of the disk is blueshifted. These observations suggest a disk geometry with an inclination angle of $i_{\rm d}=54.6\pm1.3\degr$ and a position angle of $\Omega_{\rm d}=53.0\pm0.7\degr$ (these are the averages of the consistent posteriors for the mm continuum and CO gas emission models).
Assuming Keplerian disk rotation, modeling of the CO surface brightness distribution enables an independent determination of the total stellar mass of the V892 Tau binary as $M_{\rm tot}=6.0\pm0.2\,M_{\odot}$, confirming the system is composed of two A-type stars. 
With the total stellar mass and astrometric measurements, including new constraints from high-resolution VLA observations, we are able to measure the binary orbit.

\begin{figure}[!t]
\centering
    \includegraphics[width=0.45\textwidth]{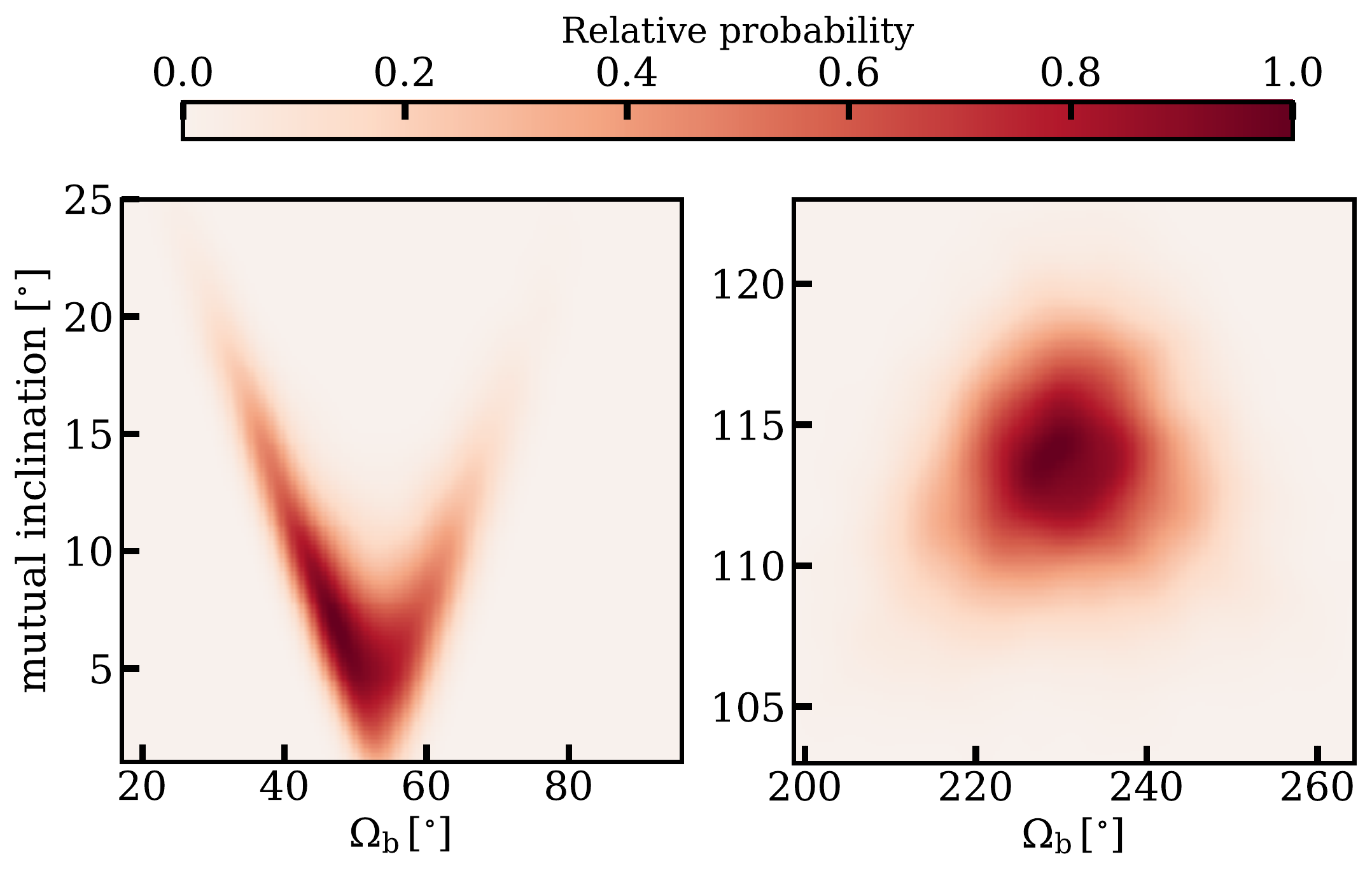} 
    \caption{The marginalized covariances between the mutual inclination of the binary and disk planes ($\Delta$) and the position angle of the binary orbit ($\Omega_{\rm b}$). Two possible solutions are obtained due to the lack of stellar radial velocity information and undetermined near/far side of the disk.   \label{fig:mutual_inc} }
\end{figure}

\subsubsection{Interaction between the Inner Binary and the Circumbinary Disk}
The mutual inclination between the binary and disk orbital planes ($\Delta$) is an important metric for understanding the effect of binary motion on the disk structure and dynamics. This quantity can be expressed as: 
\begin{equation}
    \cos \Delta = \cos i_{\rm d} \cos i_{\rm b} + \sin i_{\rm d} \sin i_{\rm b} \cos(\Omega_{\rm d} - \Omega_{\rm b}).
\end{equation}
For an individual plane, the inclination ($i$) and position angle ($\Omega$) together define its absolute orientation\footnote{We follow the same definition for inclination ($i$) and position angle ($\Omega$) as in Figure~1 of \citet{Czekala2019}.}. For both the V892 Tau binary and its circumbinary disk, their true orientations are currently unconstrained. As we are interested in the relative orientation of the two planes, we could fix the disk plane and vary the angular momentum vector of the binary orbit. We assume that the disk with $i_{\rm d}=54.6\degr$ and $\Omega_{\rm d}=53.0\degr$ has its NW side closer to the observer. Following this definition, due to the 180$\degr$ ambiguity of $\Omega_{\rm b}$, we obtain two possible solutions (see Figure~\ref{fig:mutual_inc}): 1) the two planes are nearly coplanar with $\Delta=8.0\pm4.2\degr$ when $\Omega_{\rm b}=50.5\degr$, 2) the two planes are drastically misaligned by $\Delta=113.2\pm3.0\degr$ (more close to polar orientation) when $\Omega_{\rm b}=230.4\degr$.

Theoretical models of tidal truncation predict that the inner edges ($r_{\rm in}$) of circumbinary disks would be located at 2--3$a$ for binary orbits with moderate eccentricity and when coplanar with their surrounding disks \citep{Artymowicz1994, Hirsh2020}. In the case of misalignment, $r_{\rm in}/a$ would however have a complicated dependence on the mutual inclination, orbital eccentricity, and binary mass ratio. 
Based on the calculations presented in \citet{Miranda2015} for an equal-mass binary, $r_{\rm in}/a$ varies between 1.5 and 3 when $e$ and $\Delta$ exceed some threshold values that depend on disk viscosity and thickness. For both sets of possible $\Delta$ in the V892 Tau system, an $e\sim0.3$ suggests $r_{\rm in}/a$ close to 2, though other numerical simulations suggest wider cavities than these analytical predictions (e.g., \citealt{Thun2017}). The ratio between the dust disk cavity size and the binary orbital semi-major axis is estimated to be 3.4--4, with the boundary values given by ($R_{\rm ring}-\sigma_{\rm in})/a$ and $R_{\rm ring}/a$, respectively.
The dust distribution does not directly trace the disk truncation radius, as the dust grain locations are further regulated by radial drift towards local pressure bumps \citep[e.g.,][]{Andrews2014,Cazzoletti2017}. The inner edge of the gas disk should be located closer to the binary, and is a more appropriate comparison to the model predictions. Indeed, as seen in Figure~\ref{fig:COmaps}, the $^{12}$CO emission peaks inside the dust ring, closer to what models predict.  However, in order to make direct comparisons with the models, future high-resolution gas observations with high excitation lines are required to minimise the cloud contamination.

Initially misaligned (with respect to the binary plane) circumbinary disks orbiting around eccentric binaries will undergo precession and evolve into one of two possible final configurations (e.g., \citealt{Aly2015}). For systems with small initial mutual inclination angles, tidal torques bring the disk plane into alignment with the binary orbital plane on a timescale usually much shorter than the disk viscous timescale \citep{Foucart&Lai2014,Smallwood2019}. Otherwise, the two planes would be driven towards a polar configuration, where the disk plane is perpendicular to the binary orbital plane \citep{Martin2017}. The detailed evolution process and its final outcome depend on specific system characteristics. For example, for a highly eccentric orbit, a very small initial tilt between the binary and disk planes could lead to polar alignment. Following the criteria provided by \citet{Zanazzi2018}, for a binary orbit with $e=0.27$, the mechanism leading to polar alignment will likely operate only when the initial misalignment $\Delta_{\rm init}>66\degr$, whereas the coplanar solution is achievable when $\Delta_{\rm init}<58\degr$. In the presence of eccentricity, $\Delta$ oscillates before reaching a steady state, with its amplitude and the duration of oscillations depending on the disk size, viscosity, and binary eccentricity \citep{Martin2018}. The two orbital solutions for V892 Tau presented here might both be likely, assuming the system is evolving towards the final configuration (coplanar or polar) in the damping process. However, given the short re-alignment timescale, we would favor the coplanar solution.

When radial differential precession occurs, as in the case where the internal communication timescale across the disk is longer than the disk precession timescale, the disk can warp or even break into two distinct parts \citep{Nixon2013,Facchini2013, Dogan2018}. The resulting misalignment between the inner and outer disk could sometimes leave detectable imprints, seen as shadows in scattered light images and gas kinematic features deviating from a purely Keplerian velocity field \citep[e.g.,][]{Casassus2015,Walsh2017, Juhasz2017, Benisty2018, Perez2018, Zhu2019}. Stronger disk warps are expected to develop in the cooler and radially more extended disks \citep{Martin2018}. In the warm circumbinary disk of V892 Tau, the patterned residuals from both the mm continuum emission and CO gas suggest a perturbed disk and the presence of a mild disk warp (e.g., \citealt{Facchini2018}). The asymmetries in the brightness (and presumably gas) temperatures along the disk major axis could be explained if the disk is warped with a more inclined inner region. The southwest part of the inner disk could appear puffed up and intercept more stellar radiation, resulting in a warm plateau and a cooler outer disk. Meanwhile, due to the warp (the inner-outer disk misalignment), the northeast outer disk is directly exposed to more starlight and is thus warmer than the opposite side of the disk at same radial distance (see also e.g., \citealt{Perez2018}).

\subsubsection{Interaction with the Exterior Companion}
The presence of an exterior tertiary companion can potentially affect the global evolution and kinematics of the associated disks. Gravitational interactions tend to truncate the outer disk radii, shorten the disk lifetimes, and reduce the disk masses \citep{Harris2012, Kraus2012, Long2018-cha, Akeson2019}. The close binary V892 Tau is accompanied by a low mass tertiary star at a projected separation of 4$''$ ($\sim$540\,au). The CO gas emission detected in the circumbinary disk around V892 Tau extends out to 200\,au, which is about $40\%$ of the projected stellar separation. Even though the orbital elements of the triple are unknown, this fractional radius suggests that tidal truncation of the outer disk is likely \citep{Artymowicz1994, Breslau2014}. Assuming a coplanar configuration and $r_{\rm out}/a$=0.4, we expect $e_{\rm max}\sim0.2$ for the wide binary orbit \citep{Miranda2015}. 
Tidal truncation is also the possible cause for the non-detection of dust emission around the tertiary, whose disk might have experienced a rapid dispersal.

In case of close stellar encounters, the gas disks can be strongly perturbed, producing asymmetric features like arcs, warps, spiral arms, tidal streams, and diffuse halos \citep{Cuello2020}. Pronounced emission patterns have been observed in several multiple systems recently, e.g., in RW Aur \citep{Cabrit2006,Rodriguez2018}, AS 205 \citep{Kurtovic2018}, HD 100453 \citep{Rosotti2020}, and UX Tau \citep{Menard2020}. The tentative gas spirals we noted in the redshifted velocity channels of the V892 Tau disk, as well as the other non-Keplerian features discussed in Section~\ref{sec:non-Keplerian} (e.g., twisted emission at $10.5\,\rm km\,s^{-1}$ in Figure~\ref{fig:CO_model_res}), are possible signposts of these stellar-disk interactions. However, we note that an external perturber usually triggers the formation of two widely-open spiral arms (e.g., \citealt{Gonzalez2020}). The lack of information on the large scale gas distribution prevents us from drawing a definitive conclusion on the origin of the observed emission features in the outer disk of V892 Tau.

Based on very shallow SMA $^{12}$CO observations that better sample larger spatial scales (as described in Appendix~\ref{sec:sma}), we do not see direct evidence of binary-induced gas spirals in the redshifted side of the disk (Figure~\ref{fig:sma}). 

The extended gas emission at $3\,\rm km\,s^{-1}$, beyond the Keplerian rotating disk and approaching/surrounding the NE tertiary companion, however, suggests the presence of large-scale interactions. 
The diverse outcomes of star-disk encounter interactions depend on the stellar and orbital properties, including mass ratios, orbital inclinations, and periastron distances. For example, a prograde encounter would be more disruptive than a retrograde one, and a non-coplanar encounter is expected to trigger a disk warp \citep{Clarke1993, Breslau2014}. %Xiang-Gruess2016}. 
Future ALMA observations with much better sensitivity on large spatial scales will be necessary to verify the connection between the circumbinary disk and the outer companion, and to establish the possible orbital geometry.

\begin{figure}[!t]
\centering
    \includegraphics[width=0.45\textwidth]{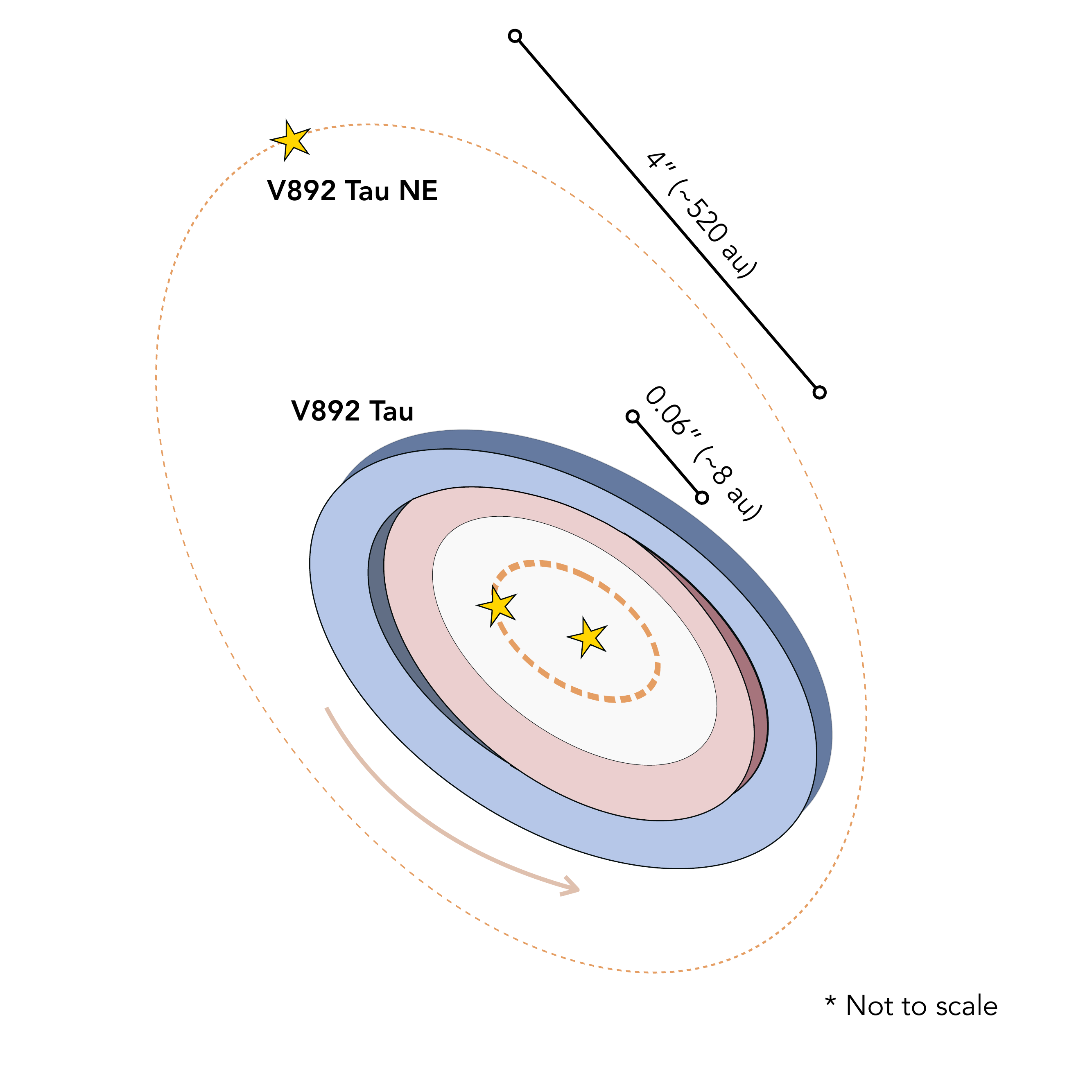} 
    \caption{A schematic view of the V892 Tau system (not-to-scale): the close inner binary, the warped circumbinary disk, and an accompanying star to the NE. The circumbinary disk is assumed to have its western side closer to us, and is suggested to be near coplanar to the inner binary plane. The outer stellar orbit is unknown, and is shown to highlight its potential impact on V892 Tau.    \label{fig:sketch} }
\end{figure}

\subsubsection{A Syetem-level View}
Figure~\ref{fig:sketch} shows the stellar and disk architecture of the V892 Tau system that we are proposing. The disk radial extensions and warped morphology are both consistent with binary-disk interactions. The coplanarity of the disk and binary orbits also agrees with theoretical expectations for rapid orbital re-alignment.

Because the direct formation of a close binary ($a<10$\,au) is inhibited by the opacity limit of fragmentation (e.g., \citealt{Larson1969}), the current stellar configuration of the V892 Tau system must have evolved significantly over the past few Myr. In order to build the inner equal-mass binary, inward migration is expected through the exchange of angular momentum between the binary and the surrounding material, including the gas envelope and the individual circumstellar disks \citep{Bate2002, Tokovinin2020}. The low mass tertiary companion V892 Tau NE at 4$''$ may have also played a critical role in the reconfiguration process (e.g., through exchange interactions, \citealt{Bate2002}).

\subsection{Comparison to Other Circumbinary Disks} \label{sec:others}
Trapping of large dust grains in the radial direction outside an inner cavity is a universal feature predicted and observed in circumbinary disks. However, along the azimuthal direction,  the distributions of mm-sized grains appear differently. The circumbinary disks around V892 Tau (this work) and GG Tau A \citep{Andrews2014} show only marginal azimuthal variations. On the contrary, a horseshoe-shaped dust distribution and substantial east-west emission contrast are observed in HD 142527 \citep{Casassus2013, Boehler2017} and AB Aur\footnote{Though the binary nature of AB Aur has not been fully established, it would be in agreement with the current data (see e.g., \citealt{Poblete2020}).} (\citealt{Tang2017}), respectively. Azimuthal dust trapping in circumbinary disks can occur due to either anticyclonic vortices induced by the Rossby wave instability \citep[e.g.,][]{Zhu2014} or asymmetric gas accumulation at the cavity edge promoted by a high mass ratio binary (e.g., \citealt{Shi2012, Ragusa2017}). In the former case, if the disk of V892 Tau has intermediate-to-high viscosity ($\alpha>10^{-4}$), a long-lived vortex cannot survive. In the latter case, the formation of a long-lasting overdense clump is independent of disk viscosity \citep{Miranda2017, Ragusa2020}. Such a feature constitutes an azimuthal pressure bump that is expected to effectively trap marginally coupled dust grains \citep{vanderMarel2021}. Both V892 Tau and GG Tau A (taking GG Tau Ab1/Ab2 as one component) are high mass ratio binaries but lack azimuthal dust trapping in their disks. The dust rings in the two systems might have low Stokes numbers due to high local gas surface densities (see discussions in \citealt{vanderMarel2021}).

Spiral density waves are also common outcomes of binary-disk interactions, especially when the binary has an eccentric orbit (e.g., \citealt{Price2018}). Such spiral structures have been observed in the inner disk of AB Aur, interior to the mm dust ring \citep{Tang2017}, and in the outer disks of GG Tau A \citep{Keppler2020}, HD 142527 \citep{Fukagawa2006, Christiaens2014}, and possibly in GW Ori \citep{Kraus2020}.  
The orbital solutions derived here for the V892 Tau binary prefer a non-zero eccentricity.  While a disk warp is suggested by the CO gas emission, spiral structures are not clearly identified. Faint spirals in the V892 Tau disk are difficult to identify from the bright Keplerian-dominated disk with our shallow observations. The exterior stellar companion could also largely complicate the disk kinematics. With the improved stellar and disk properties for the V892 Tau system, future hydrodynamical simulations would be highly desirable to explore the potential distributions of spiral structures, and to confirm or rule out the proposed disk warp.

\subsection{Planet Formation in V892 Tau}
Planets have been found in both circumstellar and circumbinary configurations \citep{Martin2018haex.bookE.156M}.
If they formed early, such planets may not survive the binary orbital decay process expected for close pairs (e.g., \citealt{Munoz2015}). However, once a stable stellar configuration is reached, subsequent planet formation can proceed if there is still sufficient disk material. The dust disk of V892 Tau is among the most massive Class II disks, with dust mass (see Section~\ref{sec:dust-model}) comparable to Class 0/I disks and sufficient to account for the mass of solids in most exoplanetary systems \citep{Tychoniec2020}. The total disk mass, as high as $\sim0.06\,M_{\odot}$, if assuming a gas-to-dust ratio of 100, is however only $\sim1\%$ of the central stellar mass. Therefore, gravitational instability may not be a feasible mechanism to form planets. Nevertheless, the abundant pebble mass and high dust density around the pressure bump make the conditions in the disk ring favorable of planet formation.

\section{Summary} \label{sec:summ}
We have presented high resolution observations of the circumbinary disk around V892 Tau at 1.3 and 8\,mm, as well as CO $J=2-1$ gas emission. These observations are used to provide an updated view of the stellar and disk properties, and to examine the binary--disk interactions. 
Our main results are as follows: 
\begin{enumerate}
\item \textit{the stellar system:} The total stellar mass of V892 Tau has been calculated from the CO gas disk rotation to be $6.0\pm0.2\,M_{\odot}$ and verified in binary orbital fits, consistent with its nature as two A type stars. Using the new astrometric measurements from the non-dust emission with 8\,mm VLA observations and previous near-infrared measurements, we constrain the binary orbit to have $a=7.1\pm0.1$\,au, $P=7.7\pm0.2$\,yr, and $e=0.27\pm0.1$. This close binary is also accompanied by an M3 star located 4$''$ to the northeast. 

\item \textit{the circumbinary disk:} The dust disk of V892 Tau has an inner cavity with mm emission peaking around $0\farcs2$. This dust ring is radially asymmetric with $\sigma_{\rm out}/\sigma_{\rm in}=2.26$, and shows slight azimuthal variations. The gas disk as traced by CO observations is very flat in the vertical direction with $z/r\sim0.1$, but exhibits substantial azimuthal asymmetries on top of the main Keplerian velocity field. The CO gas emission suggests some mild misalignment between the inner and outer disk, in which the southwest inner disk is more tilted and therefore receives more stellar radiation that leads to enhanced heating. In addition, the disk around V892 Tau is massive enough to still form giant planets.

\item \textit{binary -- disk interactions:} The radial extension of the circumbinary disk and the asymmetric dust disk ring morphology are both consistent with tidal truncation. Although lacking information on the true binary and disk orientations, we favor a near coplanar orbital configuration (with mutual inclination $\Delta=8.0\pm4.2\degr$), given the expected short re-alignment timescale. The observed gas asymmetries could be explained by the interaction between the eccentric binary and the misaligned disk ($\Delta_{\rm init}>0$). The tentatively detected gas spirals in the outer disk are likely caused by interactions with the exterior stellar companion. 

\end{enumerate}

The V892 Tau system serves as an interesting laboratory to investigate the details of star--disk interactions. Our derived binary and disk orientations should be further confined with radial velocity monitoring and high resolution gas observations.  
Future observations sensitive to large scale gas distributions would also be valuable to establish the connections with and influence of the tertiary star.  These observational results, when combined with hydrodynamical simulations, would significantly advance our knowledge of the formation and evolution of multiple stars and their accompanying planetary systems. 

\paragraph{Acknowledgments}
We thank the anonymous referee for comments that improved this paper.
F.L. thank Adam Kraus and Aaron Rizzuto for sharing their orbital solution of the V892 Tau system, and Charlie Qi for helping reducing the SMA data. F.L. is grateful to Karin \"Oberg and Wei Zhu for helpful discussions. 
F.L. and R.T. acknowledge support from the Smithsonian Institution as the Submillimeter Array (SMA) Fellow. L.P.\ acknowledges support from ANID project Basal AFB-170002 and from ANID FONDECYT Iniciaci\'on project \#11181068.
E.R. acknowledges financial support from the European Research Council (ERC) under the European Union's Horizon 2020 research and innovation programme (grant agreement No 681601). J.M.C. acknowledges support from the National Aeronautics and Space Administration under grant No.~15XRP15\_20140 issued through the Exoplanets Research Program.

This paper makes use of ALMA data of 2013.1.00498.S. ALMA is a partnership of ESO (representing its member states), NSF (USA), and NINS (Japan), together with NRC (Canada), MOST and ASIAA (Taiwan), and KASI (Republic of Korea), in cooperation with the Republic of Chile. The Joint ALMA Observatory is operated by ESO, AUI/NRAO, and NAOJ. The VLA is run by the National Radio Astronomy Observatory, a facility of the National Science Foundation operated under cooperative agreement by Associated Universities, Inc. The Submillimeter Array is a joint project between the Smithsonian Astrophysical Observatory and the Academia Sinica Institute of Astronomy and Astrophysics and is funded by the Smithsonian Institution and the Academia Sinica. 

\facilities{ALMA, VLA, SMA}

\software{\texttt{AstroPy}~\citep{Astropy2013},
\texttt{bettermoments}~\citep{Teague2018_bettermoments},
\texttt{eddy}~\citep{eddy},
\texttt{emcee}~\citep{ForemanMackey2013}, 
\texttt{CASA} (v4.5.0, v5.6.0; \citealt{McMullin2007}),
\texttt{Numpy}~\citep{harris2020array},
\texttt{matplotlib}~\citep{Hunter2007},
\texttt{orbitize!}~\citep{Blunt2017, Blunt2020}}

%\pagebreak
\appendix

\section{The continuum images at 8.0 and 9.8\,mm} \label{sec:vla}
\begin{figure*}[!t]
\centering
    \includegraphics[width=0.32\textwidth]{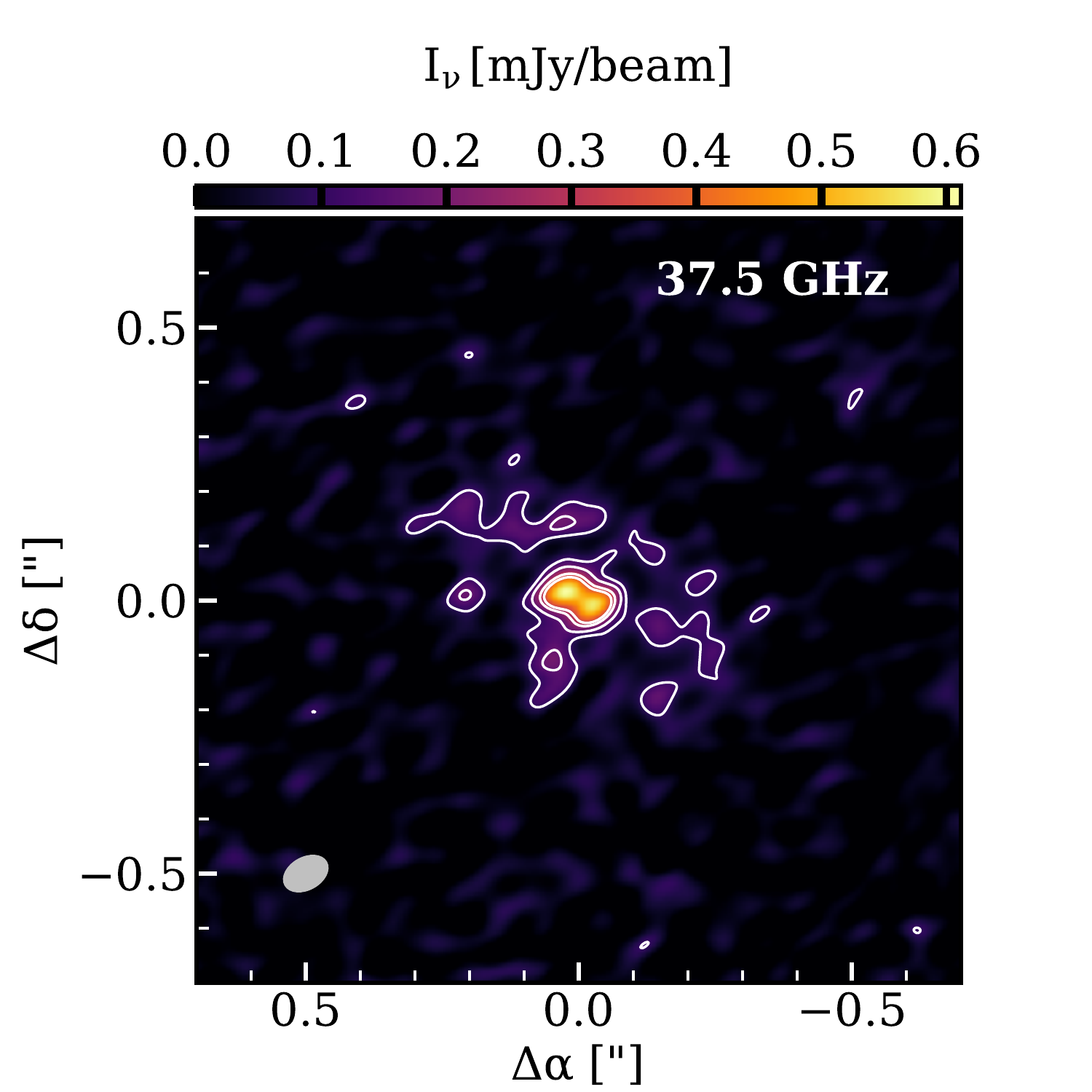}
    \includegraphics[width=0.32\textwidth]{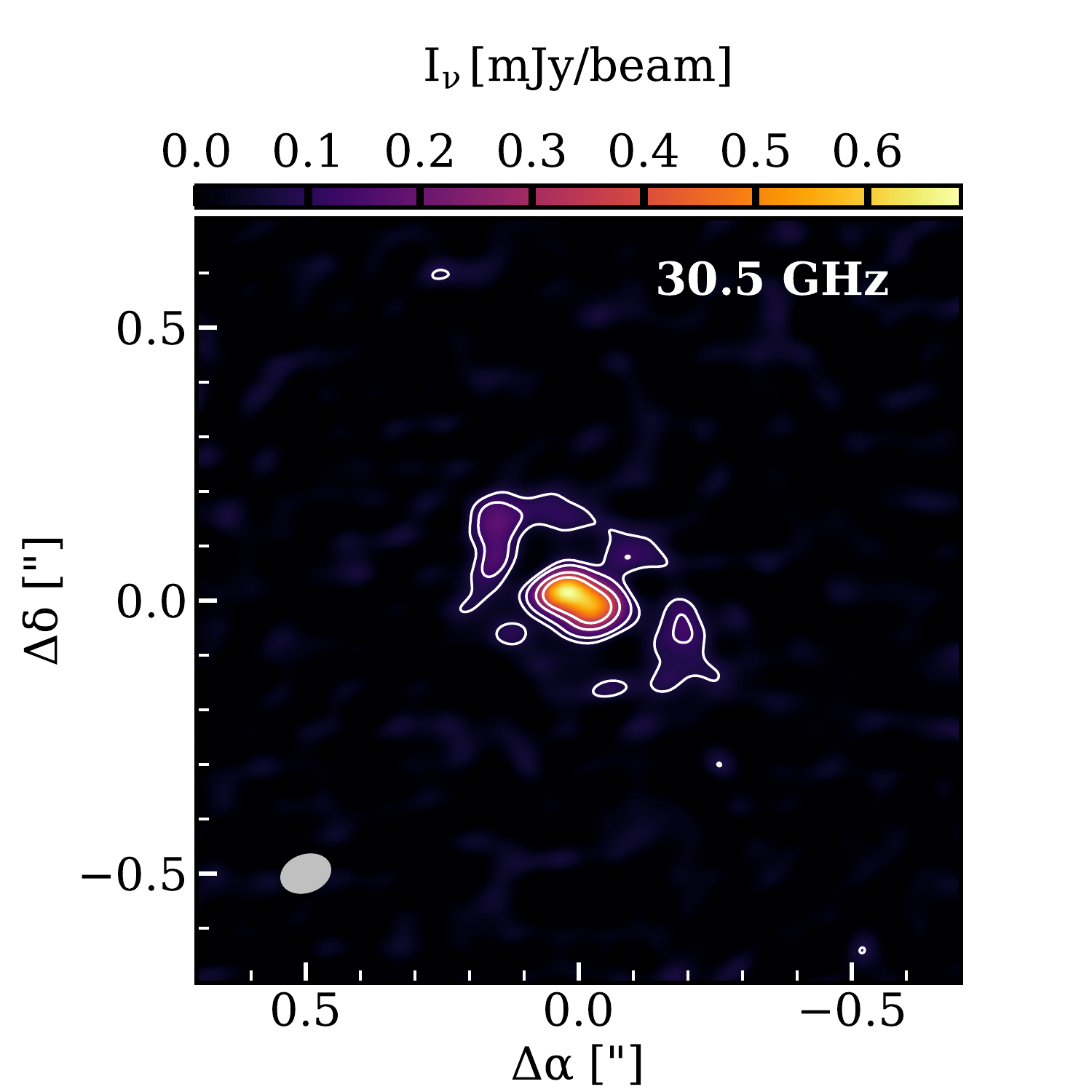}
    \includegraphics[width=0.32\textwidth]{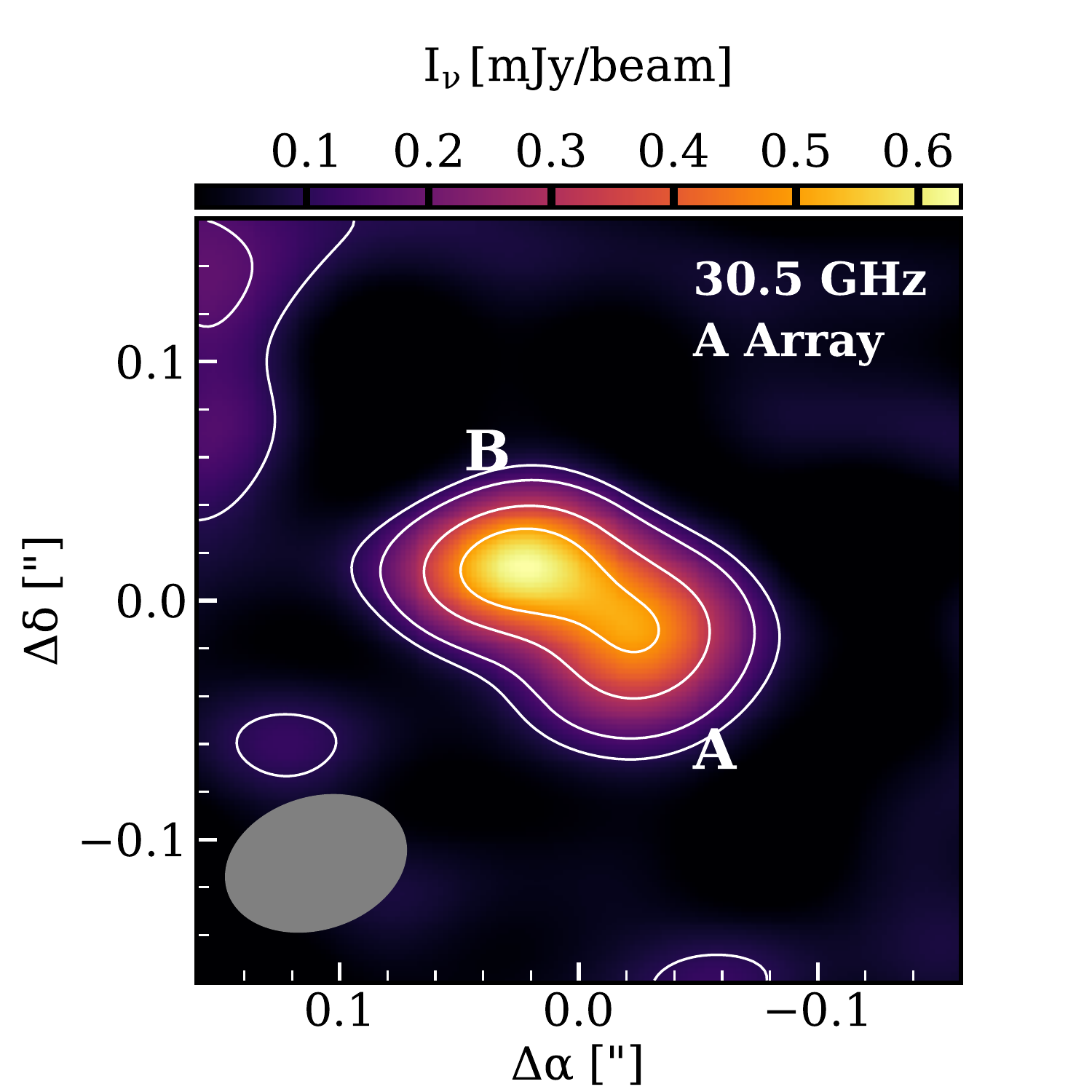} \\
    \caption{Continuum images at 8.0\,mm (left) and 9.8\,mm (middle), with three antenna configurations combined; The zoom-in view for only A array data at 9.8\,mm (right), highlighting the emission close to the two stars.  The beam sizes are displayed in the lower left corners.  \label{fig:dust-images-30ghz} }
\end{figure*}

The continuum images at 8.0\,mm (37.5GHz) and 9.8\,mm (30.5GHz) are shown in Figure~\ref{fig:dust-images-30ghz}. The overall emission distribution is very similar at both wavelengths, but with a fainter dust ring at 9.8\,mm. With the slightly larger beam size at the longer wavelength, emission from the two stars are not well resolved even for the A array data at 9.8\,mm. Nevertheless, the 9.8\,mm data help us constrain the nature of the bright emission close to the stars and provide a systematic check on the measurements of the relative stellar positions.

\section{Model for the continuum emission at 1.3\,mm}
\begin{figure*}[!t]
\centering
    \includegraphics[width=0.39\textwidth]{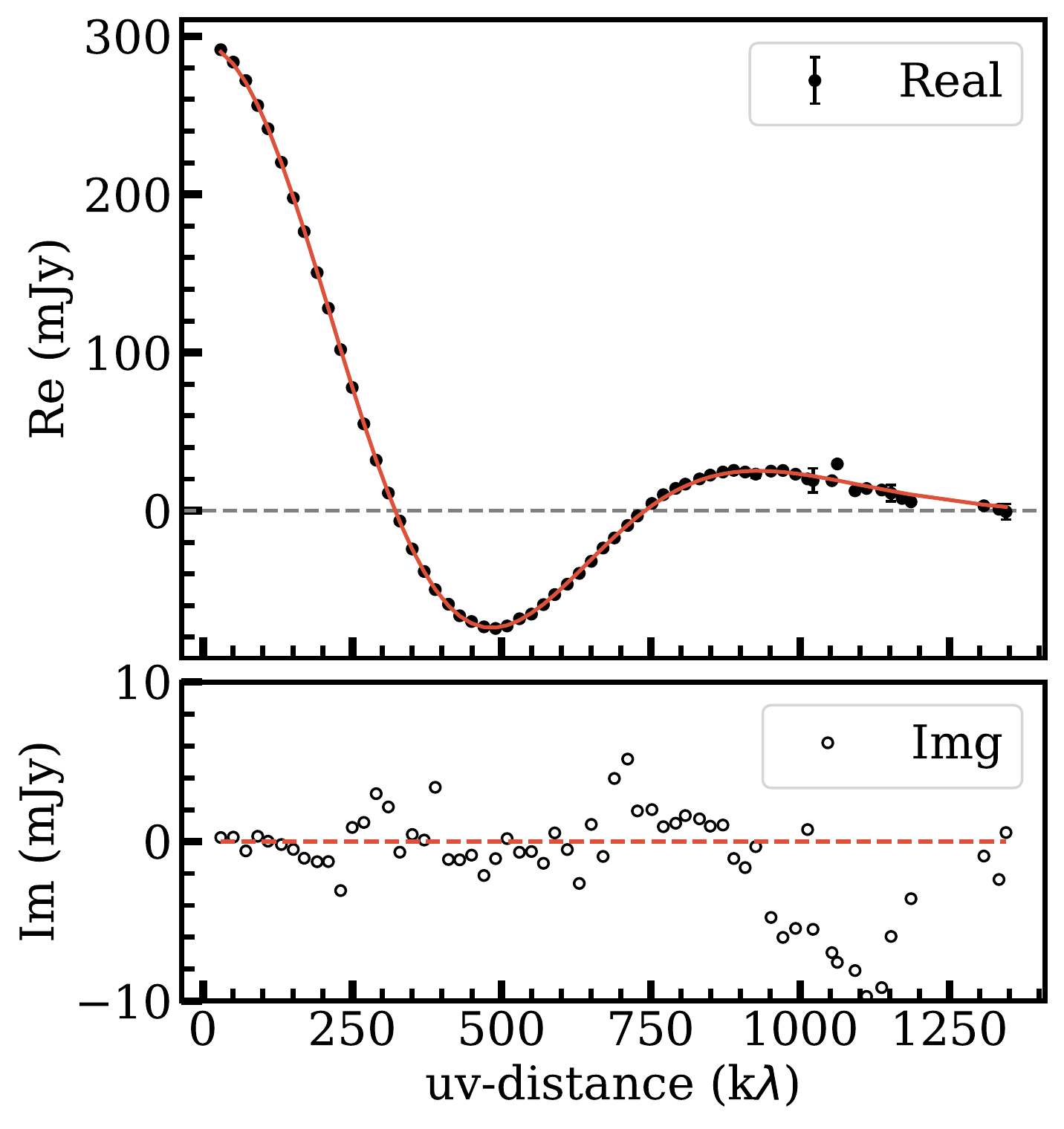}
    \includegraphics[width=0.5\textwidth]{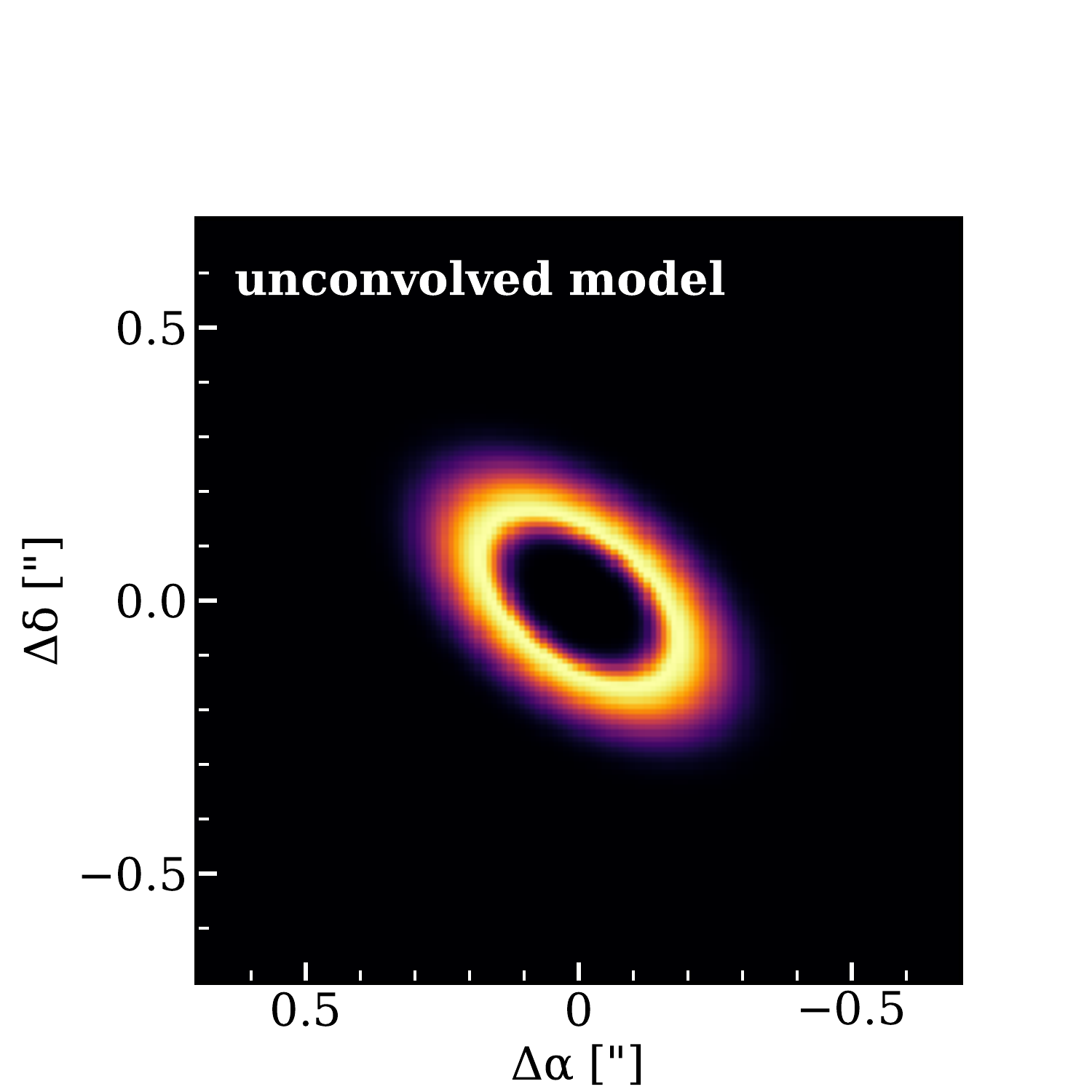} \\
    \caption{\textbf{Left:} The data and model comparison of the binned and deprojected visibility profile. The best-fit axisymmetric model is shown in red; \textbf{Right:} The unconvolved Gaussian ring model with truncated inner disk edge ($\sigma_{\rm out}/\sigma_{\rm in}=2.26$). \label{fig:alma_model} }
\end{figure*}

The unconvolved model image and the comparison with data in the deprojected visibility profile are shown in Figure~\ref{fig:alma_model}. The mismatch at long baselines suggests the presence of low-level asymmetric emission that is not accounted in our axisymmetric model.

\section{CO channel maps from ALMA observations and models}
\begin{figure*}[h]
\centering
    \includegraphics[width=0.9\textwidth]{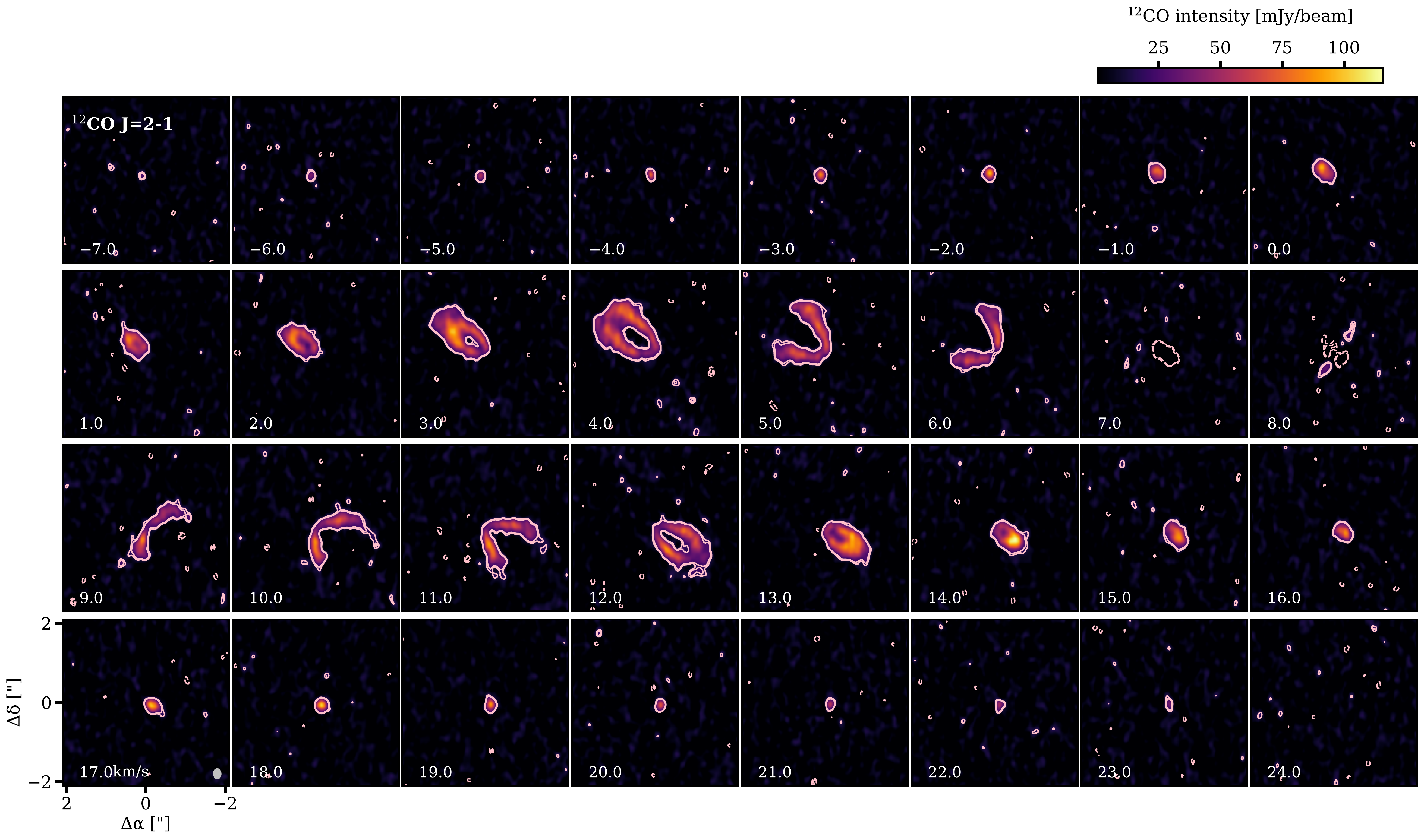} 
    \caption{The full velocity range of channel maps of $^{12}$CO $J=2-1$ emission in V892 Tau. The contours are at [3,4]$\sigma$ levels, with dashed contours for negative fluxes. The velocity is given in the left corner of each panel. \label{fig:12co-full-channel} }
\end{figure*}

\begin{figure*}[h]
\centering
    \includegraphics[width=0.9\textwidth]{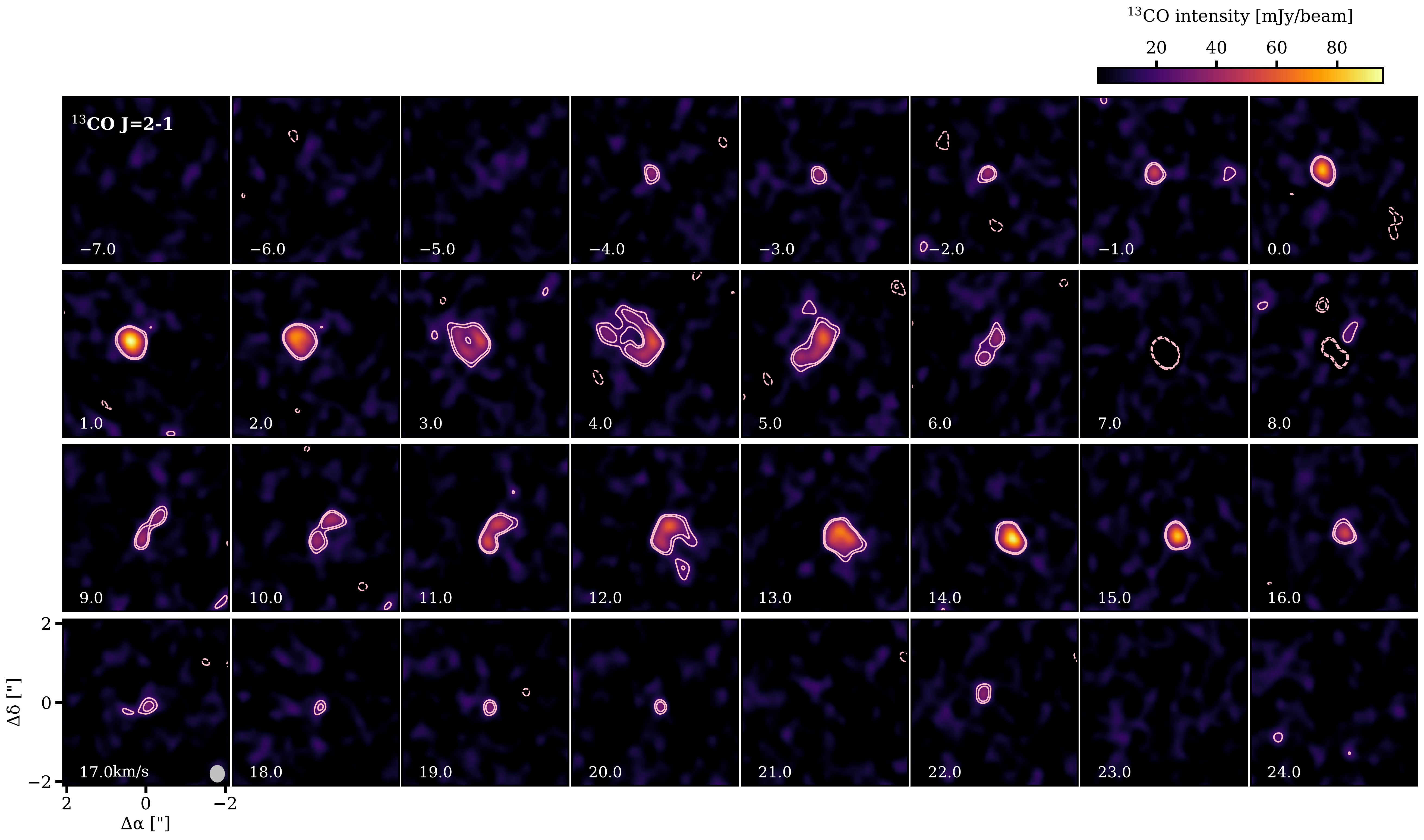} 
    \caption{As Figure~\ref{fig:12co-full-channel}, but for $^{13}$CO $J=2-1$ emission. \label{fig:13co-full-channel} }
\end{figure*}

\begin{figure*}[h]
\centering
    \includegraphics[width=0.9\textwidth]{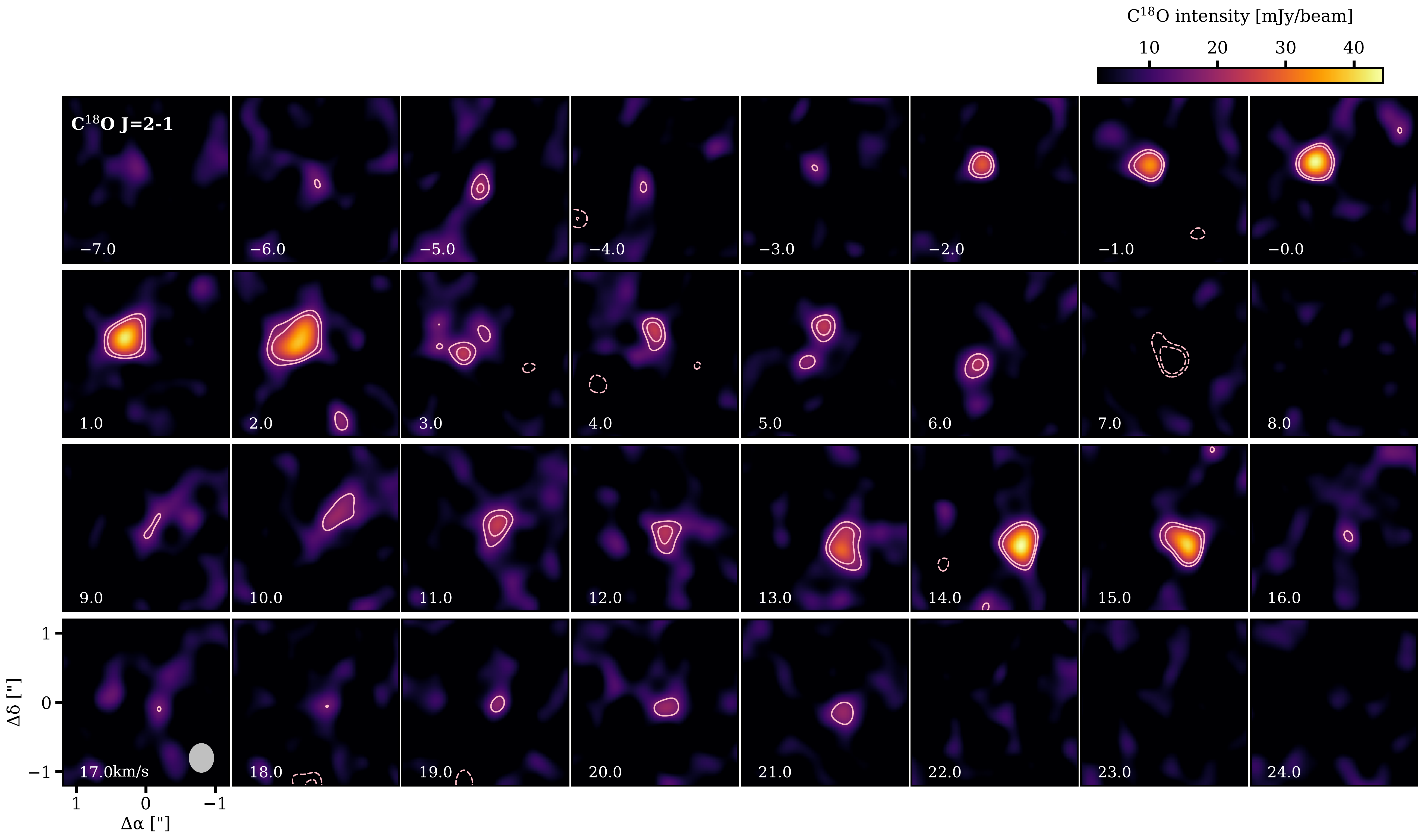} 
    \caption{As Figure~\ref{fig:12co-full-channel}, but for C$^{18}$O $J=2-1$ emission.  \label{fig:c18o-full-channel} }
\end{figure*}

We show in Figures~\ref{fig:12co-full-channel}, \ref{fig:13co-full-channel}, and \ref{fig:c18o-full-channel} the channel maps for $^{12}$CO, $^{13}$CO, and C$^{18}$O $J=2-1$ line emission, respectively, in the full velocity range from -7 to 24\,km\,s$^{-1}$ .  Some of the non-Keplerian emission features discussed in Section~\ref{sec:non-Keplerian} are also visible in $^{13}$CO. 

\begin{figure*}[h]
\centering
    \includegraphics[width=0.9\textwidth]{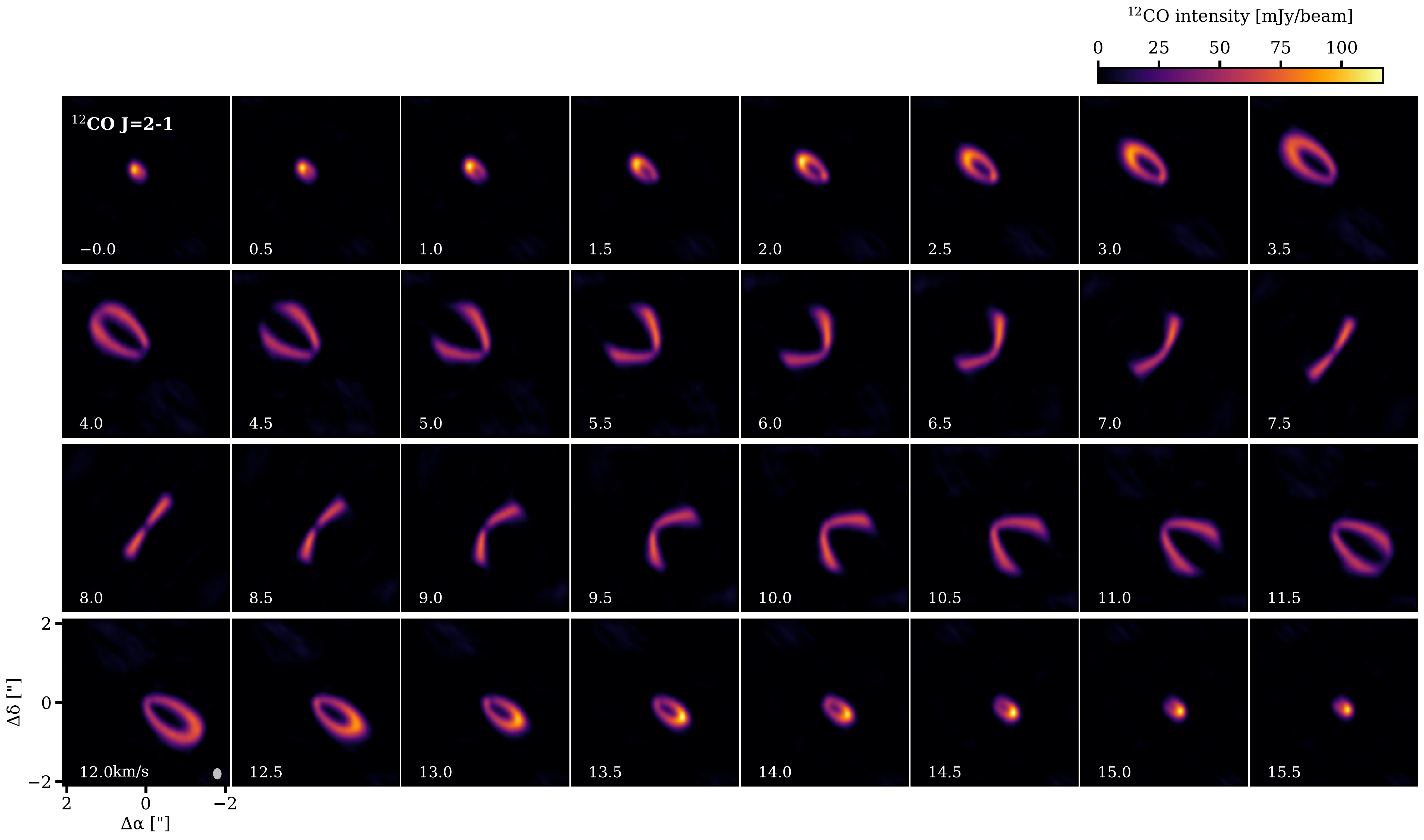} \\
    \includegraphics[width=0.9\textwidth]{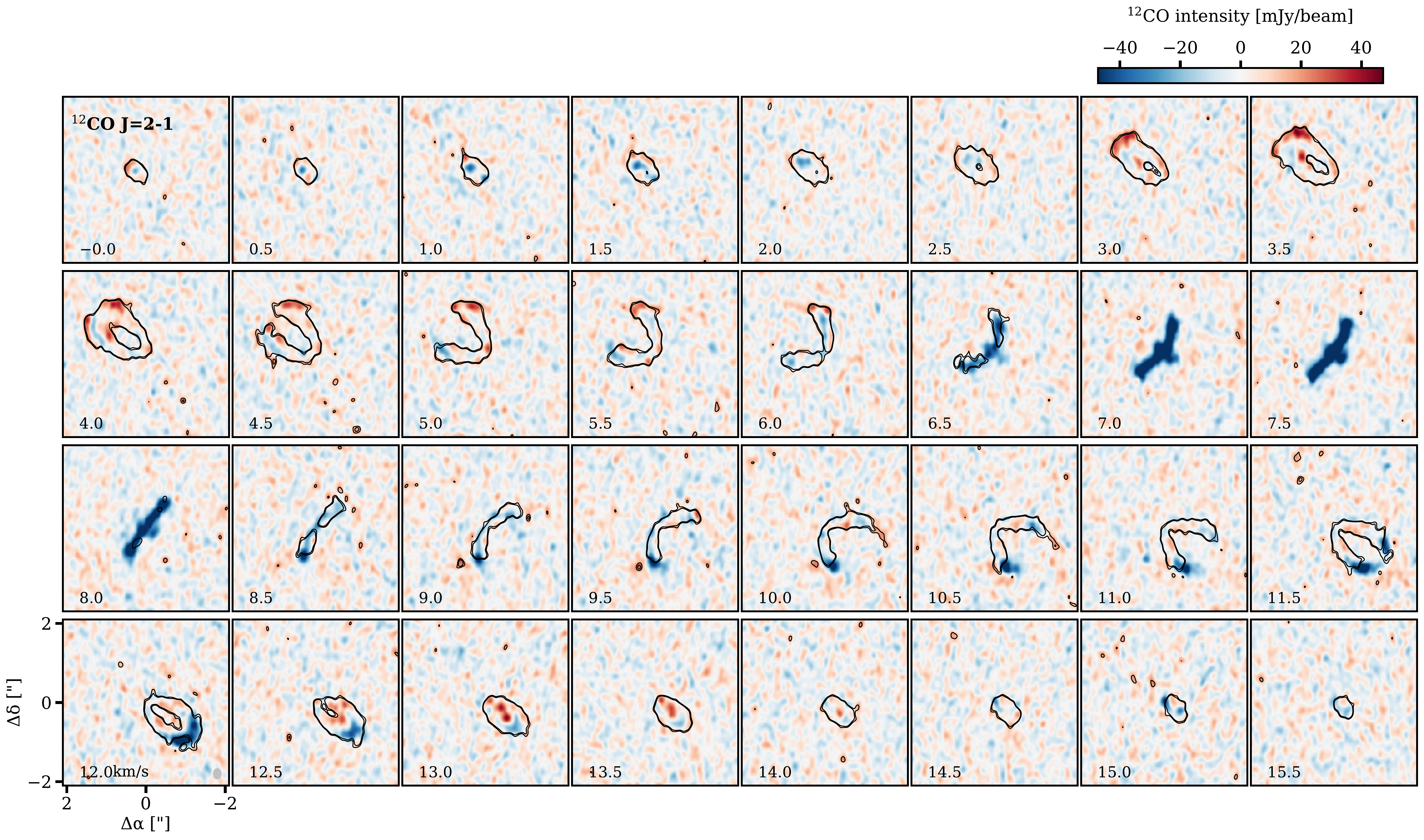} \\
    \caption{The model (upper panels) and residual (lower panels) channel maps of the $^{12}$CO emission. The color scaling in the residual maps is chosen to highlight both the positive and negative emission. In the residual maps, black contours show the data signal at the 3 and 4$\sigma$ levels. \label{fig:12co-model-channel} }
\end{figure*}

The $^{12}$CO channel maps from the Keplerian model and the associated residuals are presented in Figure~\ref{fig:12co-model-channel}.

\section{Corner maps for orbital fits}
\begin{figure*}[!t]
\centering
    \includegraphics[width=0.8\textwidth]{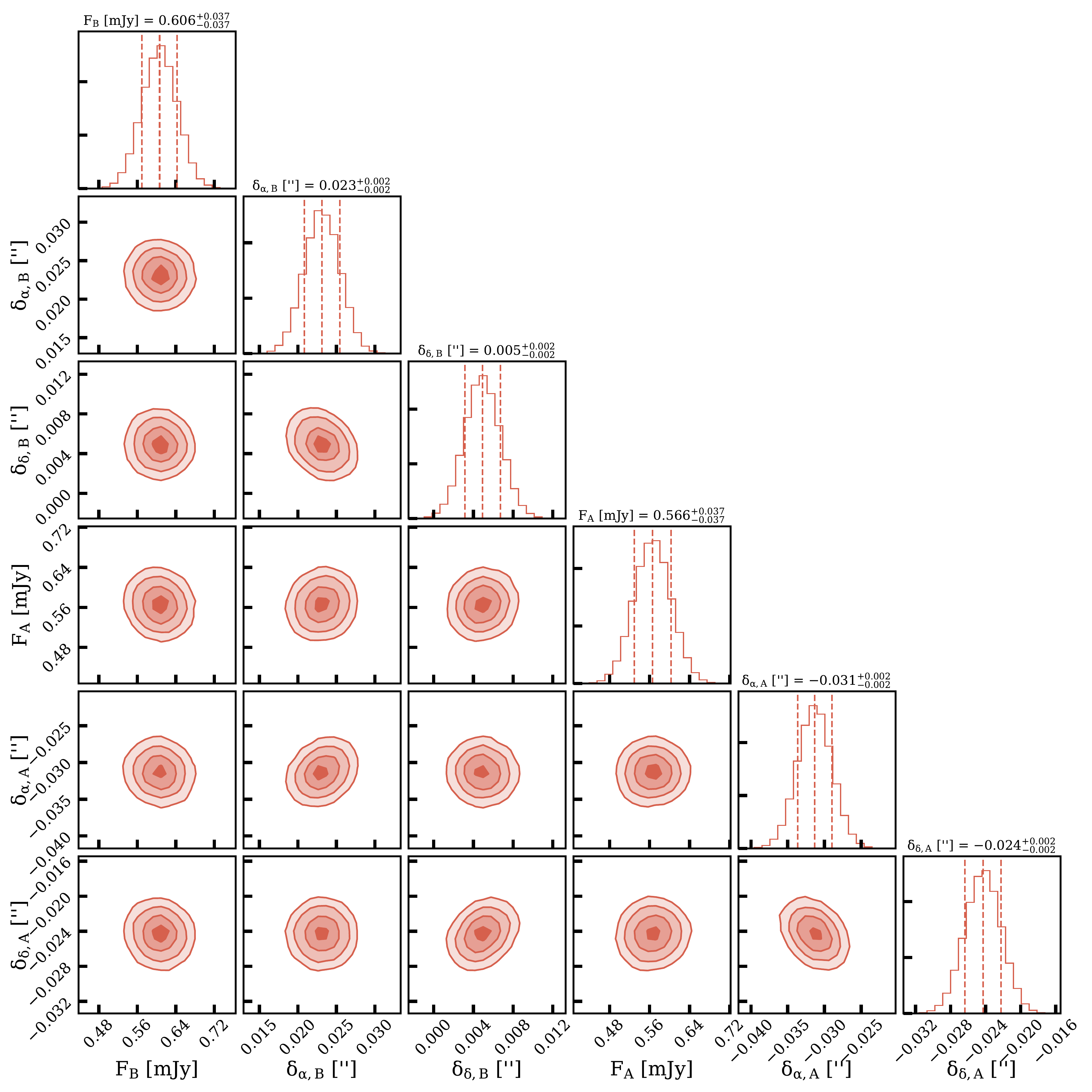} 
    \caption{Posterior distributions of the visibility fitting to the VLA 37\,GHz data with two point source models. The adopted value and the associated uncertainties for each parameter are labeled out. \label{fig:vla_corner} }
\end{figure*}

We show in Figure~\ref{fig:vla_corner} the fitting results to the VLA 37\,GHz A configuration data. The relative astrometry was determined with the locations of the two point sources and then used in the binary orbit fitting.  

\begin{figure*}[!t]
\centering
    \includegraphics[width=0.99\textwidth]{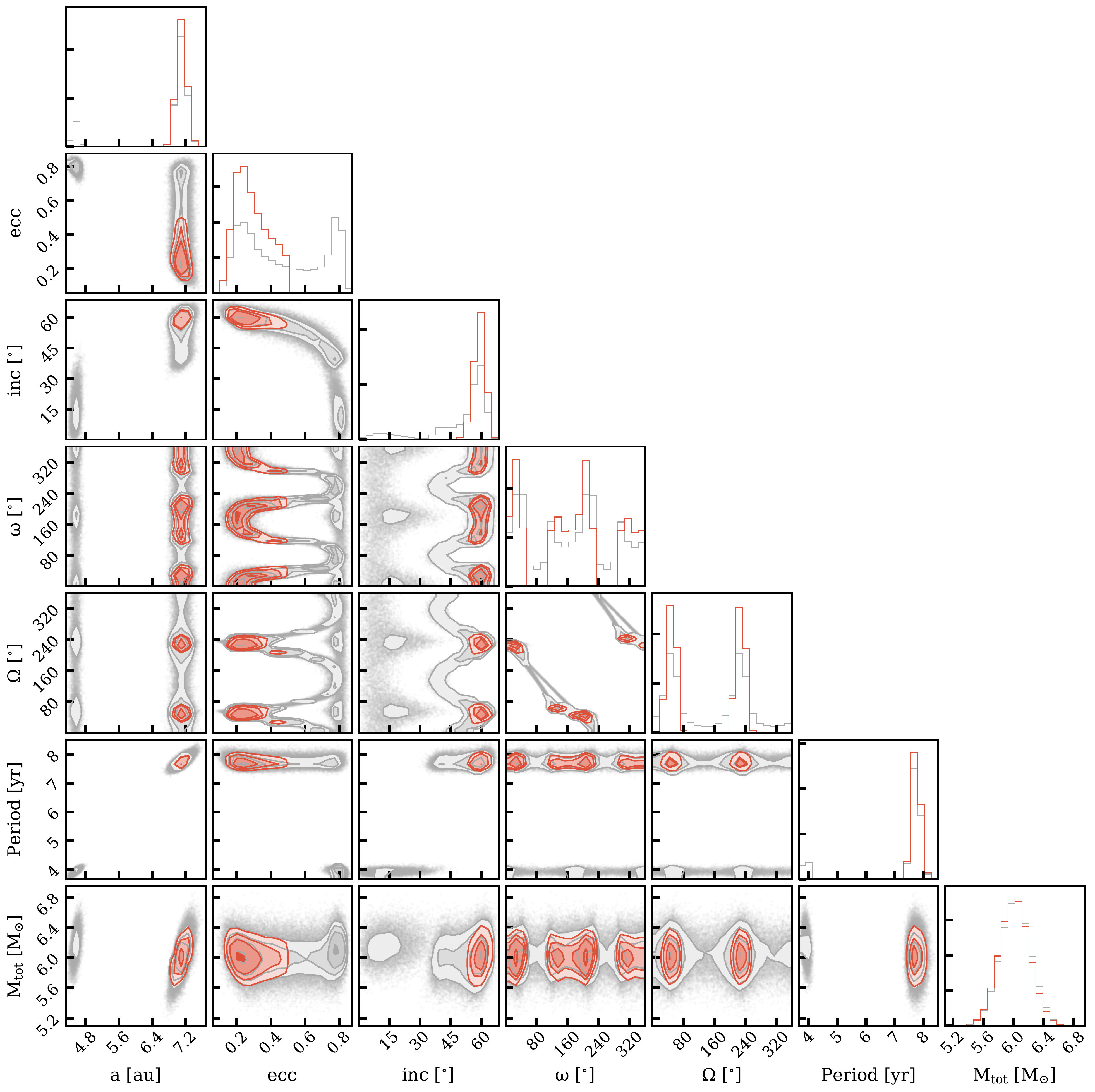} 
    \caption{Posterior distributions of the binary orbit fit parameters. Results shown in red are for fitting with a tight prior on eccentricity (uniform distribution from 0 to 0.5), while the grey results are for the full prior range.  \label{fig:orbit_corner} }
\end{figure*}

Figure~\ref{fig:orbit_corner} shows the posterior distributions of the orbital elements from 10$^5$ accepted orbits for the V892 Tau binary, made with the \texttt{orbitize!} package \citep{Blunt2017}.

\section{SMA CO gas observations for V892 Tau} \label{sec:sma}
In order to establish any connections between the circumbinary disk of V892 Tau and the V892 Tau NE star, we obtained short observations with the Submillimeter Array (SMA) in 2020 November (Project ID: 2020A-S067). These observations in the subcompact antenna configuration provide an angular resolution of $4\farcs2\times3\farcs5$, comparable to the projected separation of the V892 Tau binary and the tertiary companion (V892 Tau NE).  This shallow observation only reaches a sensitivity of $\sim3$\,K in a velocity channel of $1\rm \,km\,s^{-1}$ for the $^{12}$CO $J=2-1$ emission. The large scale ($\sim$40$''$) cloud distributions from 6--9\,km\,s$^{-1}$ are also revealed from these SMA data.

\begin{figure*}[!t]
\centering
    \includegraphics[width=0.99\textwidth]{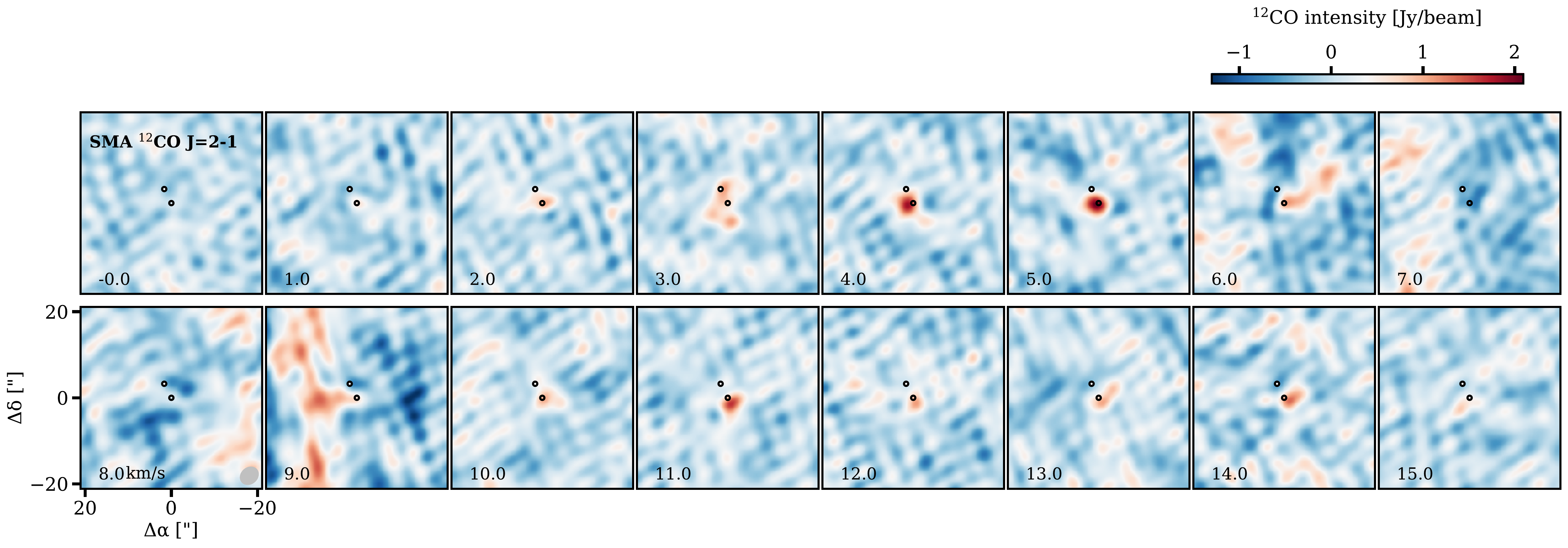} 
    \caption{The channel maps of $^{12}$CO $J=2-1$ emission for the V892 Tau system from SMA observations. The velocity is given in the left corner of each panel. The V892 Tau binary and the 4$''$ NE star locations are marked by black circles. 
    \label{fig:sma} }
\end{figure*}

\bibliography{ms}{}
\bibliographystyle{aasjournal}


\end{CJK*}
\end{document}